\documentclass[useAMS,usenatbib,SFB@referee]{mn2e}

\def\lesssim{\mathrel{\hbox{\rlap{\hbox{\lower4pt\hbox{$\sim$}}}\hbox{$<$}}}}
\def\gtrsim{\mathrel{\hbox{\rlap{\hbox{\lower4pt\hbox{$\sim$}}}\hbox{$>$}}}}
\def\geqq{\mathrel{\hbox{\rlap{\hbox{\lower4pt\hbox{$=$}}}\hbox{$>$}}}}

\usepackage[dvips]{graphicx}

\input epsf

\graphicspath{{figuras/}}       

\title[Local Star formation triggered by SN]{Local Star formation triggered by SN shocks in magnetized diffuse neutral clouds}

\author[M. R. M. Le\~ao et al.]{M. R. M. Le\~ao$^{1}$\thanks{mrmleao@astro.iag.usp.br}, E. M. de Gouveia Dal Pino$^{1}$\thanks{dalpino@astro.iag.usp.br}, D. Falceta-Gon\c{c}alves$^{2,3}$\thanks{diego.goncalves@unicsul.br},\newauthor
C. Melioli$^{1,4}$\thanks{cmelioli@astro.iag.usp.br}, F. G. Geraissate$^{1}$\thanks{geraissate@astro.iag.usp.br} \\
$^{1}$ Universidade de S\~ao Paulo, IAG, Rua do Mat\~ao 1226, Cidade Universit\'aria, S\~ao Paulo 05508-900, Brazil\\
$^{2}$ N\'ucleo de Astrof\'{\i}sica Te\'orica, Universidade Cruzeiro do Sul - Rua Galv\~ao Bueno 868, CEP 01506-000, S\~ao Paulo, Brazil \\
$^{3}$ Astronomy Department, University of Wisconsin, Madison, 475 N. Charter St., WI 53711, USA\\
$^{4}$ Dipartimento di Astronomia, Universit\'a di Bologna, via Ranzani 1, 40126 Bologna, Italy}

\begin{document}

\date{Accepted ??? ???. Received ??? ???; in original form ??? ??? ???}

\pagerange{\pageref{firstpage}--\pageref{lastpage}}
\pubyear{2008}

\maketitle

\label{firstpage}

\begin{abstract}

In this work, considering the impact of a supernova remnant (SNR) with a neutral magnetized cloud we derived analytically a set of conditions that are favorable for driving gravitational instability in the cloud and thus star formation. Using these conditions, we have built diagrams of the SNR radius, $R_{SNR}$, versus the initial cloud density, $n_c$, that constrain a domain in the parameter space where star formation is allowed.  This work is an extension to previous study performed without considering magnetic fields (Melioli et al. 2006). The diagrams are also tested with fully 3-D MHD radiative cooling simulations involving a SNR and a self-gravitating cloud and we find that the numerical analysis is consistent with the results predicted by the  diagrams.  While the inclusion of a homogeneous magnetic field approximately perpendicular to the impact velocity of the SNR with an intensity $\sim 1 \; \mu$G within the cloud results only a small shrinking of the star formation zone in the diagram relative to that without magnetic field, a larger magnetic field ($\sim 10\;\mu$G) causes a significant shrinking, as expected. Though derived from simple analytical considerations these diagrams provide a useful tool for identifying sites where star formation could be triggered by the impact of a SN blast wave. Applications of them to a few regions of our own galaxy (e.g., the large CO shell in the direction of Cassiopeia, and the Edge Cloud 2 in the direction of the Scorpious constellation) have revealed that  star formation in those sites could have been triggered by shock waves from  SNRs  for specific values of the initial neutral cloud density and the SNR radius. Finally, we have evaluated the effective star formation efficiency for this sort of interaction and found that it is generally smaller than the observed values in our own Galaxy (sfe $\sim$ 0.01$-$0.3). This result is consistent with previous  work in the literature and also suggests that the mechanism presently investigated, though very powerful to drive structure formation, supersonic turbulence and eventually, local star formation, does not seem to be sufficient to drive $global$ star formation in normal star forming galaxies, not even when the magnetic field in the neutral clouds is neglected.

\end{abstract}

\begin{keywords}
stars: star formation --- ISM: clouds - supernova remnants.
\end{keywords}

\section{Introdution}

Essentially all present day star formation takes place in molecular clouds (MCs; e.g. Blitz 1993; Williams, Blitz \& McKee 2000). It is likely that the MCs are relatively transient, dynamically evolving structures produced by compressive motions in the diffuse HI medium of either gravitational or turbulent origin, or some combination of both (e.g., Hartmann et al. 2001; Ballestero-Paredes et al. 2007). In fact, observations of supersonic line-widths in the MCs support the presence of supersonic turbulence in these clouds with a wealth of structures on all length-scales (Larson 1981; Blitz \& Williams 1999; Elmegreen \& Scalo 2004; Lazarian \& Esquivel 2003). Recent numerical simulations in periodic boxes have shed some light on the role of the turbulence in the evolution of the MCs and the star formation within them. They suggest that the $continuous$ injection of supersonic motions, maintained by internal or external driving mechanisms (see below), can support a cloud  $globally$ against gravitational collapse so that the net effect of turbulence seems to inhibit collapse and this would explain the observed low overall star formation efficiencies in the Galaxy (Klessen et al. 2000; Mac Low \& Klessen 2004; Vazquez-Semandeni et al. 2005). On the other hand, the supersonic turbulence is also able to produce density enhancements in the gas  that may allow $local$ collapse into stars in both nonmagnetized (Klessen et al. 2000;  Elmegreen \& Scalo 2004) and magnetized media (Heitsch et al. 2001; Nakamura \& Li 2005).

What are the possible sources of this turbulence in the MCs? Suggested candidates for an internal driving mechanism of turbulence include feedback from both low-mass and massive stars. These later, in particular, are major structuring agents in the ISM  in general (McCray \& Snow 1979),  initially through  the production of powerful winds and  intense ionizing radiation and, at the end of their lives through the  explosions as supernovae (SNe). It is worth noting however, that  MCs with and without star formation have similar kinematic properties (Williams, Blitz \& McKee 2000). External candidates include galactic spiral shocks (Roberts 1969; Bonnell et al. 2006) and again SNe shocks (Wada \& Norman 2001; Elmegreen \& Scalo 2004). These processes seem to have sufficient energy to explain the kinematics of the ISM and can generate the observed velocity dispersion-sizescale relation (Kornreich \& Scalo 2000). Other mechanisms, such as magnetorotational instabilities, and even the expansion of HII regions and fluctuations in the ultraviolet (UV) field apparently inject energy into the ambient medium at a rate which is about an order of magnitude lower than the energy that is required to explain the random motions of the ISM at several scales, nonetheless the relative importance of all these injection mechanisms upon star formation is still a matter of debate (see, e.g., Joung \& Mac Low 2006; Ballesteros-Paredes et al. 2007; Mac Low 2008 for reviews).

In this work, we focus on one of these driving mechanisms $-$
supernova explosions produce large blast waves that strike the
interstellar clouds, compressing and sometimes destroying them, but
the compression by the shock may also trigger local star formation
(Elmegreen \& Lada 1977; see also Nakamura \& Li 2005 and references
therein).   There has been extensive analytical and numerical  work
exploring the role  of the SNe  in generating supersonic turbulence
and  multi-phase structures in the ISM (e.g.,  Cox \& Smith 1974;
McKee \& Cowie 1975; McKee \& Ostriker 1977; Kornreich \& Scalo
2000; Mac Low \& Klessen 2004; Vazquez-Semadeni et al. 2000; de
Avillez \& Breitschwerdt 2005) and even at the galactic scale,  in
the generation of large-scale outflows like galactic fountains and
winds (e.g., de Gouveia Dal Pino \& Medina-Tanco 1999; de Avillez
2000; Heckman et al. 2001; de Avillez \& Berry 2001; Melioli et al.
2008a, 2008b). The collective effect of the SNe is likely to be the
dominant contributor to the observed supersonic turbulence in the
Galaxy (Norman \& Ferrara 1996; Mac Low \& Klessen 2004), however,
it is possible that it inhibits $global$ star formation rather than
triggering it (Joung \& Mac Low 2006). Here, instead of examining
the global effects of multiple SNe explosions upon the evolution of
the ISM and the MCs, we will explore  the $local$ effects of these
interactions. To this aim, we will consider a supernova remnant
(SNR) either in its adiabatic or in its radiative phase impacting
with an initially homogeneous diffuse neutral cloud and show that it
is possible to derive analytically a set of conditions that can
constrain a domain in the relevant parameter space where these
interactions may lead to the formation of gravitationally unstable,
collapsing structures. In Melioli et al. (2006; hereafter Paper I),
we gave a first step into this direction by examining interactions
involving a SNR and a non-magnetized cloud and then applying the
results to a SF region in the local ISM where  the young stellar
association of $\beta$ Pictoris was born. Presently, besides
including the effects of the magnetic fields, we will extend this
analysis to few other SF regions with some indication of recent-past
interactions with SN shock fronts (e.g., the Edge Cloud 2,  Yasui et
al. 2006, Ruffle et al. 2007;  and the "Great CO Shell", Reynoso \&
Mangum 2001). Finally, we will also test our analytically derived
$SF$ $domain$ with 3D MHD simulations of SNR interactions with
self-gravitating neutral clouds.

In Section \ref{SNR-Cloud interactions}, we consider the equations that describe the interaction between an expanding SNR and a cloud. In Section \ref{Conditions to collapse}, we obtain analytically the set of constraints from these interactions with both non-magnetized and magnetized clouds that may lead to star formation (SF) and then build diagrams where these constraints delineate a domain in the parameter space which is appropriate for SF. In Section \ref{MHD simulations}, we describe radiative cooling 3D MHD simulations of these interactions which  confirm the results obtained from the analytical study, and \ref{Applications} we apply these analytical results to few examples of  regions of the ISM that present some indication of SF due to a recent SNR-cloud interaction.  In Section \ref{SFE}, we briefly discuss the role  of these interactions to the global efficiency of star formation in the Galaxy and in Section \ref{Conclusions}, we draw our conclusions.

\section{SNR-Cloud Interactions: an analytical description}
\label{SNR-Cloud interactions}

As in paper I, we start by considering the equations that are relevant in the study of the interaction between a SNR and a cloud. A type II SN explosion generates a spherical shock wave that sweeps the interstellar medium (ISM), leading to the formation of a SNR. The interaction between a SNR and a cloud may compress the gas sufficiently to drive the collapse of the cloud. In order to describe analytically this interaction  we will consider a diffuse neutral cloud with initially homogeneous density and constant temperature. After the impact, an internal forward shock propagates into the cloud  with a velocity $v_{cs}$. The ram pressure of the blast wave, $\sim n_{sh} v_{SNR}^2$, must be comparable to the ram pressure behind the shock in the cloud, $\sim n_{c} v_{cs}^2$ and this results  the well known relation for $v_{cs}$ in the case of a planar shock:

\begin{equation}
v_{cs} \sim v_{SNR}\left(\frac{n_{sh}}{n_{c}}\right)^{1/2}
\label{eq:vcs}
\end{equation}

\noindent where $\left(\frac{n_{sh}}{n_{c}}\right) = \chi$ is the density contrast between the SNR shell and the cloud, {\bf $n_{sh}$ is the shell density and $n_c$ is the gas cloud density}. During its evolution a SNR undergoes two main regimes: an adiabatic (or Sedov-Taylor) and later a radiative regime. When the SNR is still in the adiabatic phase, we may use its expansion velocity (as derived in Eq. 4 of Paper I) to find

\begin{eqnarray}
     v_{cs,a} \sim 68 \left(\frac{n_{sh}}{n_{c}} \right)^{1/2} \left(\frac{E_{51}}{n}\right)^{1/2} \frac{1}{R_{SNR,50}^{3/2}} \nonumber
\end{eqnarray}

\begin{eqnarray}
    \sim 43 \frac{E_{51}^{1/2}}{R_{SNR,50}^{3/2} \; n_{c,10}^{1/2}} \;\;\; \rm km/s
\label{eq:velcsAd}
\end{eqnarray}

\noindent  where $n_{c,10}$ is the cloud density in units of 10 $\rm cm^{-3}$, $E_{51}$ is the SN energy in units of $10^{51}\;\rm erg$, $R_{SNR,50}$ is the SNR shell radius in units of 50 pc. When the SNR enters the the radiative phase, the expansion velocity of the SNR is described by Eq. (7) of Paper I and this results a forward shock into the cloud:

\begin{equation}
    v_{cs,r} \sim 47 \frac{E_{51}^{0.8}\; f_{10}^{1/2}}{R_{SNR,50}^{5/2}\; n_{c,10}^{1/2}
    n^{0.41}} \;\;\; \rm km/s
\label{eq:velcsRad}
\end{equation}

\noindent  where $f_{10}$ ($f=\left(\frac{n_{sh}}{n}\right)$) is the density contrast between the SNR shell and the ISM density in units of 10 and $n$ is the ambient medium density.

The equations above are valid  for a planar shock only. Since it is not rare that both the SNR and the cloud have dimensions which are of the same order, the effects of curvature of the shock should be considered. In Ap. \ref{Apendice}, we have derived a correction factor (I) that must be multiplied to $v_{cs}$ for the case of spherical cloud$-$SNR interactions, i.e.:

\begin{eqnarray}
{\hat{v}}_{cs} = v_{SNR} \ \left({{n_{sh}} \over {n_c}}\right)^{0.5} \ I
\label{eq:vel}
\end{eqnarray}

\noindent where I is given by eq.  (\ref{eq:integral}) of Ap.
\ref{Apendice}.

\subsection{The inclusion of the magnetic field in the cloud}

Zeeman measurements indicate that the magnetic flux-to-mass ratios in molecular cloud cores are close to the critical value  for magnetic support (e.g., Crutcher 1999, 2005, 2008), while in neutral clouds these ratios  are  observed to be supercritical. In other words, the typical observed magnetic fields in these clouds ($\sim 6 \mu$G - $10 \mu$G) are smaller than the critical value ($B_{cr} \simeq 2 \pi G^{1/2}$N, where N is the cloud surface density) necessary to suppress gravitational
instability in them (e.g., Nakano \& Nakamura 1978). This could be an indication that the presence of the magnetic field in these neutral clouds is not relevant when considering their interactions with SNRs. However, the impact will compress the magnetic field behind the shock and this may affect the evolution of the shocked material and therefore, its conditions for gravitational collapse.

Let us then consider a magnetized diffuse cloud. For simplicity, let us assume that its magnetic field is initially approximately uniform and,
in order to maximize the effects of the field we shall also assume that it is initially normal to the SNR shock velocity at the impact. The observed magnetic fields are actually randomly distributed. This means that only a fraction of its average strength will be effectively normal to the SNR velocity and work against the impact. We will find below that when we take an effective value for the cloud magnetic field $B_c \simeq 1 \mu$G, its effect upon the SF diagrams is in fact not too relevant. However, when we take values $\sim 5-10\; \mu$G, the allowed SF domain in the diagrams shrank considerably.

{\bf Depending on the physical conditions of the cloud, the propagation of the shock front into it can be either adiabatic or radiative. For the shocked gas at temperatures $T\leq 10^4$ K, we find that the radiative cooling time is shorter than the crushing time (see e.g. Melioli, de Gouveia Dal Pino \& Raga 2005) and therefore, we can assume that the forward shock wave propagating into the cloud is approximately radiative. The Rankine-Hugoniot relations for a radiative strong shock (with M $\geq$ 10) (see, e.g. Draine \& McKee 1993), are }

\begin{equation}
    T_{c,sh}=T_c
\label{eq:tempmag}
\end{equation}

\begin{equation}
    n_{c,sh,B}= y\; n_{c}
\label{eq:RHdensmag}
\end{equation}

\begin{equation}
    B_{c,sh}= y \; B_{c}
\label{eq:campochocado}
\end{equation}

\begin{equation}
    y=\frac{4}{2M^{-2} + M_A^{-2} + [(2M^{-2} + M_A^{-2})^2 + 8 M_A^{-2}]^{1/2}}
\label{eq:y}
\end{equation}

\noindent where $T_{c,sh}$, $n_{c,sh,B}$ and $B_{c,sh}$ are the temperature,  density, and  magnetic field of the shocked cloud gas, respectively, $M$ is the Mach number given by Eqs. (\ref{eq:machad}) and (\ref{eq:machrad}) for an adiabatic  and for a radiative-phase SNR, respectively, and $M_A$ is the Afv\'enic Mach number. For an interaction with a SNR in the adiabatic phase, using equation (\ref{eq:velcsAd}), $M_A$ becomes

\begin{equation}
    M_{A,a}=\frac{\hat{v}_{cs}}{v_A}= 68.5\;\frac{E_{51}^{1/2}I_5}{B_{c,6}\; R_{SNR,50}^{3/2}}
\label{eq:machAlfad}
\end{equation}

\noindent  where $v_A$ is the Alfv\'en speed in the cloud, $B_{c,6}$ is the pre-shock magnetic field of the cloud in units of $10^{-6}\; G$ and $I_5$ is the factor I (eq. \ref{eq:integral}) calculated for $R_{SNR}/r_c=5$. For an interaction with a SNR in the radiative phase, using Eq. (\ref{eq:velcsRad}), $M_A$ is given by:

\begin{equation}
    M_{A,r}=75\;\frac{f_{10}^{1/2} \;E_{51}^{0.8}\;I_5}{B_{c,6} \;R_{SNR,50}^{5/2}\; n^{0.41}}
\label{eq:machAlfrad}
\end{equation}

\noindent Replacing (\ref{eq:machAlfad}) and (\ref{eq:machad}) in Eq. (\ref{eq:RHdensmag}), we obtain the density of the  shocked gas in the magnetized cloud after the interaction with an adiabatic SNR:

\begin{equation}
     n_{c,sh,B,a} \sim \frac{4\times10^4\;E_{51}\;n_{c,10}\;I_5^2}{R_{SNR,50}^3\;\left[ F + \sqrt{F^2+G}\right]}
\label{eq:densadmag}
\end{equation}

\noindent where \[F=1.2\;T_{c,100}\;n_{c,10} + 0.21\;B_{c,6}^2\] and

\[G=1700\frac{B_{c,6}^2\;E_{51}\;I_5^2}{R_{SNR,50}^3}\]

\noindent Using (\ref{eq:machAlfrad}) and (\ref{eq:machrad}), we obtain the density of the shocked cloud gas after the interaction with a radiative SNR:

\begin{equation}
     n_{c,sh,B,r} \sim \frac{4\times10^4 \;E_{51}^{1.6}\;n_{c,10}\;f_{10}\;I_5^2}{R_{SNR,50}^5\; n^{0.82} \left[ H +
     \sqrt{H^2 + J}\right]}
\label{eq:densradmag}
\end{equation}

\noindent where \[H=1\;T_{c,100}\;n_{c,10} + 0.18\;B_{c,6}^2\]
and \[J= 1430 \frac{B_{c,6}^2\;E_{51}^{1.6}\;f_{10}\;I_5^2}{R_{SNR,50}^5\; n^{0.82}}\]

When $1\ll M_A \ll M^2$, the equations above become (e.g., Draine \&
McKee 1993):

\begin{equation}
    n_{c,sh,a} = 2^{1/2} M_A\; n_c \sim 970\; n_{c,10}\;\frac{E_{51}^{1/2}I_5}{B_{c,6}\; R_{SNR,50}^{3/2}}
\end{equation}

\noindent  and

\begin{equation}
    n_{c,sh,r} = 2^{1/2} M_A\; n_c \sim 1060 \; n_{c,10}\;\frac{f_{10}^{1/2} \;E_{51}^{0.8}\;I_5}{B_{c,6} \;R_{SNR,50}^{5/2}\; n^{0.41}}
\end{equation}

\noindent In the limit  $1\ll M^2 \ll M_A$, which implies a dynamically negligible  magnetic field behind the shock, we recover the equations without magnetic field derived in Ap. \ref{Apendice} (Eq. \ref{eq:densadiabatica} and \ref{eq:densradiativa}; see also Paper I).

\section{Conditions to drive gravitational collapse in the cloud}
\label{Conditions to collapse}

We can now  derive a set of constraints that will determine the conditions for driving gravitational instability in the cloud right after the interaction with a SNR. A first constraint will simply determine the Jeans mass limit for the compressed cloud material. A second one will establish the condition on the SNR forward shock front at which it will $not$ be too strong to destroy the cloud completely making the gas to disperse in the interstellar medium before becoming gravitationally unstable. A third constraint will establish the penetration extent of the same shock front inside the cloud before being stalled due to radiative losses. The shock must have energy enough to compress as much cloud material as possible before fading.  Let us derive these three constraints.

\subsection{The Jeans Mass Constraint}
\label{Jeans}

\subsubsection{In the absence of magnetic field}

In Paper I, we have derived the Jeans mass limit for the shocked gas in interactions involving non-magnetized clouds. The more precise equations obtained in Ap. \ref{Apendice} for spherical SNR$-$cloud interactions produce slight modifications in the Jeans mass derived in Eqs. (22) and (23) of paper I.  When the interacting SNR is still in the adiabatic regime, the corresponding Jeans mass of the shocked cloud material in the absence of magnetic field is given by:

\begin{equation}
m_{J,a}\sim 750 \frac{T_{c,100}^2\;R_{SNR,50}^{1.5}}{I_5\;E_{51}^{0.5}} \;\; M_{\odot}
\label{eq:mjeansad}
\end{equation}

\noindent and if the SNR is in the radiative regime:

\begin{equation}
m_{J,r}\sim 685 \frac{T_{c,100}^2\;R_{SNR,50}^{2.5}\;n^{0.41}}{I_5\;f_{10}^{0.5}\;E_{51}^{0.8}}\;\;M_{\odot}
\label{eq:mjeansrad}
\end{equation}

{\bf In terms of the SNR radius, the conditions above over the shocked gas of the cloud imply:}

\begin{equation}
    R_{SNR,a}\lesssim 55.8\;\frac{E_{51}^{1/3}\;I_5^{2/3}\; n_{c,10}^{2/3}\;r_{c,10}^{2}}{T_{c,100}^{4/3}}\ \ \ {\rm pc}
\label{eq:parametro1}
\end{equation}

\noindent and
\begin{equation}
    R_{SNR,r}\lesssim 55.1 \; \frac{E_{51}^{0.33}\;I_5^{0.4}\;\;f_{10}^{0.2}\;n_{c,10}^{0.4}\;r_{c,10}^{2.4}}{T_{c,100}^{0.8}\;\;\;n^{0.17}}\ \ \ {\rm pc}
\label{eq:parametro1rad}
\end{equation}

\noindent respectively.

\subsubsection{In the presence of magnetic field}

When including the magnetic field in the cloud, the corresponding minimum mass (Jeans mass) that the shocked material must have in order to suffer gravitational collapse is given using the Virial Theorem by

\begin{equation}
m_{J,B} \simeq \frac{6.63\times 10^{24}}{n_{c,sh}^{1/2}}\left[ \frac{B_{c,sh}^2}{8\pi\; n_{c,sh}} + 3 k_B\; T_{c,sh} \right]^{3/2} \; M_{\odot}
\label{eq:mJmagsimples}
\end{equation}

In terms of the cloud pre-shock and the SNR parameters this condition reads (using Eqs. \ref{eq:tempmag} - \ref{eq:campochocado}):

\begin{equation}
m_{J,B} \simeq 2100\left[\frac{4y B_{c,6}^2}{n_{c,10}} + 4.14  T_{c,100}\right]^{3/2}\frac{1}{(y\;n_{c,10})^{1/2}} \;\; M_{\odot}
\label{eq:mJmag}
\end{equation}

We can obtain $R_{SNR}$ as function of $n_{c}$ and then obtain an approximate condition for gravitational collapse solving the condition $m_c\geq m_{J,B}$, where $m_c$ is the cloud mass, or:

\begin{equation}
 n_{c,10}^2\; r_{c,10}^2\; y^{1/3} \geq 5.63 \: n_{c,10} \: T_{c,100} + 0.54\: B_{c,6}^2\; y
 \label{eq:parametro1mag}
\end{equation}

\noindent where $y$ is given by Eq. (\ref{eq:y}). Replacing $y$ in the equation above and substituting Eqs. (\ref{eq:machad}) and (\ref{eq:machAlfad}) we obtain numerically the Jeans limit or an upper limit for $R_{SNR,a}$ for the shocked gas of a magnetized cloud due to the impact with a SNR in the adiabatic regime and substituting (\ref{eq:machrad}) and (\ref{eq:machAlfrad}) we obtain the same conditions for an interaction with a SNR in the radiative regime (see Section 3.4 below).

We notice that in the limit that $\frac{B^2}{8\pi}\ll \rho\;c_s^2$ the equation above (\ref{eq:parametro1mag}) recovers the solutions (\ref{eq:parametro1}) and (\ref{eq:parametro1rad}), for the adiabatic and radiative phases, respectively.

\subsection{Constraint for non-destruction of the cloud due to a SNR impact}
\label{non-destruction}

\subsubsection{In the absence of magnetic field}

As remarked before, if the SNR-cloud interaction is too strong this
may cause the complete destruction of the cloud even if the shocked
material had $a$ $priori$ a total mass superior to the Jeans limit.
{\bf The stability of a cloud against destruction right after the
impact of a wind or a SNR due to the growth of Rayleigh-Taylor and
Kelvin-Helmholtz instabilities has been explored in detail by
several authors  (see e.g., Murray et al. 1993; Poludnenko et al.
2002; Melioli, de Gouveia Dal Pino \& Raga 2005, and references
therein). In order to obtain the condition for a shocked cloud not
to be destroyed, we must compare the gravitational free-fall
time-scale of the shocked gas with the destruction time-scale due to
the impact and the consequent development of those instabilities,
$t_d$.} More precisely, in order to have collapse, a gravitationally
unstable mode (with  typical time $t_{un}$) must grow fast enough to
become  non-linear within the time-scale of the cloud-SNR shock
interaction (see, e.g., Nakamura et al. 2006).  {\bf Following
Nakamura et al., the condition for non-destruction then reads:
$t_{un} < t_d$, where $t_d$ is a few times the crushing time,
$t_{cc} = 2 r_c/{\hat{v}}_{cs} $ (eqs. A3 and A4), i.e., the time
that the internal forward shock takes to cross the cloud. With the
help of numerical simulations that have taken into account the
effects of non-equilibrium radiative cooling in the interactions
between clouds and SNRs, Melioli, de Gouveia Dal Pino \& Raga (2005)
have shown that the cloud destruction time is $t_d \sim$ 4-6
$t_{cc}$, when radiative cooling is present.} Hence, in order to
have collapse, we should expect that  $t_{un} \lesssim 5 t_{cc}$.
This implies a Mach number: \footnote{We notice that in paper I, we
have assumed a condition for collapse which was given by  $t_{un}
\lesssim 3 t_{cc}$. In fact, this is a more appropriate condition
for shocks where the radiative cooling is not much important.
Presently, we have corrected this condition by $t_{un} \lesssim 5
t_{cc}$, as $t_{d} \sim 5 t_{cc}$ is a more representative average
value for the cloud destruction time in the presence of a strong
radiative cooling shocked cloud (Melioli, de Gouveia Dal Pino \&
Raga 2005).}

{\bf
\begin{equation}
M \lesssim 60.5 \; \frac{\mu_0^{7/9}}{\zeta^2}\;.
\end{equation}

\noindent where $\zeta \approx 0.8$ is the shock compression
factor in the transverse direction (Nakamura et al. 2006) and
\begin{equation}
\mu_0=m_c\frac{(G^3 \rho_c)^{1/2}}{c_s^3}\;
\end{equation}

\noindent is the non-dimensional cloud mass, and $m_c$ is the cloud
mass. Thus, we have: }

\begin{equation}
M \lesssim 47 \left(\frac{n_{c,10}}{T_{c,100}}\right)^{1.16} r_{c,10}^{2.3}
\end{equation}

\noindent We note that in Paper I, the exponent of $r_c$ in the equation above was incorrectly given by 3. This relation implies:

\begin{equation}
R_{SNR,a} \gtrsim 45 \ {{E_{51}^{0.33} \ T_{c,100}^{0.44} \ I_5^{0.66}} \over{n_{c,10} \ r_{c,10}^{1.56}}} \ \ \ \ {\rm pc}
\label{eq:parametro2}
\end{equation}

\noindent for an interaction with a SNR in the adiabatic regime, and

\begin{equation}
R_{SNR,r} \gtrsim 48.7 \ {{E_{51}^{0.33} \ f_{10}^{0.2} \ T_{c,100}^{0.26}\ I_5^{0.4}} \over {n_{c,10}^{0.7} \ n^{0.17} \ r_{c,10}^{0.93}}} \ \ \ \ {\rm pc}
\label{eq:parametro2rad}
\end{equation}

\noindent for an interaction with a SNR in the radiative phase.

\subsubsection{In the presence of magnetic field}

When the cloud is magnetized with an average magnetic field normal to the shock velocity then the shock strength must be decreased by the magnetic pressure and tension forces. Using the same condition for collapse as described in the previous paragraph, but including the magnetic field in the cloud, we obtain the following condition over the Mach number:

\begin{eqnarray}
\frac{M (1 + y \beta/3)^{3/8}}{y^{5/8}}\lesssim \; 18.7\left(\frac{n_{c,10}}{T_{c,100}}\right)^{7/8}r_{c,10}^{7/4}
\label{eq:parametro2mag}
\end{eqnarray}

\noindent where

\[\beta=\frac{B_c^2}{8\pi \rho_c c_s^2}\]

\noindent Substituting the relations found for $y$ (Eq. \ref{eq:y}), $M$ and $M_A$ for a collision with an adiabatic (Eqs. \ref{eq:machAlfad} and \ref{eq:machad}) and with a radiative SNR (Eqs. \ref{eq:machAlfrad} and \ref{eq:machrad}) into the equation above (Eq. \ref{eq:parametro2mag}) we can find numerically the new constraints over $R_{SNR,a}$ and $R_{SNR,r}$, respectively, in order to not destroy the magnetized cloud at the impact and allow its gravitational collapse (see below).

\subsection{Penetration extent of the SNR shock front into the cloud}
\label{penetration}

Besides the constraints derived in the previous sections (3.1 and 3.2), we still need a third condition upon the impact. As remarked before, the shock should have strength enough to penetrate into the cloud and compress as much material as possible for this to undergo gravitational collapse.

\subsubsection{In the absence of magnetic field}

{\bf When the shock propagates into the cloud, it will decelerate
due to continuous radiative losses and to the resistance of the
unshocked cloud material. We can estimate the approximate time at
which the shock will stall within the cloud ($t_{st}$) using energy
conservation arguments. At  $t=t_{st}$, the velocity of the shocked
gas in the cloud should become $\approx  c_s$, so that the Mach
number will decay from its initial value right after the impact to
$M \sim 1$. As in paper I, from the energy conservation behind the
shock, we find approximately that:}

\begin{eqnarray}
    \frac{5}{2}\;\frac{k_B\;T_{c,sh}(0)}{\mu m_H} +\frac{1}{2}\;v_{sh}^2(0)  \nonumber
\end{eqnarray}

\begin{eqnarray}
    \simeq \frac{5}{2}\;\frac{k_B\;T_{c,sh}(t_{st})}{\mu m_H} + \frac{1}{2}\;v_{sh}^2(t_{st}) + \frac{\Lambda[T_{c,sh}(t_{st})]\;n_{c,sh}(t_{st})}{\mu m_H}\;t_{st}
\end{eqnarray}

{\bf \noindent where the left-hand side is the total energy behind
the shock in the cloud immediately after the impact and the
right-hand side is the total energy behind the shock at the time
$t_{st}$ when it stalls. The initial temperature and density of the
shocked cloud material in the left hand-side of the equation are
approximately given by the adiabatic values.  On the right hand
side, at $t_{st}$, the shocked temperature and density  are obtained
from the Rankine-Hugoniot relations for a radiative shock with $M
\simeq 1$, i.e., $T_{c,sh}(t_{st}) \simeq T_c$, and
$n_{c,sh}(t_{st}) \simeq n_c$.  The radiative cooling function at
$t_{st}$, $\Lambda [T_{sh} (t_{st})] \simeq \Lambda (T_{c}) $, can
be approximated by that of an optically thin gas (see e.g. Dalgarno
\& McCray 1972). The substitution of these conditions into the
equation above results:} \footnote{We notice that in Paper I, there
was a typo in the equation for $t_{st}$ (Eq. (31) of that paper).
The multiplying factor in that equation ($9/16$) has been now
correctly replaced by {\bf 31/32}.}

\begin{equation}
    t_{st} \simeq {31 \over 32} {{\mu  m_H} \over {n_c \Lambda(T_{c})}} \hat{v}_{cs}^2\ \ \ {\rm s.}
\label{eq:t_st}
\end{equation}

\noindent {\bf where $\hat{v}^2_{cs}$ is given by eq. (\ref{eq:vel})
or, in terms of the initial Mach number, by eq. (\ref{eq:machad}) or
(\ref{eq:machrad}), and $\Lambda(T_c)$ is calculated for $T_c = 100$
K.}

As in paper I, we use this time to compute the maximum distance that the shock front (initiated by a given SNR) can travel into the cloud before being stopped. This distance is then compared with the {\bf diameter of the cloud ($\hat{v}_{cs} \; t_{st}\geq 2 r_c$)} in order to establish the maximum size (that is, the minimum energy) that the SNR should have in order to generate a shock  able to compress the cloud as much as possible before being stalled. In the case of an adiabatic SNR this gives:

\begin{equation}
    R_{SNR,a} \lesssim 170 {{E_{51}^{0.33} I_5^{0.66}} \over {(r_{c,10}\Lambda_{27})^{2/9} n_c^{0.5}}} \ \ \ {\rm pc}
 \label{eq:parametro3mag}
\end{equation}

\noindent And in the case of a SNR in the radiative regime, we find:

\begin{equation}
    R_{SNR,r}\lesssim 108 \frac{E_{51}^{0.32}I_5^{2/5}f_{10}^{1/5}}{n_{c,10}^{1/3}\;n^{0.16}\;(r_{c,10}\Lambda_{27})^{2/15}}\;\;pc
    \label{eq:parametro3magRad}
\end{equation}

\noindent where $\Lambda_{27}$ is the cooling function ($\Lambda$)
in units of $10^{-27}$  erg cm$^{3}$ s$^{-1}$, {\bf which is the
approximate value for $T_c=100$ K and an ionization fraction
$10^{-4}$ (see below).}

\subsubsection{In the presence of magnetic field}

When considering the presence of the normal magnetic field in the cloud, our estimates from energy conservation imply approximately:

\begin{eqnarray}
    \frac{5}{2}\;\frac{k_B\;T_{c,sh}(0)}{\mu m_H} +\frac{1}{2}\;v_{sh}^2(0) + \frac{B_{c,sh}^2(0)}{4\pi\mu m_H n_{c,sh}(0)} \nonumber
\end{eqnarray}

\begin{eqnarray}
    \simeq \frac{5}{2}\;\frac{k_B\;T_{c,sh}(t_{st})}{\mu m_H} + \frac{1}{2}\;v_{sh}^2(t_{st}) + \frac{\Lambda[T_{c,sh}(t_{st})]\;n_{c,sh}(t_{st})}{\mu m_H}\;t_{st} \nonumber
\end{eqnarray}

\begin{eqnarray}
    + \frac{B_{c,sh}(t_{st})^2}{4\pi \mu m_H n_{c,sh}(t_{st})}
    \label{eq:consenergmag}
\end{eqnarray}

{\bf \noindent Where again  the left-hand side gives the total
energy behind the shock in the cloud immediately after the impact
with the initial temperature, density and magnetic field of the
shocked cloud material being approximately given by the adiabatic
values. The right-hand side gives the total energy behind the shock
at $t_{st}$  when $v_{sh}$ decreases to $v_{sh}(t_{st})^2  \simeq
\gamma c_s^2 + v_A^2$. At this time, $n_{c,sh}(t_{st})$,
$T_{c,sh}(t_{st})$, and $B_{c,sh}(t_{st})$ are obtained from the
Rankine-Hugoniot relations for a magnetized radiative shock (eqs.
\ref{eq:tempmag} to \ref{eq:campochocado}), for $y \simeq 1$. So
that again $\Lambda [T_{sh} (t_{st})] \simeq \Lambda (T_{c}) $.} The
substitution of these conditions into the equation above results
approximately the same estimate for $t_{st}$ as in Eq.
(\ref{eq:t_st}) for the non-magnetic case and, therefore, the same
upper limits over $R_{SNR}$ in order to the shock to sweep the cloud
as much as possible before being stalled (Eqs.
\ref{eq:parametro3mag} and \ref{eq:parametro3magRad}). This is due
to the fact that for the typical fields observed in these clouds
($\sim 10^{-4} - 10^{-6}\;G$), the magnetic energy terms in Eq.
(\ref{eq:consenergmag}) are negligible compared to the others.

\subsection{Diagrams for Star Formation}

The three constraints derived above in Sections 3.1, 3.2, and 3.3
for both non-magnetized and magnetized clouds interacting with SNRs
either in the adiabatic or in the radiative phase can be plotted
together in a diagram showing the SNR radius versus the initial
(un-shocked) cloud density for different values of the cloud radius,
as performed in Paper I. Figures 1 and 2 show examples of these
diagrams for non-magnetized and magnetized clouds with $B=1\;\mu$G,
respectively, with an initial temperature $T_c =$ 100 K and radius
varying between $r_c=$ 1 pc and 20 pc.\footnote{{\bf We notice that
in the diagrams of Figs. 1 and 2, and in most of the diagrams of
this work, we have assumed an ambient density $n = 0.05 $ cm$^{-3}$.
We have chosen this low density value  in order to try to better
reproduce the hot phase, low density medium commonly expected around
a SNR, particularly in its adiabatic phase.}} The three constraints
establish a shaded zone in the parameter space of the diagrams where
conditions are appropriate for gravitational collapse of the shocked
cloud material. Only cloud$-$SNR interactions with initial physical
conditions ($r_c$, $n_c$, and $R_{SNR}$)lying within the shaded
region (between the solid, dotted and dashed lines) may lead to a
process of star formation. We have chosen a SNR in the adiabatic
phase in these diagrams because it has more stored energy than one
with the same characteristics in the radiative phase (see, however,
Figure \ref{fig:compdiag20pcRadN2c} an example of an interaction
involving a radiative SNR). As in paper I, we should remark that
according to the equations (\ref{eq:parametro1},
\ref{eq:parametro1rad}, \ref{eq:parametro2}, \ref{eq:parametro2rad},
\ref{eq:parametro3mag} and \ref{eq:parametro3magRad}), these
interactions are not very sensitive to the initial temperature of
the cloud and this explains why we have taken only a characteristic
value for it. We further notice that a cloud with a temperature in
the range of 10$-$50 K and a radius larger than 10 pc is already
Jeans unstable over a large range of densities ( $>$ 5 $cm^{-3}$
when the magnetic field is neglected) and does not require, in
principle, a shock wave to trigger SF. Besides, it will be more
difficulty for a SNR shock front to destroy a cloud at these
temperatures.

Figure 1, which describes interactions with a non-magnetized cloud, was already presented in Paper I. However, the modifications in the equations described in Sections \ref{Jeans}$-$\ref{penetration} above have resulted slight modifications in the diagrams. Few remarks are in order with regard to this Figure:

\begin{figure}
    \centering
        \includegraphics[width=7cm]{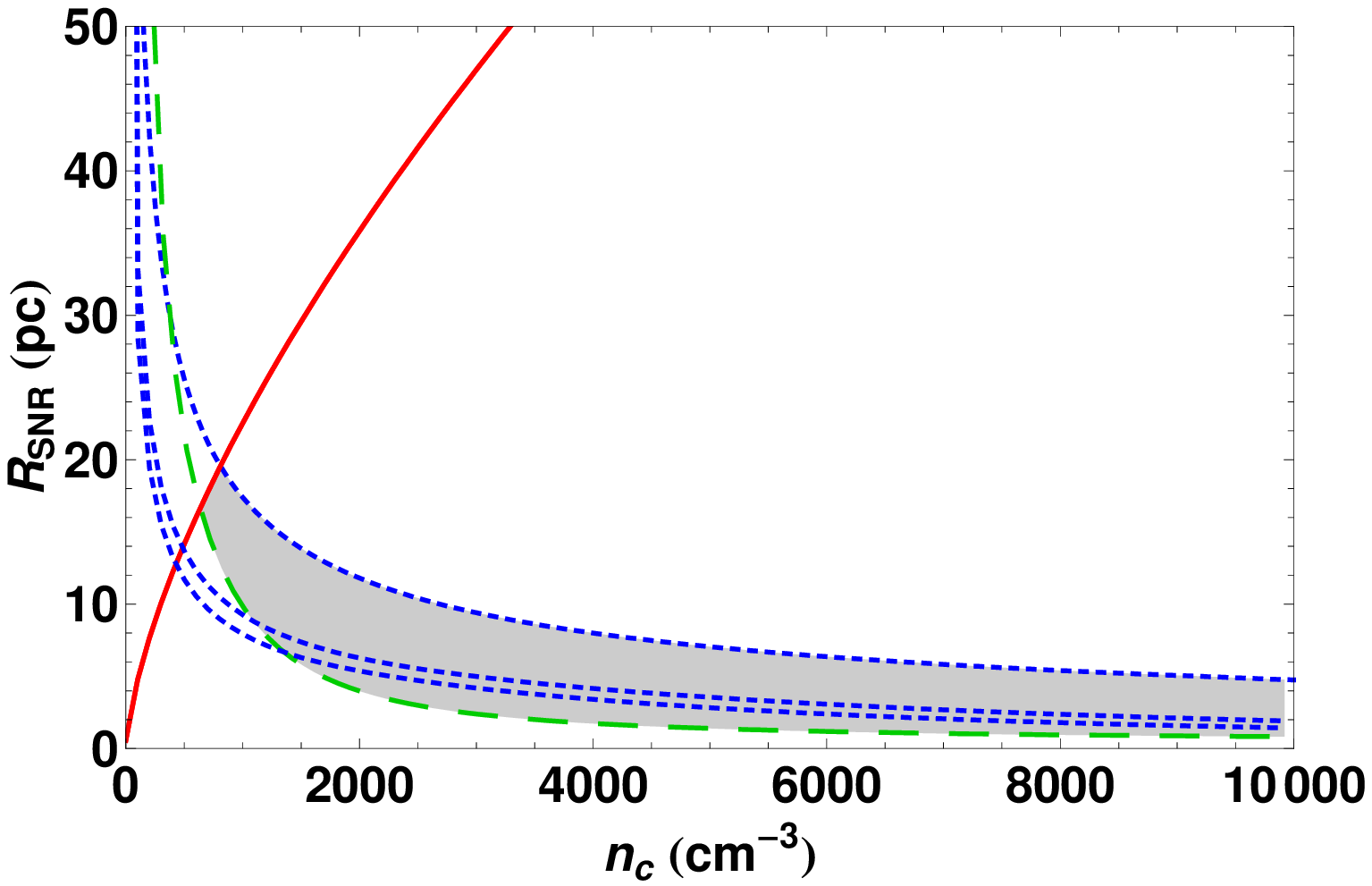}
        \includegraphics[width=6.7cm]{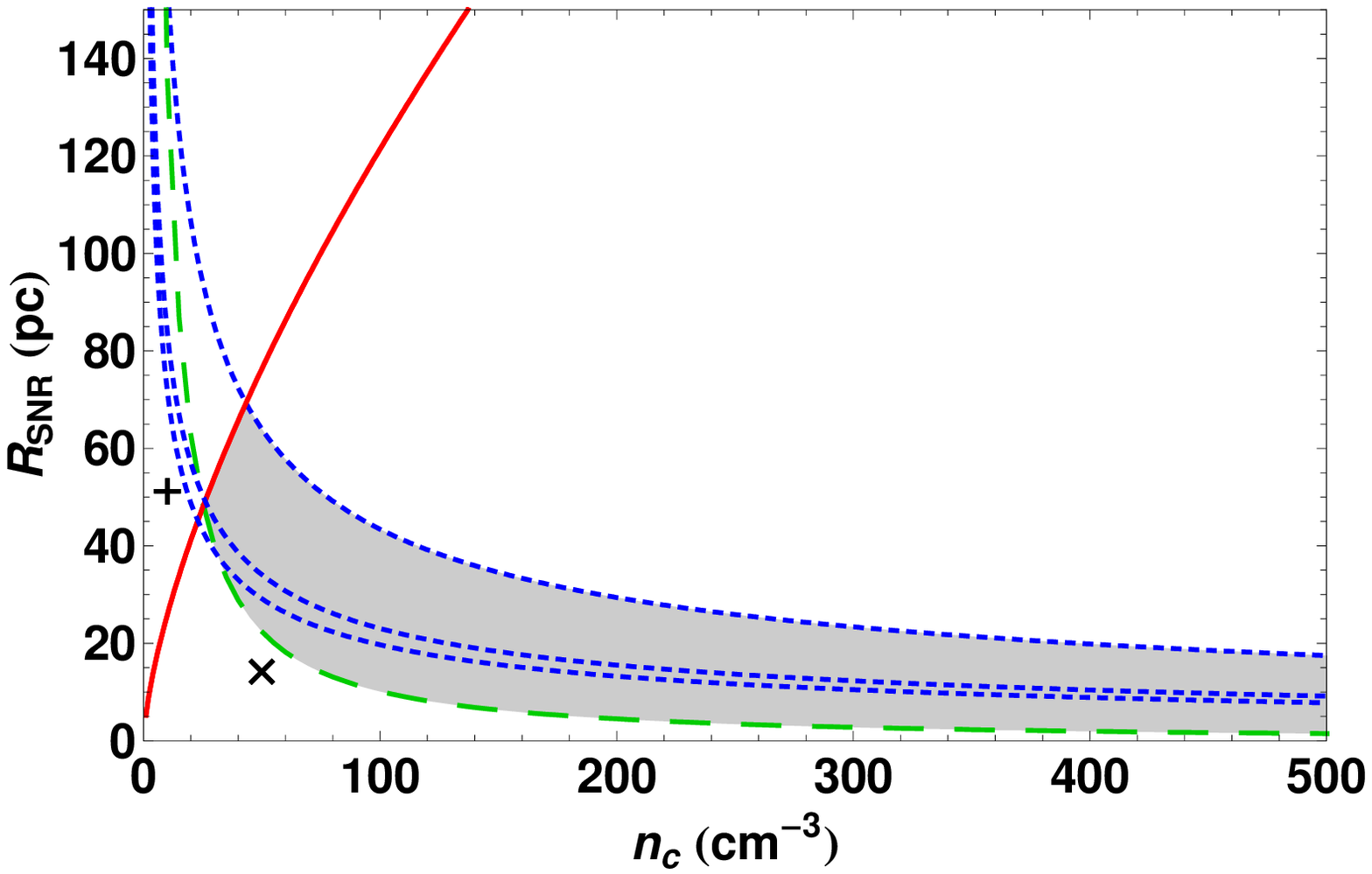}
        \includegraphics[width=6.7cm]{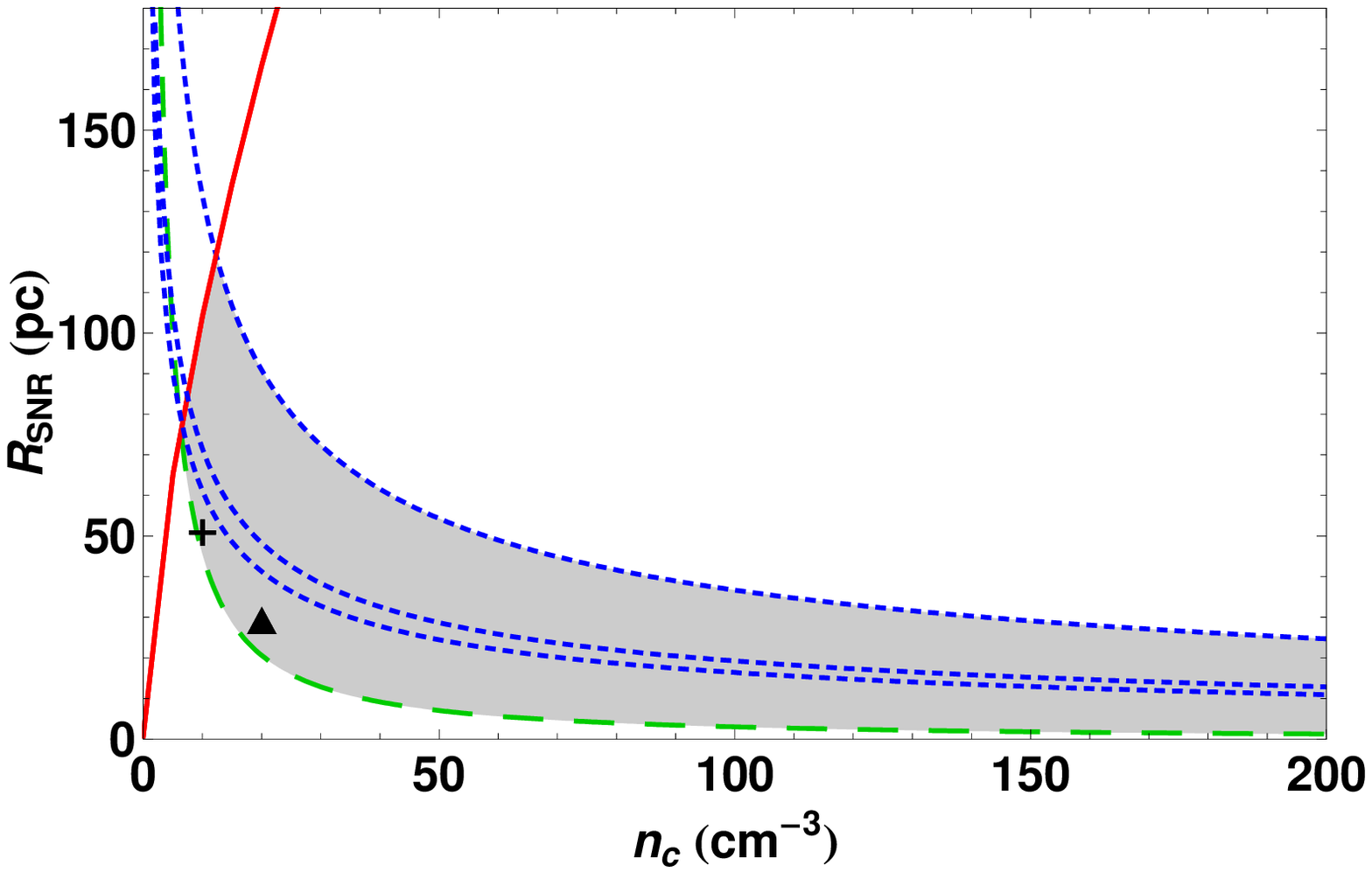}
        \includegraphics[width=6.7cm]{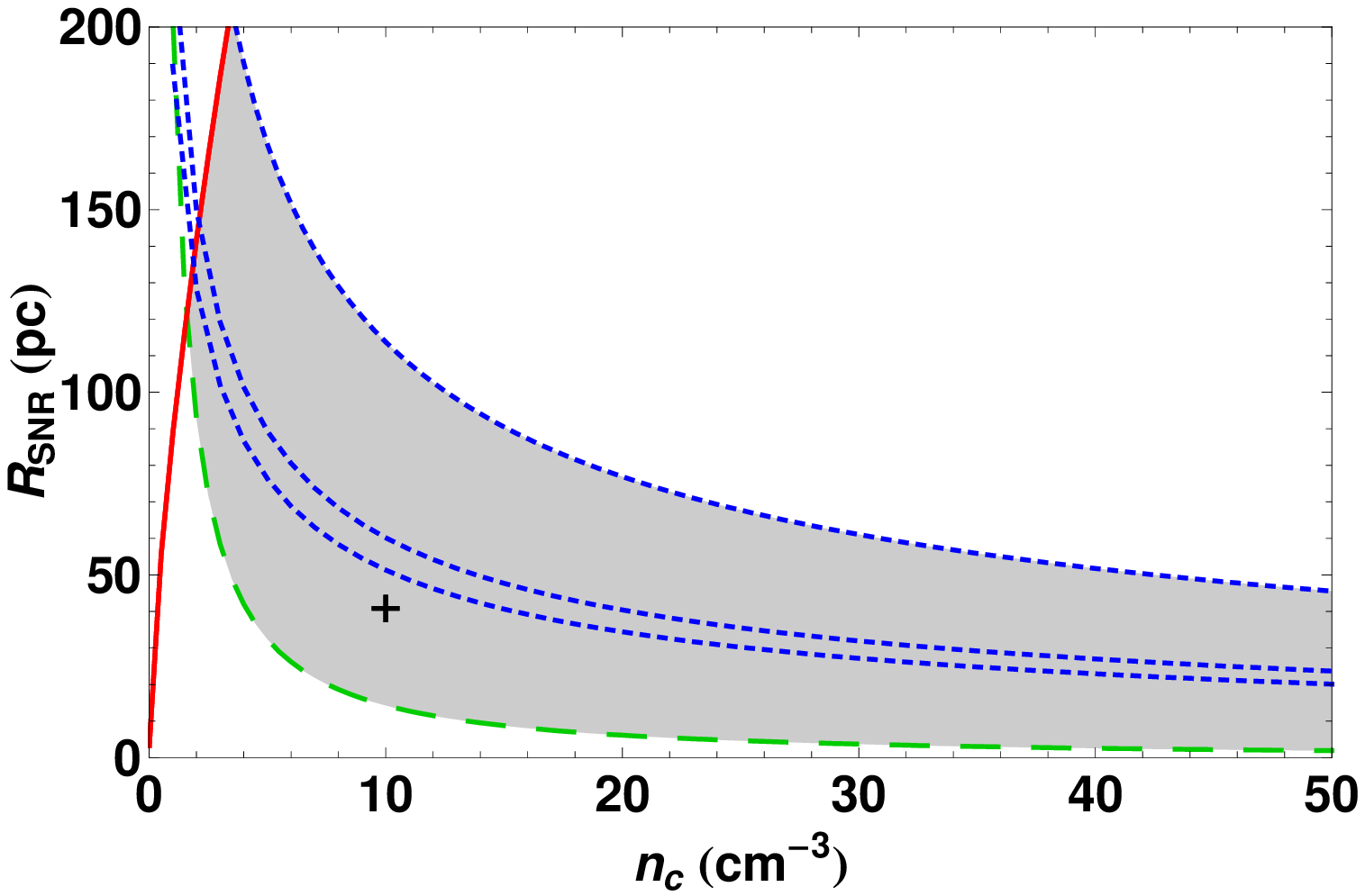}
\caption{Constraints on the SNR radius  versus cloud density for 4
different cloud radius. From top to bottom: first panel, $r_c$ = 1
pc; second panel, $r_c$ = 5 pc; third panel, $r_c$ = 10 pc; and
forth panel, $r_c$ = 20 pc. Dashed (green) line: upper limit for
complete cloud destruction after an encounter with an adiabatic SNR
derived from Eq. (\ref{eq:parametro2}); solid (red) line: upper
limit for the shocked cloud to reach the Jeans mass derived from Eq.
(\ref{eq:parametro1}); dotted (blue) lines: upper limits for the
shock front to travel into the cloud before being decelerated to
subsonic velocities derived from Eq. (\ref{eq:parametro3mag}) for
different values of the cooling function $\Lambda(T_{sh})= 10^{-25}$
erg cm$^{3}$ s$^{-1}$ (lower curve), $5 \times 10^{-26}$ erg
cm$^{3}$ s$^{-1}$ (middle curve), and $3 \times 10^{-27}$ erg
cm$^{3}$ s$^{-1}$ (upper curve). The shaded area (between the solid,
dashed and dotted lines) defines the region where star formation can
be induced by a SNR-cloud interaction. The crosses and the triangle
in the panels indicate the initial conditions assumed for the clouds
in the numerical simulations described in Section 4.2 of Paper I
(see text for details).} \label{fig:diagrama}
\end{figure}

\begin{enumerate}

\item In paper I, the dotted (blue) curves of the diagrams were built for one value only of the radiative cooling function of the shocked material, i.e. $\Lambda(T_{sh}) \simeq 10^{-27}$ erg cm$^{3}$ s$^{-1}$ which is valid for a diffuse gas with temperature  100 K and ionization fraction $ \le$ 10$^{-4}$. Considering that the constraint established by the dotted (blue) curve is sensitive to $\Lambda (T_{sh})$ (through Eq. \ref{eq:parametro3mag}) which in turn, can vary by two orders of magnitude depending on the value of the ionization fraction of the cloud gas, we have presently plotted in the diagrams three different dotted (blue) curves in order to cover a more realistic range of possible ionization fractions from 0.1 to 10$^{-4}$, corresponding to $\Lambda(T_{sh})= 10^{-25}$ erg cm$^{3}$ s$^{-1}$  (lower dotted curve), $5 \times 10^{-26}$ erg cm$^{3}$ s$^{-1}$ (middle dotted curve), and  $3 \times 10^{-27}$ erg cm$^{3}$ s$^{-1}$ (upper dotted curve), respectively (see Dalgarno \& McCray 1972). The middle dotted (blue) curve corresponds to the average value of $\Lambda$ in the range above, $5 \times 10^{-26}$ erg cm$^{3}$ s$^{-1}$, and could be taken as a reference.

\item In the solution presented in Paper I for the cloud with $r_c=$ 1 pc (top panel of Figure \ref{fig:diagrama}), there was no allowed SF shaded zone. According to the present corrections and modifications, we see that a thin shaded "star-formation unstable" zone appears now when the cooling function $\Lambda$ has values which are smaller than $ 10^{-25}$ erg cm$^{3}$ s$^{-1}$, or ionization fractions $\lesssim 10^{-1}$.

\item The cross labeled  in the bottom panel of  Figure 1 for an $r_c=$ 20 pc diffuse cloud corresponds to the initial conditions of the numerical simulations presented in Figure 6 of paper I (i.e., for a SNR at a distance $R_{SNR} \sim$ 42 pc from the surface of the cloud). In the paper I, that cross lies outside the unstable shaded zone just above the upper limit for a complete shock penetration into the cloud (the middle dotted, blue line of the diagram). With the present modifications, the cross now lies near the upper limit of the shaded unstable zone for values of the cooling function $\Lambda \lesssim 10^{-25}$ erg cm$^{3}$ s$^{-1}$, or ionization fractions $ \lesssim 10^{-1}$. This result remarks how sensitive the analytical diagrams are to the choice of $\Lambda$ (or the ionization fraction) for a given initial  cloud temperature. According to the radiative cooling chemo-hydrodynamical simulations of Figure 6 of Paper I (which corresponds to the cross in the diagram of Fig. \ref{fig:diagrama}), the SNR shock front really stalls within the cloud before being able to cross it completely and, after all the shocked cloud material does not reach the conditions to become Jeans unstable, as predicted, but the dense cold shell that develops may fragment and later generate dense cores, as suggested in Paper I. This points to an ambiguity of the results due to their sensitivity to $\Lambda$ and the real ionization fraction state of the gas. We should also remark that the constraint established by the dotted (blue) curves in the diagrams is actually only an upper limit for the condition of penetration of the shock into the cloud. The condition that the mach number of the shocked material goes to $M \sim 1$ implies the pressure equilibrium between the shocked and the unshocked cloud material at the time $t_{st}$ when the shock stalls within the cloud. A quick exam of the numerical simulations of Figure 6 of Paper I, however, shows that the shock front stalls even before this balance is attained. This implies that the time $t_{st}$ could be possibly smaller and therefore, the dotted (blue) curves in the diagrams should lie below the location predicted by Eq. (\ref{eq:parametro3mag}).

\end{enumerate}

In spite of the important alterations above in the diagrams of Figure \ref{fig:diagrama}, the main results and conclusions of Paper I for interactions involving SNRs with $non-magnetized$ clouds, particularly those regarding the young stellar system $\beta$-Pictoris, have  remained unchanged (see below, however, the implications for this system when the magnetic field is incorporated into the cloud).

\begin{figure}
    \centering
        \includegraphics[width=6.5cm]{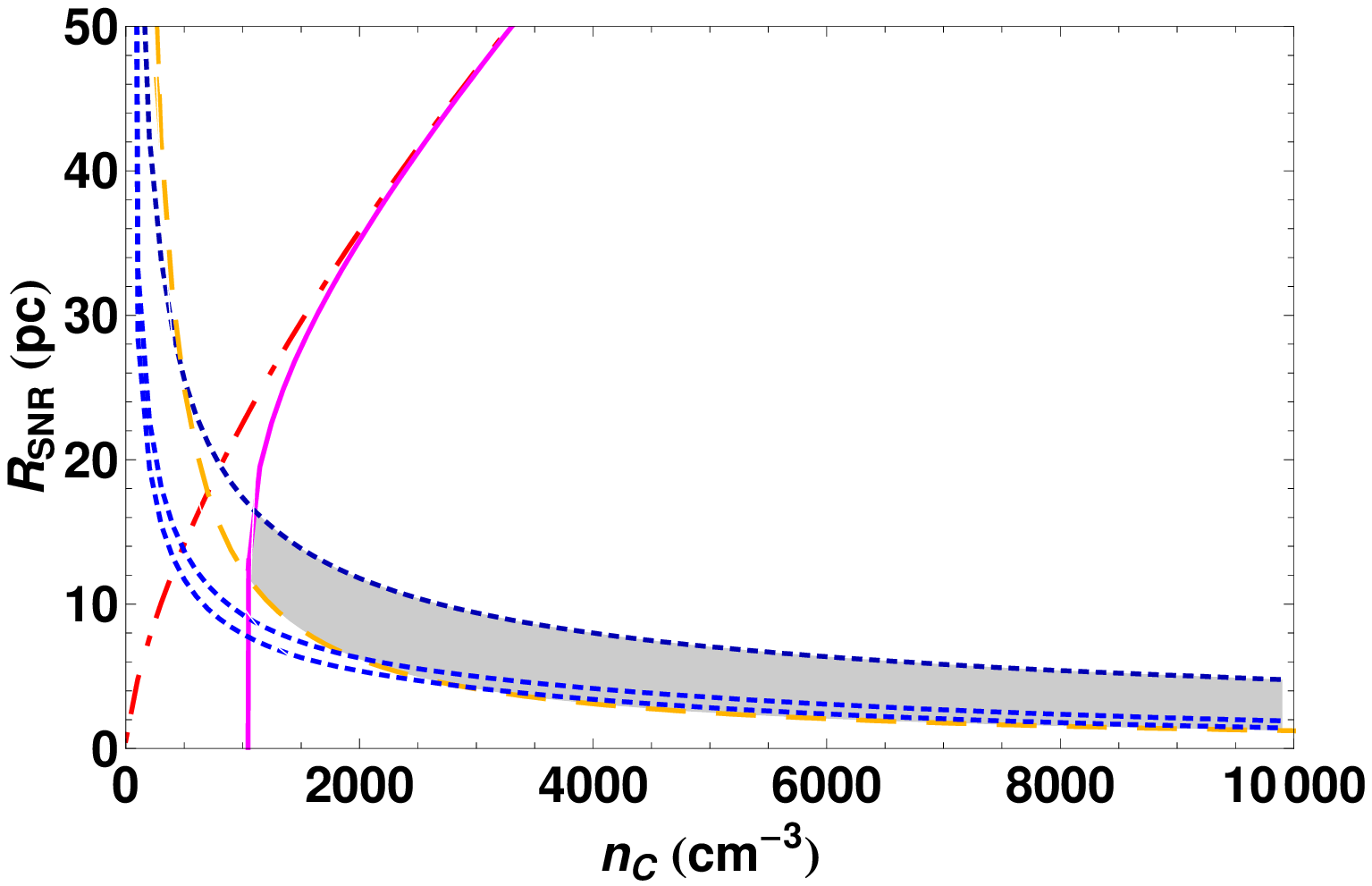}
        \includegraphics[width=6.5cm]{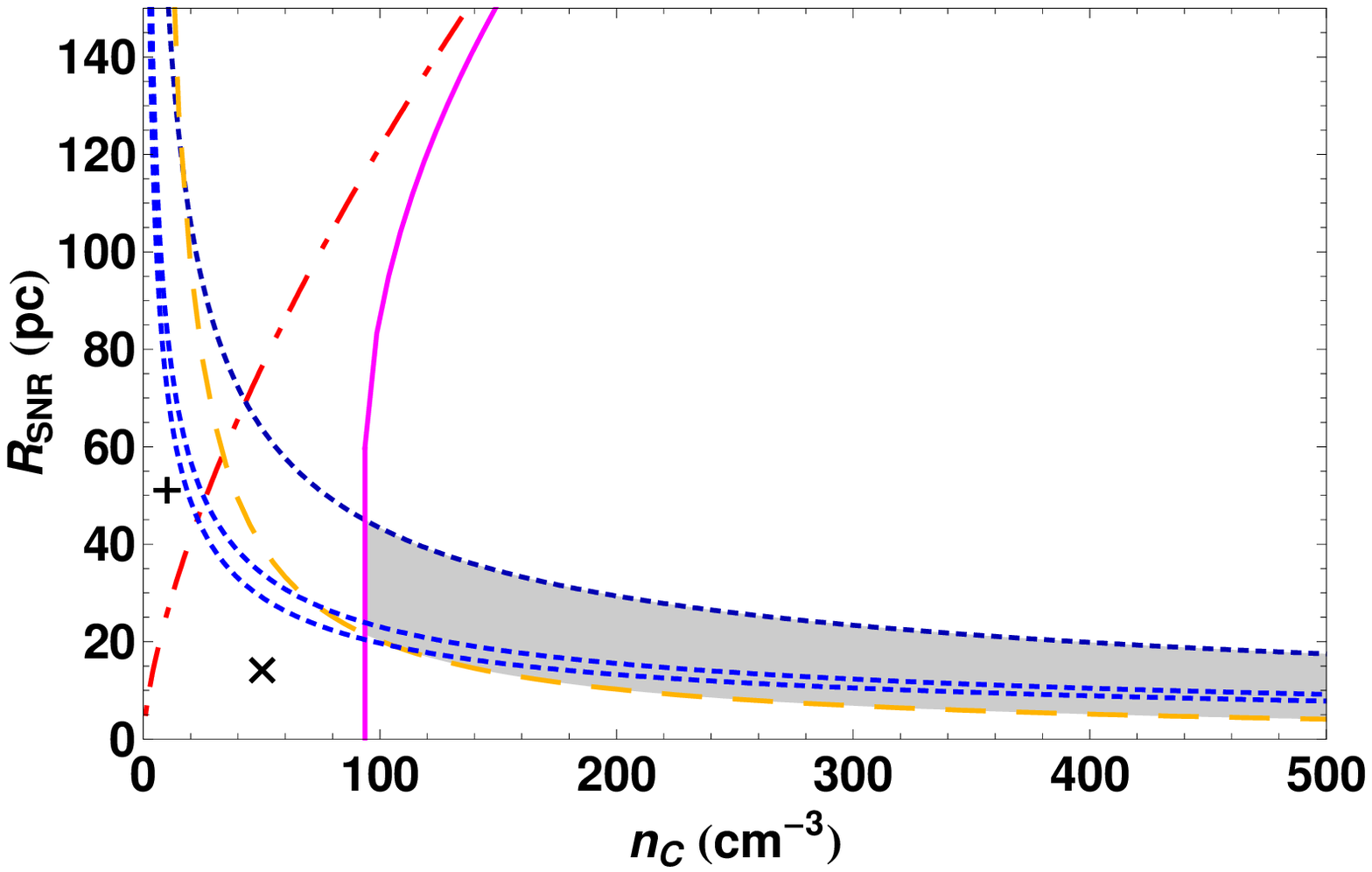}
        \includegraphics[width=6.5cm]{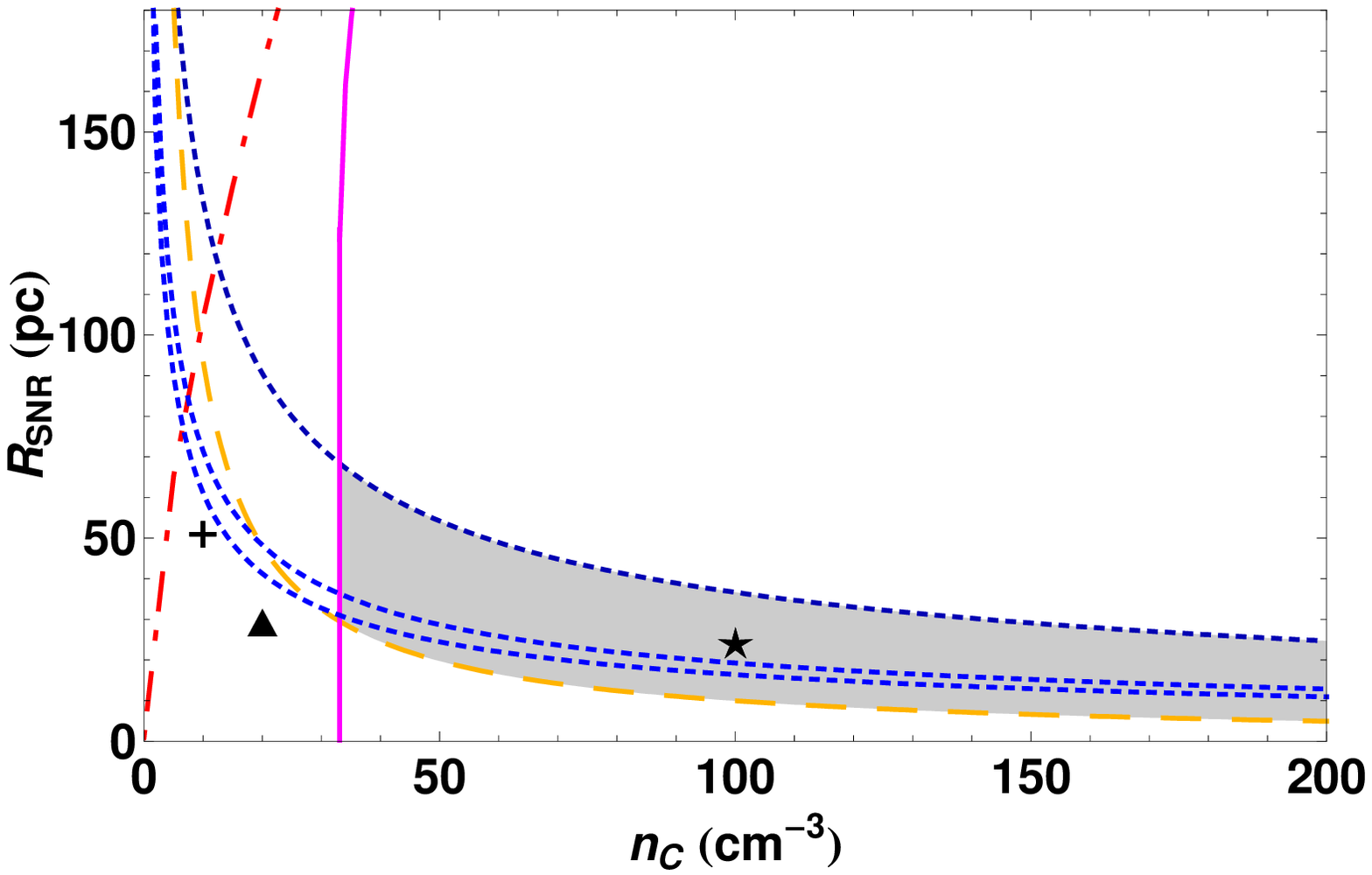}
        \includegraphics[width=6.5cm]{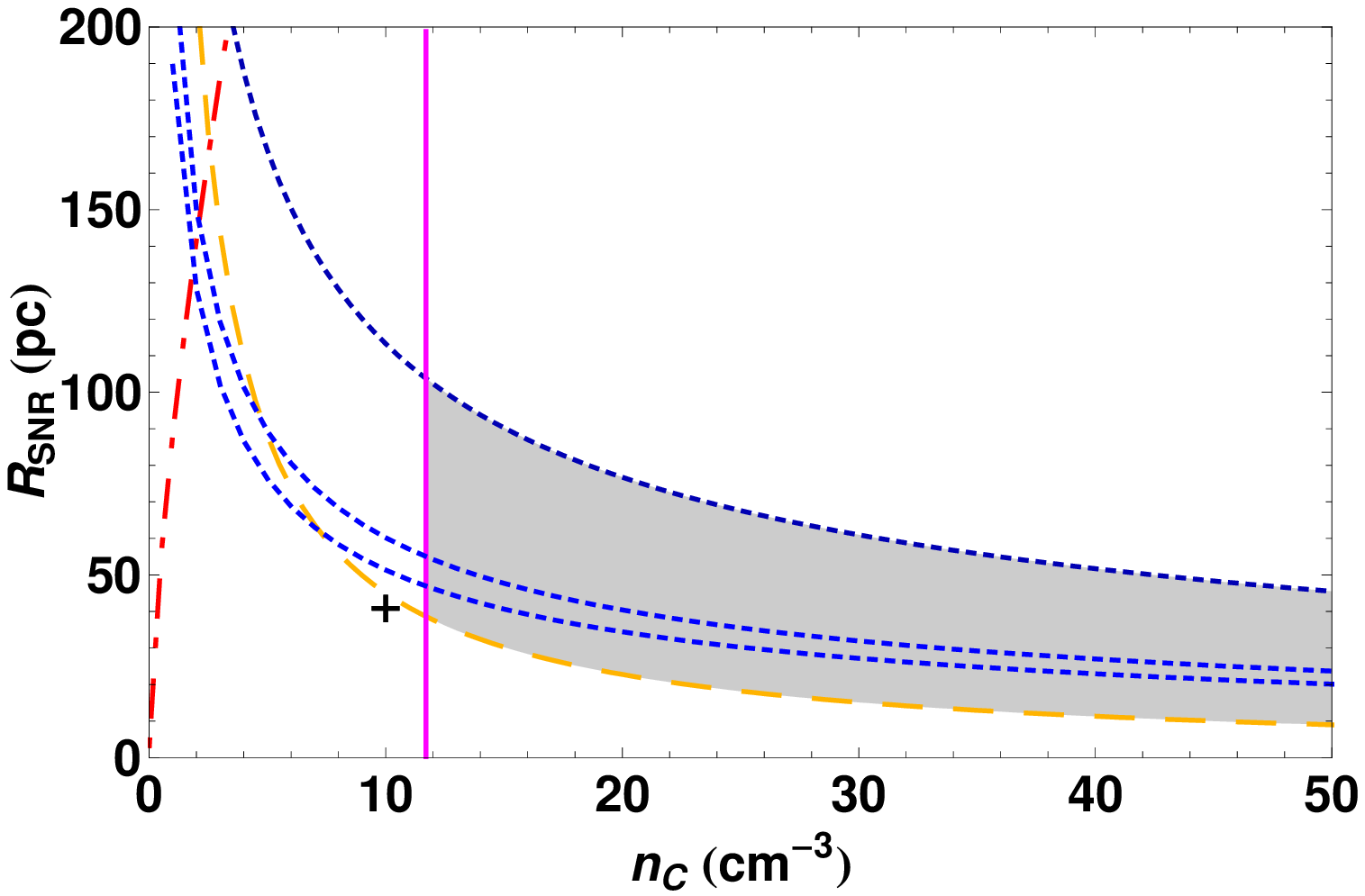}
\caption{Constraints on the SNR radius versus cloud density for 4 different cloud radius {\bf as in Figure \ref{fig:diagrama}, but} in the presence of a magnetized cloud with a normal B to the shock front B= 1 $\mu$G. Dashed (yellow) line: upper limit for complete cloud destruction after an encounter with an adiabatic SNR derived from Eq.  (\ref{eq:parametro2}); solid (magenta) line: upper limit for the shocked cloud to reach the Jeans mass derived from Eq. (\ref{eq:parametro1}); dotted (blue) lines: upper limits for the shock front to travel into the cloud before being decelerated to subsonic velocities derived from Eq. (\ref{eq:parametro3mag}) for different values of the cooling function $\Lambda(T_{sh}(t_{st}))=  10^{-25}$ erg cm$^{3}$ s$^{-1}$ (lower curve), $5 \times 10^{-26}$ erg cm$^{3}$ s$^{-1}$ (middle curve), and $3 \times 10^{-27}$ erg cm$^{3}$ s$^{-1}$ (upper curve). The dashed-dotted line (red) corresponds to the Jeans mass constraint for null magnetic field (see text for details). The shaded area defines the region where star formation can be induced by a SNR-cloud interaction (between the solid, dashed and dotted lines). The crosses {\bf and the triangle} in the panels indicate the initial conditions assumed for the clouds in the numerical simulations described in Section 4.2 of Paper I. {\bf The star corresponds to the initial conditions of the MHD simulation of Figure \ref{fig:dmag100}.}}
\label{fig:diag1muG}
\end{figure}

\begin{figure}
    \centering
        \includegraphics[width=0.8\columnwidth]{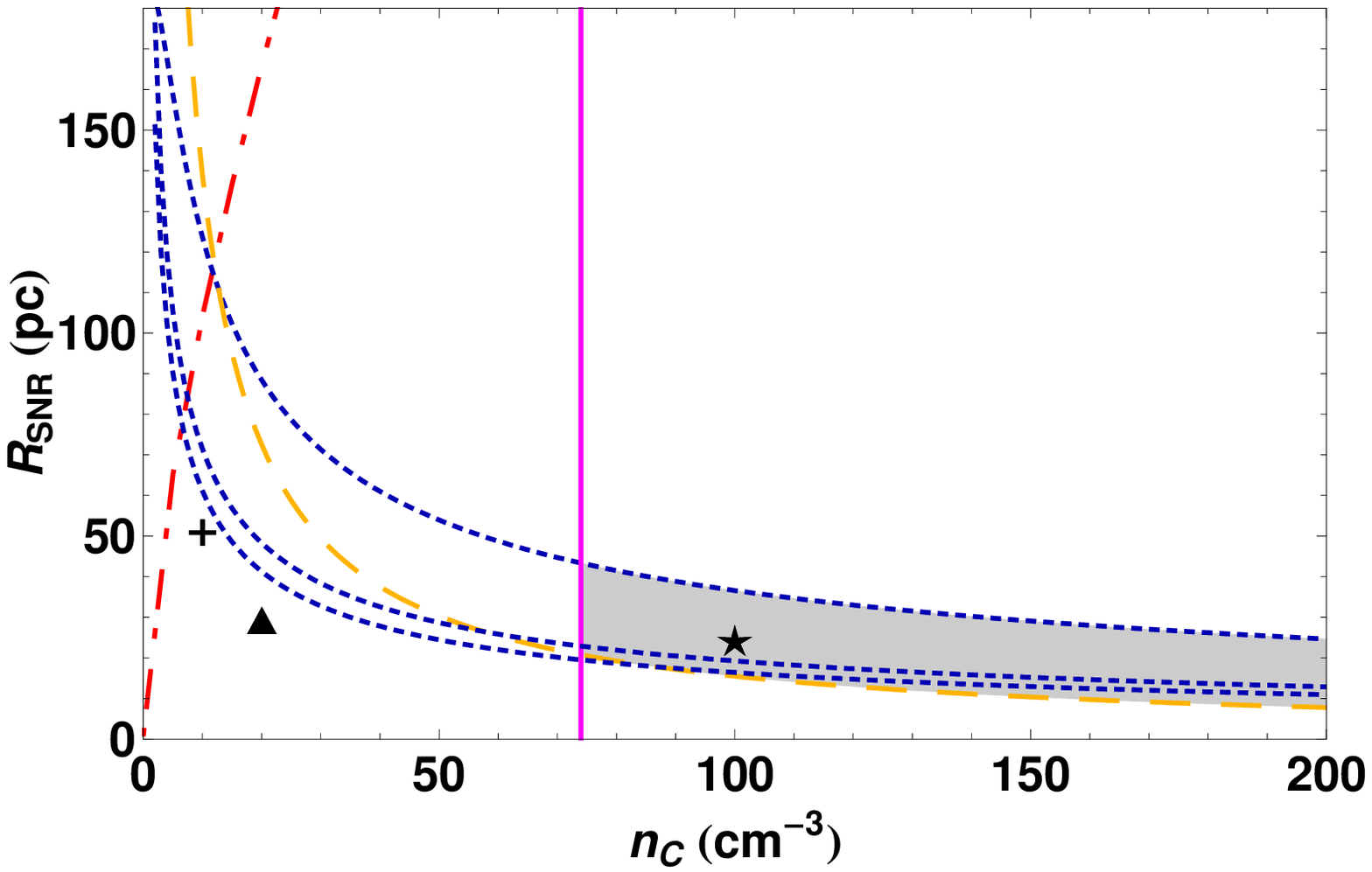}
        \includegraphics[width=0.8\columnwidth]{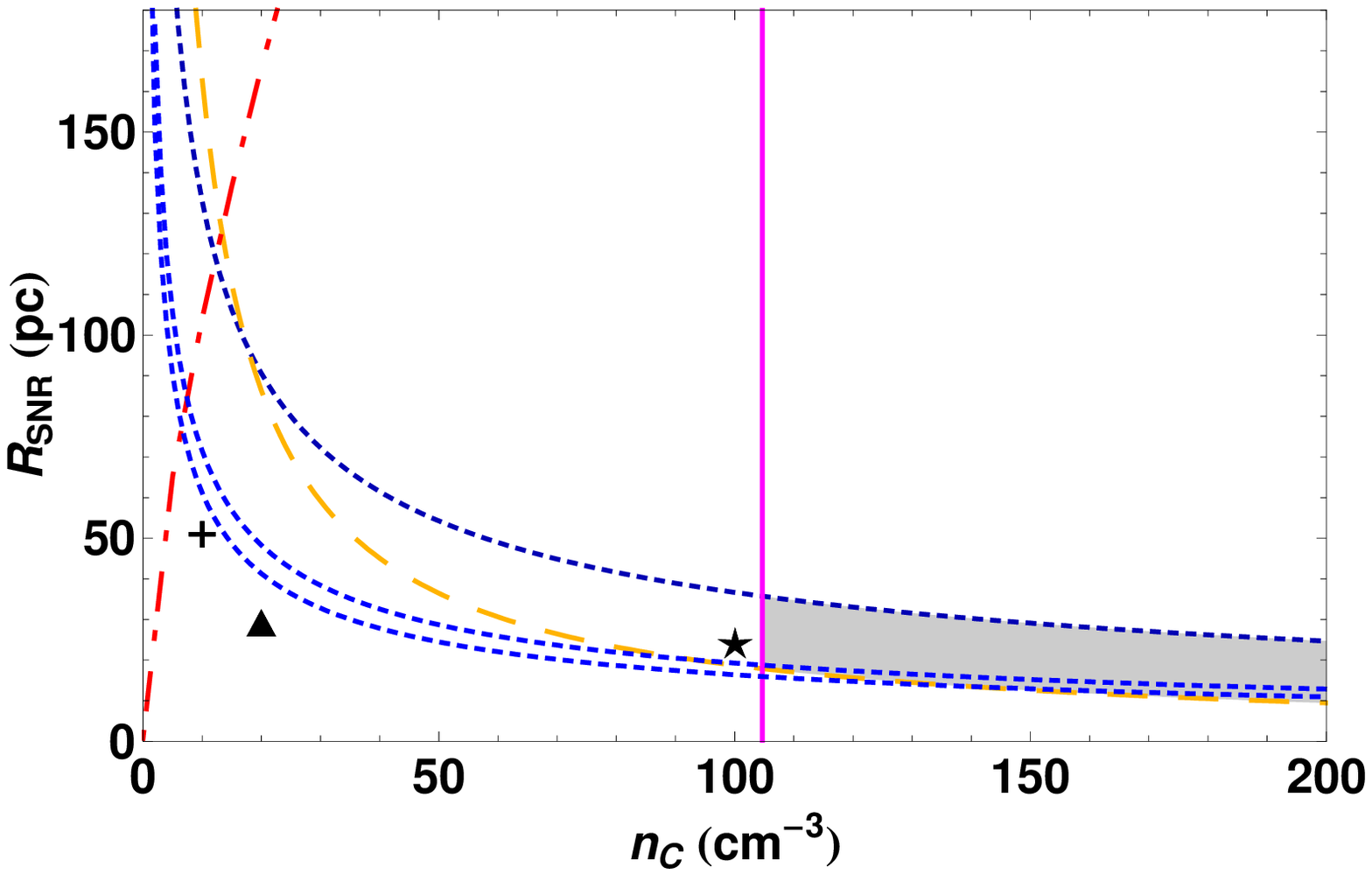}
\caption{The same as in Figure \ref{fig:diag1muG} for a cloud with $r_c=$ 10 pc and now considering {\bf $B = 5 \;\mu$G} (upper panel) and {\bf $B = 10 \;\mu$G} (bottom panel).}
\label{fig:diag10muG}
\end{figure}

Figure \ref{fig:diag1muG} shows the same diagrams as in Figure
\ref{fig:diagrama} except that they include the effects of the
magnetic field in the cloud, as described by Eqs.
\ref{eq:parametro1mag}, \ref{eq:parametro2mag} and
\ref{eq:parametro3mag}.  We notice that the presence of a normal
magnetic field to the shock front with an intensity of 1 $\mu$G
inhibits slightly the domain of SF in the diagrams, as expected. The
magnetic field plays a dominant role over the Jeans constraint (the
{\bf solid (red) line in the diagrams}) that causes a drift of the
allowed (shaded) zone of SF to the right in the diagrams (i.e., to
larger cloud densities) when compared to the diagrams without
magnetic fields of Fig. \ref{fig:diagrama}. {\bf This drift however
must be interpreted with care. When deriving the Jeans constraint in
the presence of a non-null uniform magnetic field  normal to the
shock front (eq. \ref{eq:mJmagsimples}) we assumed an
one-dimensional collapse. However, although the presence of $\vec B$
may affect the initial compression and collapse of the shocked
material, the later evolution and collapse of this material in
three-dimensional space will occur mostly in the direction parallel
to $\vec B$ and thus will not be any further affected by it. For
this reason, we should expect that the actual domain of an unstable
magnetized cloud in the diagrams of Figure 2 will be more extended
to the left of the solid (red) line and will be ultimately bounded
by the dotted-dashed (red) line that gives the Jeans constraint for
a null (or parallel) magnetic field, just like in Figure 1.  This
part of the diagrams with non-null $\vec B$ will be particularly
important when comparing them with the 3-D numerical simulations of
cloud-SNR interactions.} When larger  magnetic field intensities are
considered (5-10 $\mu$G) there is a significant shrinking of the
allowed SF zone in the diagrams (see Figure \ref{fig:diag10muG}).
This can be understood in terms of the mass to magnetic flux ratio
of the cloud before the impact. It is given by $N/B =
885(n_{c,10}r_{c,10})/B_{c,6}$. This implies $B/B_{cr}\sim 0.7
B_{c,6}/(n_{c,10}r_{c,10})$, which is larger than 1 for $n_{c,10}
\leq 34.8$ cm$^{-3}$ and $n_{c,10} \leq 69.6$ cm$^{-3}$ for an
unshocked cloud with $B=5\;\mu$G and $B=10\;\mu$G of Fig.
\ref{fig:diag10muG}, respectively, before the interaction.

As in Figure \ref{fig:diagrama}, the symbols (i.e., the crosses and
the triangle) in Fig. \ref{fig:diag1muG} indicate the initial
conditions assumed for the SNR-clouds interactions examined in the
numerical simulations described in Paper I for unmagnetized
SNR-cloud interactions. We see that when the magnetic field is
included, both the crosses and the triangle lie outside the SF
domain of the diagrams. This means that for the initial conditions
corresponding to these points in the diagram, SF is unlikely to
occur. In paper I, the application of the results of the diagram of
Figure \ref{fig:diagrama} for an unmagnetized cloud with $r_c=$ 10
pc and $n_c \gtrsim $ 10 cm$^{-3}$ to the young stellar association of
$\beta-$Pictoris ({\bf see the region near the} cross in the third
panel of Fig. \ref{fig:diagrama}) had led us to conclude that this
stellar association could have originated from recent past
interaction between a cloud and an SNR with a radius of
approximately 52 pc. However, the inclusion of an effective magnetic
field in the cloud with an intensity of only 1 $\mu$G has put the
same cross corresponding to the initial conditions for formation of
this young stellar association outside of the SF zone (see the cross
in the third panel of Fig. \ref{fig:diag1muG})\footnote{We must
remember, however, the comment of the previous paragraph.}.

\section{SNR-cloud interaction: MHD numerical simulations of self-gravitating clouds}
\label{MHD simulations}

{\bf As remarked before,} there has been  several numerical studies of the impact of shock
fronts on  interstellar clouds (e.g., Sgro 1975; McKee \& Cowie
1975; Woodward 1976; Nittmann et al. 1982; Tenorio-Tagle \& Rozyczka
1986; Hartquist et al. 1986; Bedogni \& Woodward 1990;  Mac Low et
al. 1994; Klein, McKee \& Colella 1994; Anderson et al. 1994; Dai \&
Woodward 1995; Xu \& Stone 1995;  Jun, Jones \& Norman 1996; Redman,
Williams \& Dyson 1998; Jun \& Jones 1999; Lim \& Raga 1999; de
Gouveia Dal Pino 1999; {\bf Miniati, Jones \& Ryu 1999;} Poludnenko, Frank \& Blackman 2002; Fragile
et al. 2004; Steffen \& L\'opez 2004, Raga, de Gouveia Dal Pino et
al. 2002; Fragile et al. 2005; Marcolini et. al 2005; Melioli, de
Gouveia Dal Pino \& Raga 2005; Nakamura et al. 2006) most of which
were mainly concerned with the effects of these interactions upon
the destruction of the cloud. In particular, the studies that
incorporated the effects of the radiative cooling revealed the
relevance of it in delaying the destruction of the cloud and the
mixing of its materials with the ISM (e.g., Melioli, de Gouveia Dal
Pino \& Raga  2005).

In paper I, in order to check the  predictions of our  semi-analytic
diagrams  built for interactions involving SNR shocks and
non-magnetized clouds, we  performed  3-D hydrodynamical radiative
cooling simulations following the initial steps of these
interactions. Here, we  repeat this analysis but also take into
account  the effects of the magnetic fields and the self-gravity, in
order to follow the late evolution of the shocked material within
the magnetized clouds and check whether it suffers gravitational
collapse or not, in consistence with our diagrams.

To this aim, we have employed the grid code Godunov-MHD presented in Kowal \& Lazarian (2007), and tested in Falceta-Gon\c{c}alves et al. (2008), which solves the gas dynamical equations in conservative form as follows:

\begin{equation}
\frac{\partial \rho}{\partial t} + \mathbf{\nabla} \cdot (\rho{\bf v}) = 0,
\end{equation}

\begin{equation}
\frac{\partial \rho {\bf v}}{\partial t} + \mathbf{\nabla} \cdot \left[ \rho{\bf v v} + \left( p+\frac{B^2}{8 \pi} \right) {\bf I} - \frac{1}{4 \pi}{\bf B B} \right] = -\rho \mathbf{\nabla \Phi},
\end{equation}

\begin{equation}
\frac{\partial \mathbf{B}}{\partial t} - \mathbf{\nabla \times (v \times B)} = 0,
\end{equation}

\noindent with $\mathbf{\nabla \cdot B} = 0$, where $\rho$, ${\bf v}$ and $p$ are the plasma density, velocity and pressure, respectively, ${\bf B}$ is the magnetic field and $\mathbf{\nabla ^2 \Phi}=4 \pi G  \rho$. The equations are solved using a second-order Godunov scheme, with an  HLLD Riemann solver to properly consider the MHD characteristic speeds. For the  self-gravity term, we used the FACR (Fourier Analysis Cyclic Reduction) Poisson solver at each time step. In Paper I, given the importance of the radiative cooling behind the shocks, we simulated explicitly its effects  in the hydrodynamical simulations then presented. Presently, since the main focus in the simulations is to study  the role of the magnetic field and self-gravity on the  evolution of the clouds, we solved the equations under the  approximation of strong radiative cooling. The set of  equations above is closed by the equation of state $p \propto \rho^{\gamma}$, setting an effective $\gamma_{eff} = 1.2$ to simulate the strong radiative cooling.

\begin{table*}
\centering
\caption{Description of the simulations}
\begin{tabular}{ccccccc}
\hline\hline
$n_c$ (cm$^{-3}$) & $r_c$ (pc) & $R_{SNR} (pc)$ & code & $B$ ($\mu$G) & Result & Prediction \\
\hline
    10  & 20 & 40 & Hydro & - & collapse & collapse  \\
    10  & 20 & 40 & MHD   & 1 & stable   & stable    \\
    50  & 5  & 15 & MHD   & 1 & stable   & stable    \\
    10  & 10 & 50 & MHD   & 1 & stable   & stable    \\
    100 & 10 & 25 & MHD   & 1 & collapse & collapse  \\
\hline\hline
\end{tabular}
\label{tab:tabela}
\end{table*}

The computational box has dimensions 100 pc $\times$ 100 pc $\times$ 100 pc, corresponding to a fixed mesh of 256$^3$ grid points. A SNR is generated by the explosion of a SN with energy $E_0  = 10^{51}$ erg initially injected at the left-bottom corner of the box.  Several runs were carried out with different initially uniform cloud densities ($n_c$), radii ($r_c$),  and distance between the SNR and the cloud's surface  ($R_{SNR}$). The initial conditions are described in Table 1. We have selected two of these simulations to show in detail, as  follows.

\begin{figure*}
\centering
\includegraphics[width=6.7cm]{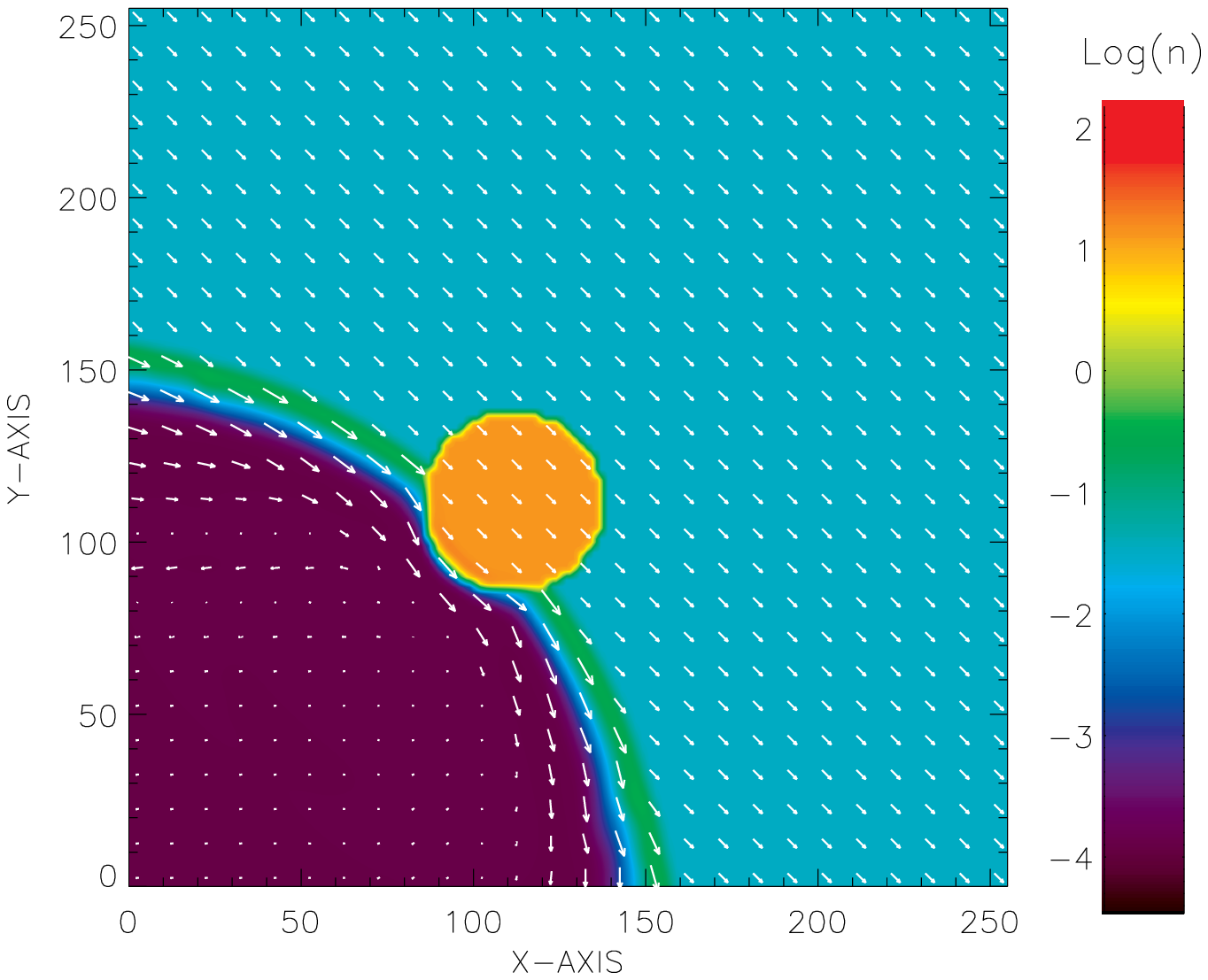}
\includegraphics[width=6.7cm]{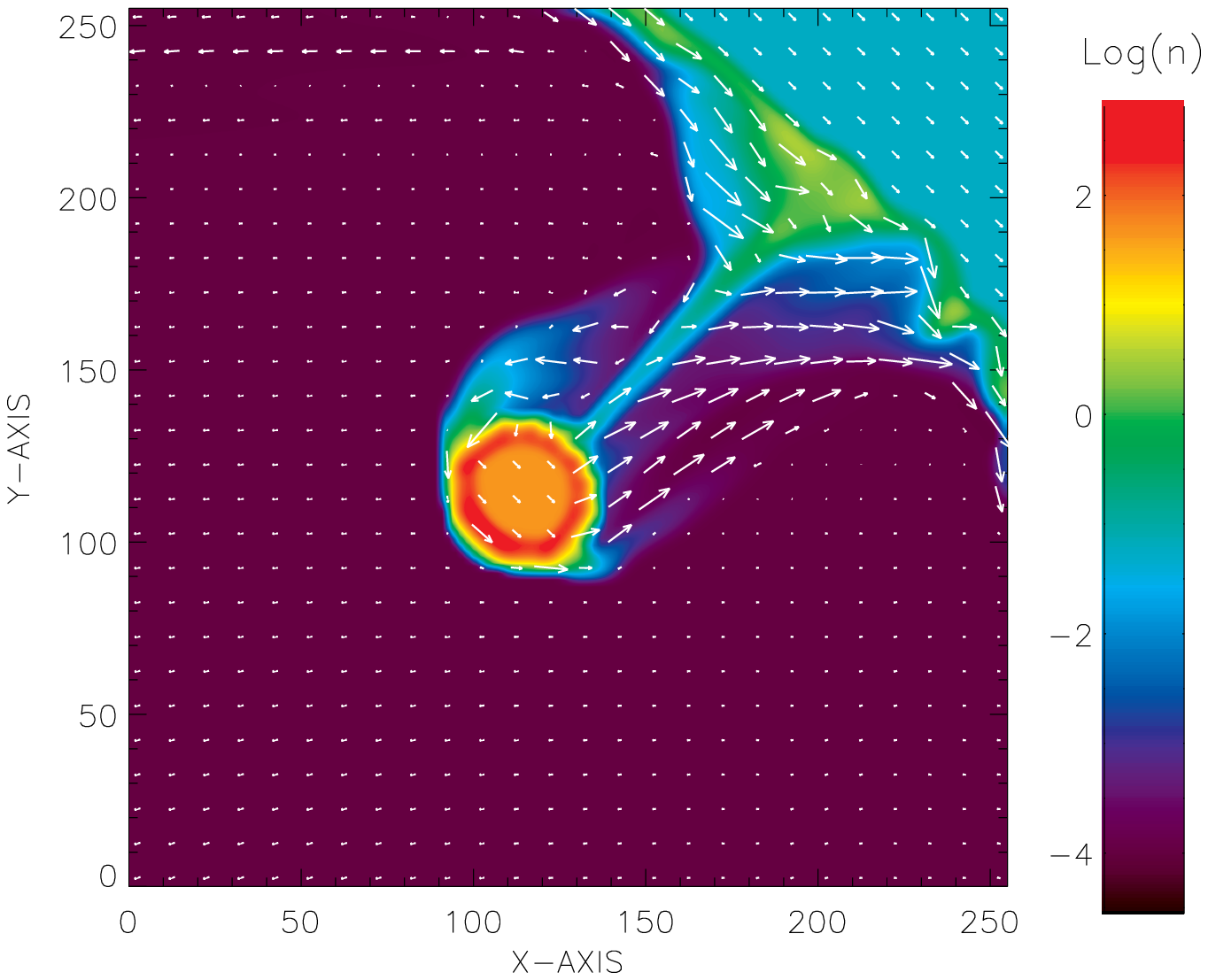}\\
\includegraphics[width=6.7cm]{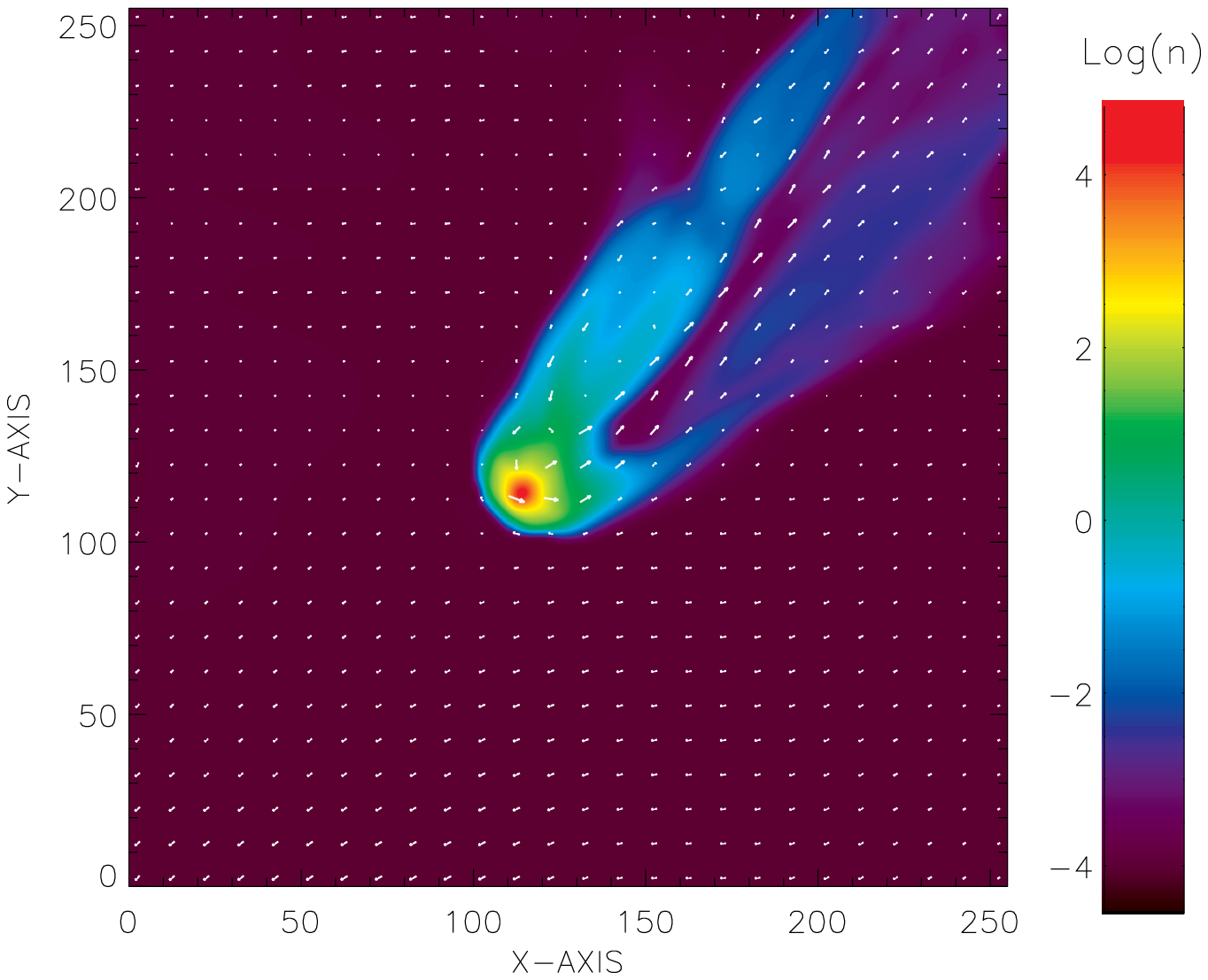}
\includegraphics[width=6.7cm]{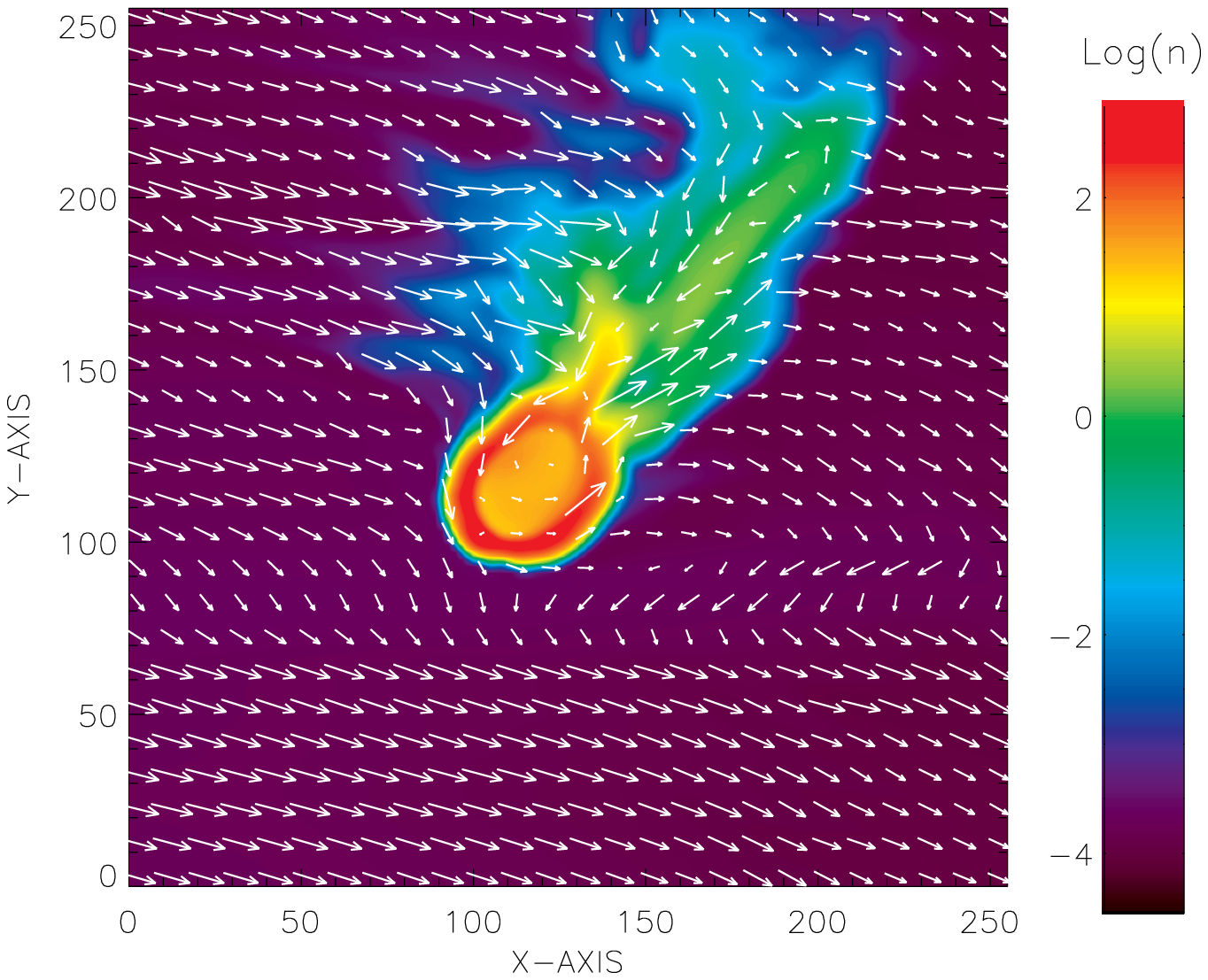}
\caption{Colour-scale maps of the mid-plane density distribution and magnetic field vectors  evolution for the  model with  $n_c=10$ cm$^{-3}$, $r_c=10$ pc, $R_{SNR}=50$ pc, and $B=1 \mu$G. The evolutionary  times are $t=3 \times 10^5$yr (top left-hand panel); $8 \times 10^5$yr (top right-hand panel); $1.8 \times 10^6$yr (bottom left-hand panel); and $3.3 \times  10^6$yr (bottom right-hand panel).The ISM where the SNR expands has a number density $n = 0.05$ cm$^{-3}$, and a temperature $10^4$ K.   The field vectors are normalized by the maximum at each frame.}
\label{fig:dmag10}
\end{figure*}

Fig. \ref{fig:dmag10} depicts the maps of the density and magnetic
field vectors at the mid x-y plane that intercepts both the  cloud
and de SNR for a cloud with $n_c= 10$ cm$^{-3}$, $r_c =10$ pc,
$R_{snr}=50$ pc. These initial conditions represent the cross at the
third panel of Fig. \ref{fig:diag1muG} (from top to bottom) which
lies well outside of the SF shaded zone in the diagram and predicts
cloud destruction due to the impact. In the simulation of Fig. 4,
the SNR shock front compresses the cloud increasing the magnetic
energy density. After the  interaction (at $t \simeq 1.8 \times 10^6$
s, a tail of gas is swept behind the cloud by the expanding SNR, {\bf similarly to the results obtained, e.g., by
Murray et al. (1993)}.
The magnetic energy  density along the tail is also increased.  We
see that, {\bf due to the compression,} the cloud initially becomes gravitationally unstable and
starts to collapse. At $1.8 \times 10^6$yr, the core is $\sim 10^3$
times denser. {\bf The
contraction of the cloud also causes the increase of the magnetic energy
that acts against further collapse. Later, the simulated cloud
rebounds and undergoes expansion and evaporation, as seen
at $3.3 \times 10^6$yr. This re-expansion may be occurring mostly because of the inefficient
cooling of the simulation. Since we set $\gamma_{eff} = 1.2$ the gas
may be not being realistically cooled, as one should expect for a real ISM collapsing cloud.
In such situations, $\gamma_{eff}$ may be even smaller than unity, i.e. the
shocked material may present temperatures lower than right before the shock. On the other hand,
Spaans \& Silk (2000) modeled the chemistry, thermal balance and radiative
transfer for different conditions of the ISM, and showed that a polytropic pressure
equation may be used as first approximation. The effective polytropic index is
$\gamma_{eff} \sim 0.8 - 1.4$, depending on the local conditions.
Nonetheless, even though the real cloud would not re-expand in the presence of  a  stronger radiative cooling, the non-collapsing condition is still possible due to  the increase in the magnetic energy. In any case, we
are currently implementing a more realistic method to calculate the cooling function
in our Godunov-MHD code, based on the interpolation method for a table of
$\Lambda(T)$ parameter (see Stone et al. (2008)). We intend to compare the presented results
with  more realistic calculations in a forthcoming work.}

{\bf In Paper I, where non-equilibrium radiative cooling has been properly taken into account, but no magnetic field or self-gravity were considered, the hydrodynamical simulations (Fig. 4 of that paper) suggest that the  cloud evaporates. As the magnetic field is introduced  (Fig. \ref{fig:dmag10} of the present paper)  the increasing magnetic pressure at the later stages of the cloud evolution prevents the collapse, as predicted by the SF diagram of Fig. \ref{fig:diag1muG}.}

\begin{figure*}
\centering
\includegraphics[width=6.7cm]{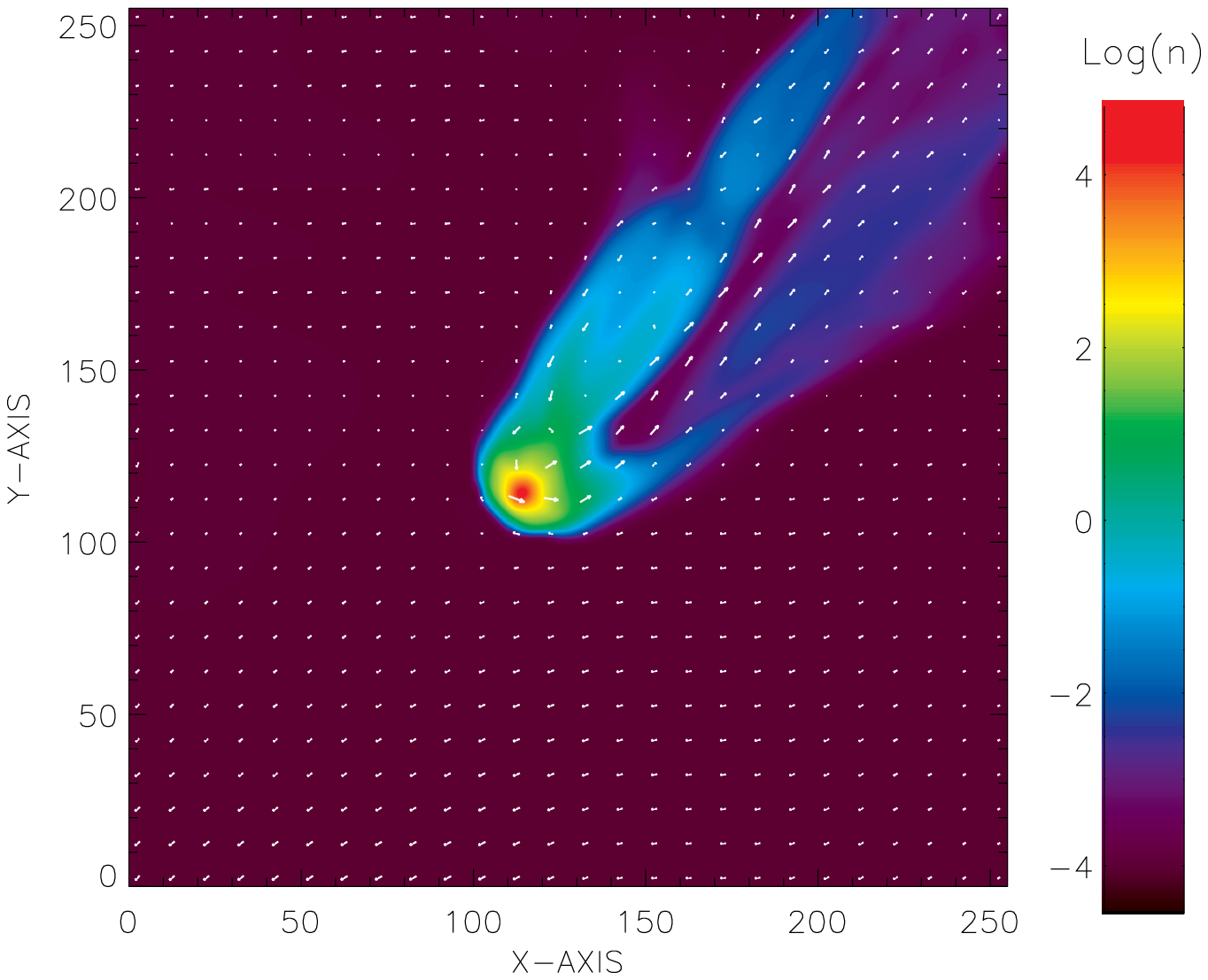}
\includegraphics[width=6.7cm]{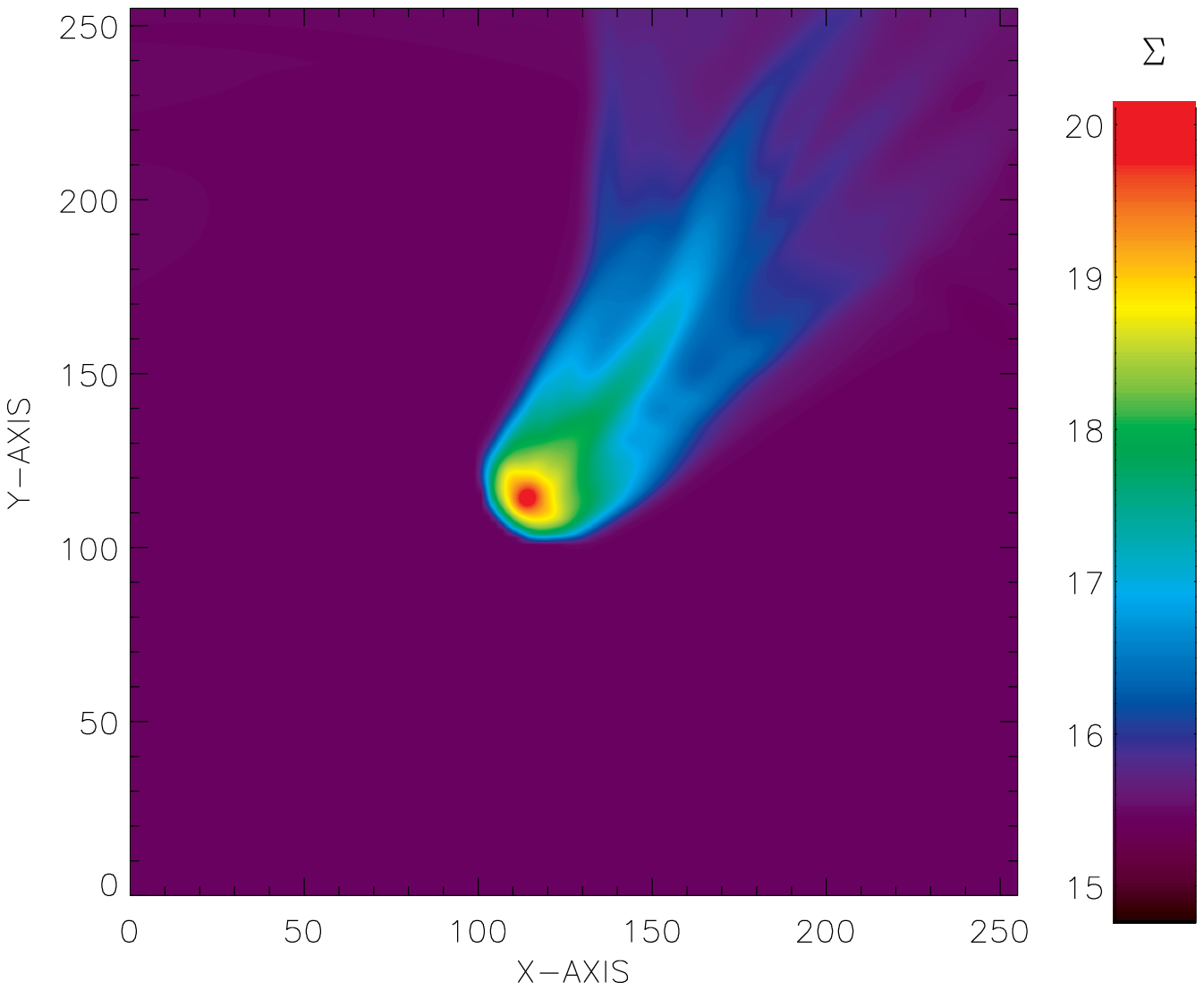}\\
\includegraphics[width=6.7cm]{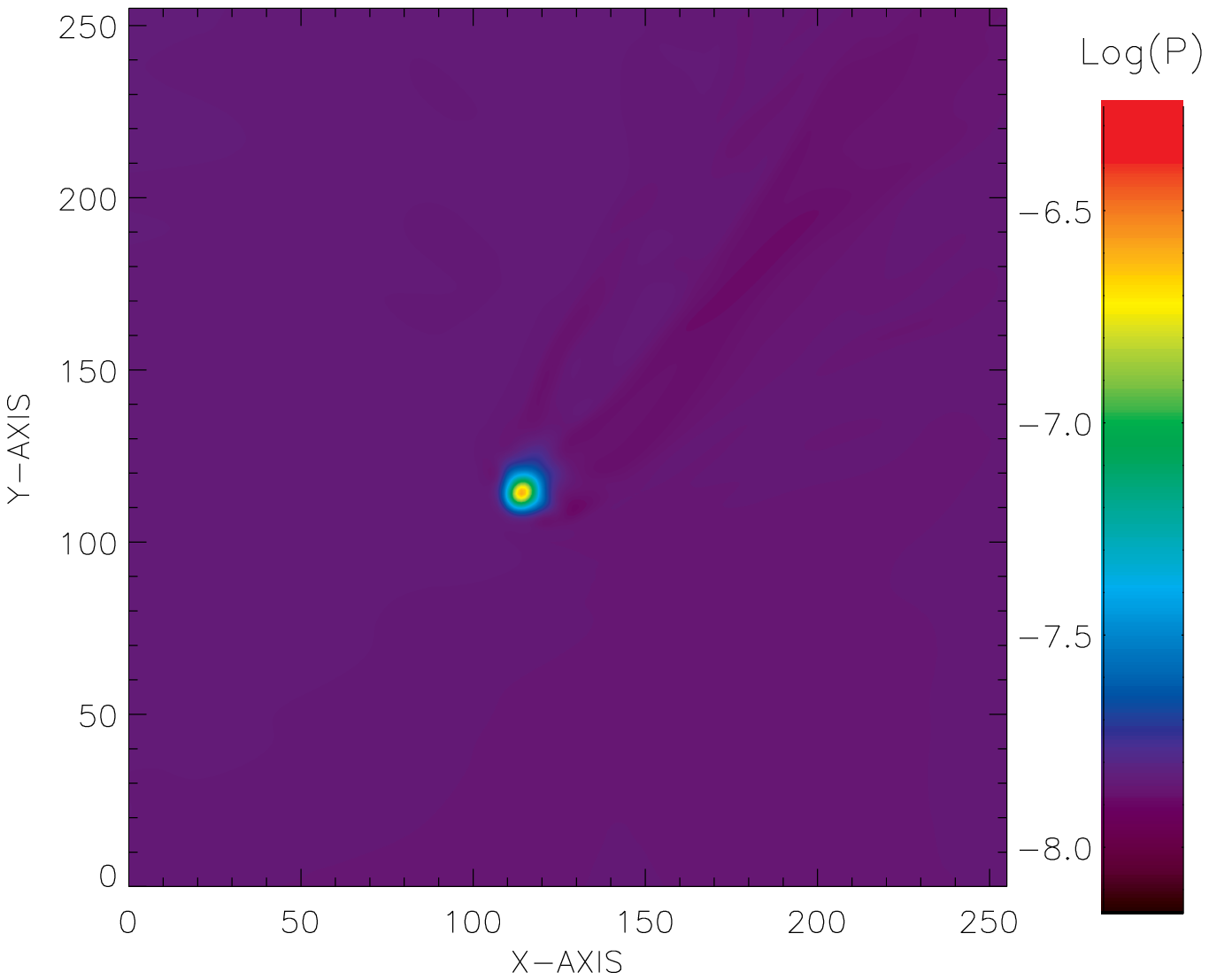}
\includegraphics[width=6.7cm]{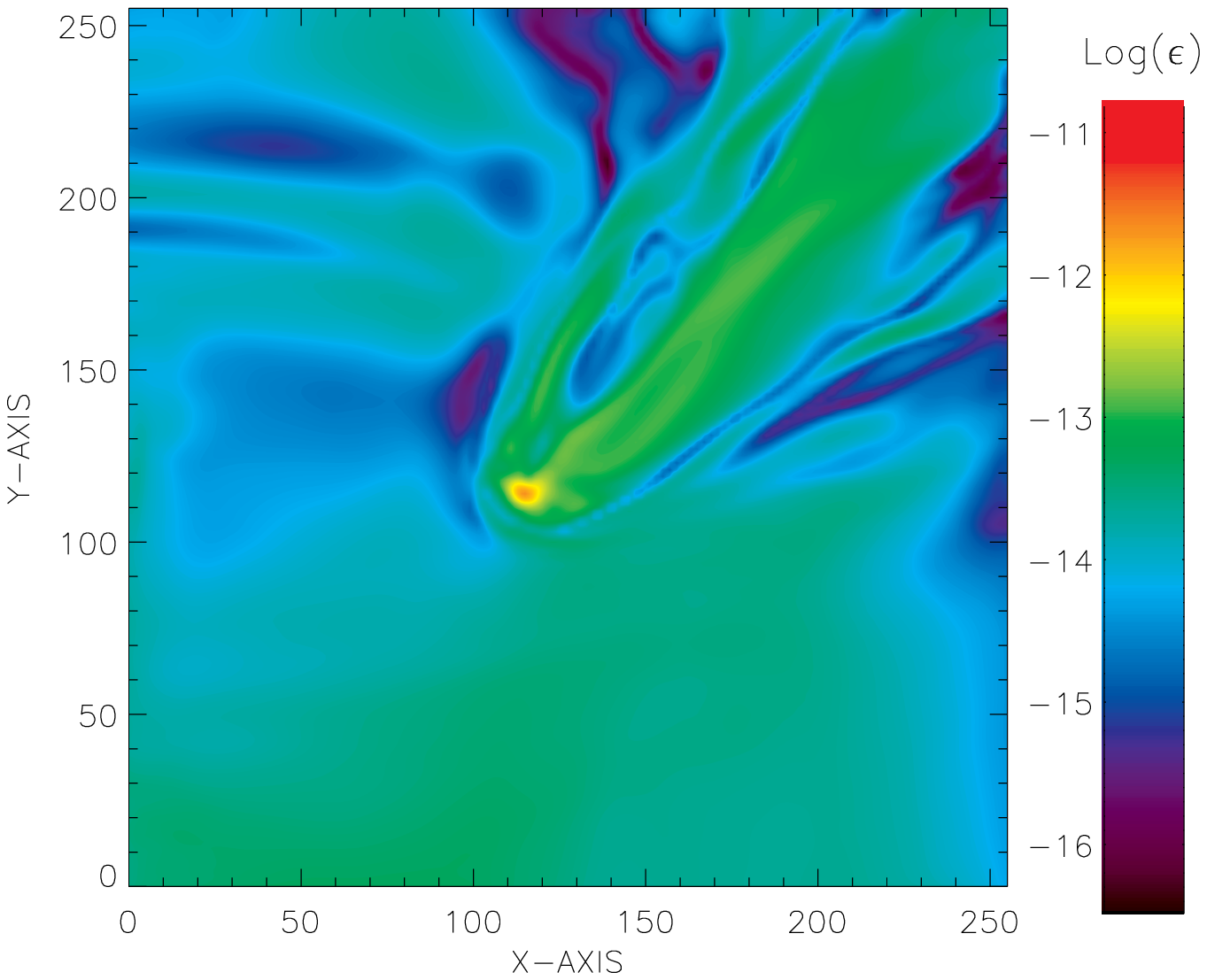}
\caption{Colour-scale maps of the mid-plane density with magnetic field vectors,  column density, gas pressure and  magnetic energy density for the same model as in Figure \ref{fig:dmag10} at $t=1.8
\times 10^6$yr.} \label{fig:model1}
\end{figure*}

{\bf Figure \ref{fig:model1} shows the distribution of several physical  parameters of the model of Fig. 4 at $t = 1.8 \times 10^6$ yr. The column density is $N > 10^{20}$ cm$^{-2}$ in the core. As remarked above, we see that the dense core formed during the initial collapse results an increase of the total gas and magnetic pressures with a maximum temperature $T > 10^4$ K, while the magnetic energy density is an order of magnitude larger than that of the surrounding medium. The large increase of the gas pressure is probably the main responsible for the  re-expansion of the cloud. This is highly dependent on the  radiative cooling. In a more realistic situation, the  cooling timescale $t_{cool} \sim kT/n\Lambda(T)$ should be generally  shorter than the collapsing timescale $t_g \sim (G \rho)^{-1}$,  which means that the core temperature should be smaller than $10^4$K.  Nevertheless, the magnetic field also plays a crucial role in stabilizing the cloud and supporting the cloud against collapse, as discussed below in Fig. \ref{fig:masstoflux}.}

\begin{figure*}
\centering
\includegraphics[width=6.7cm]{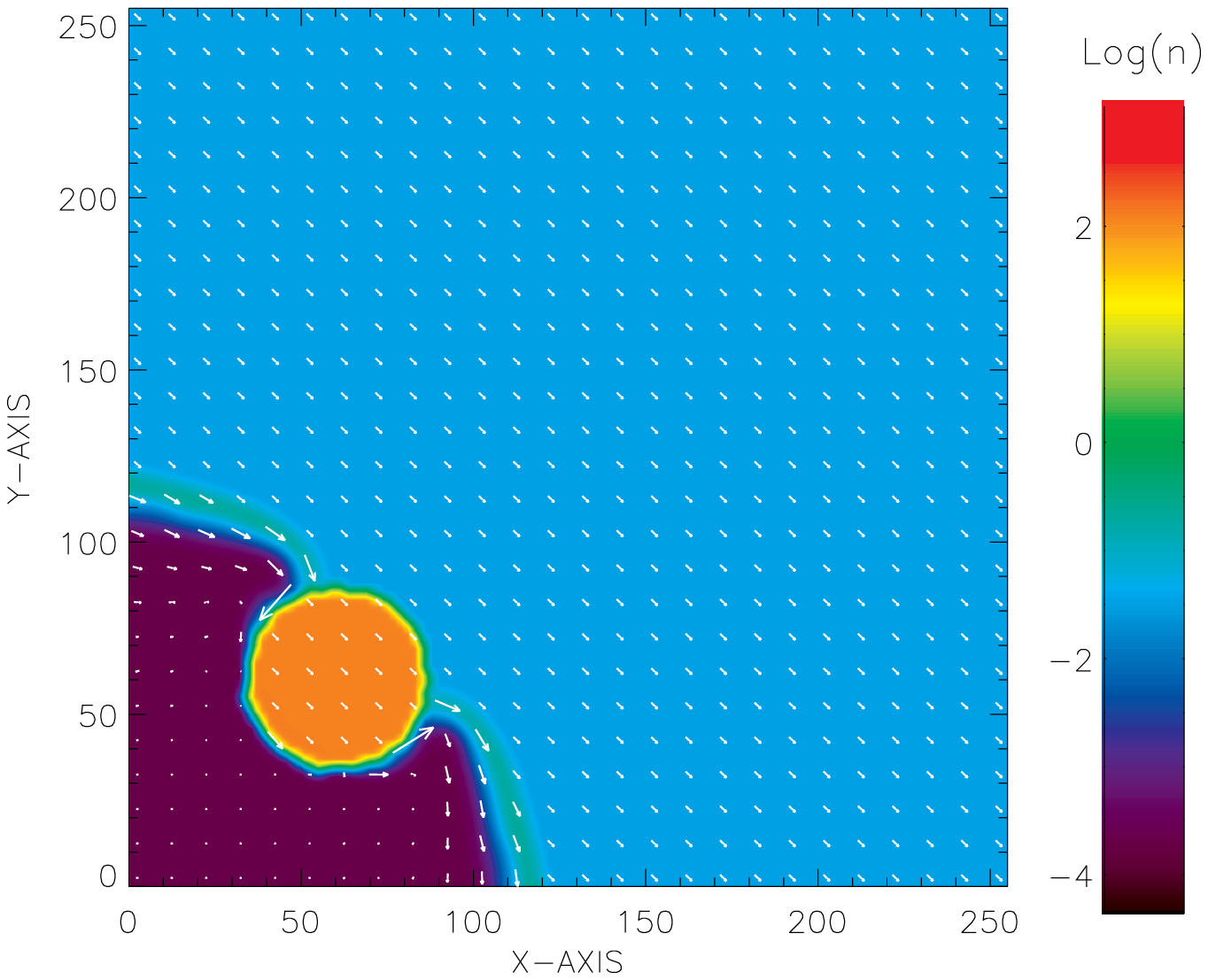}
\includegraphics[width=6.7cm]{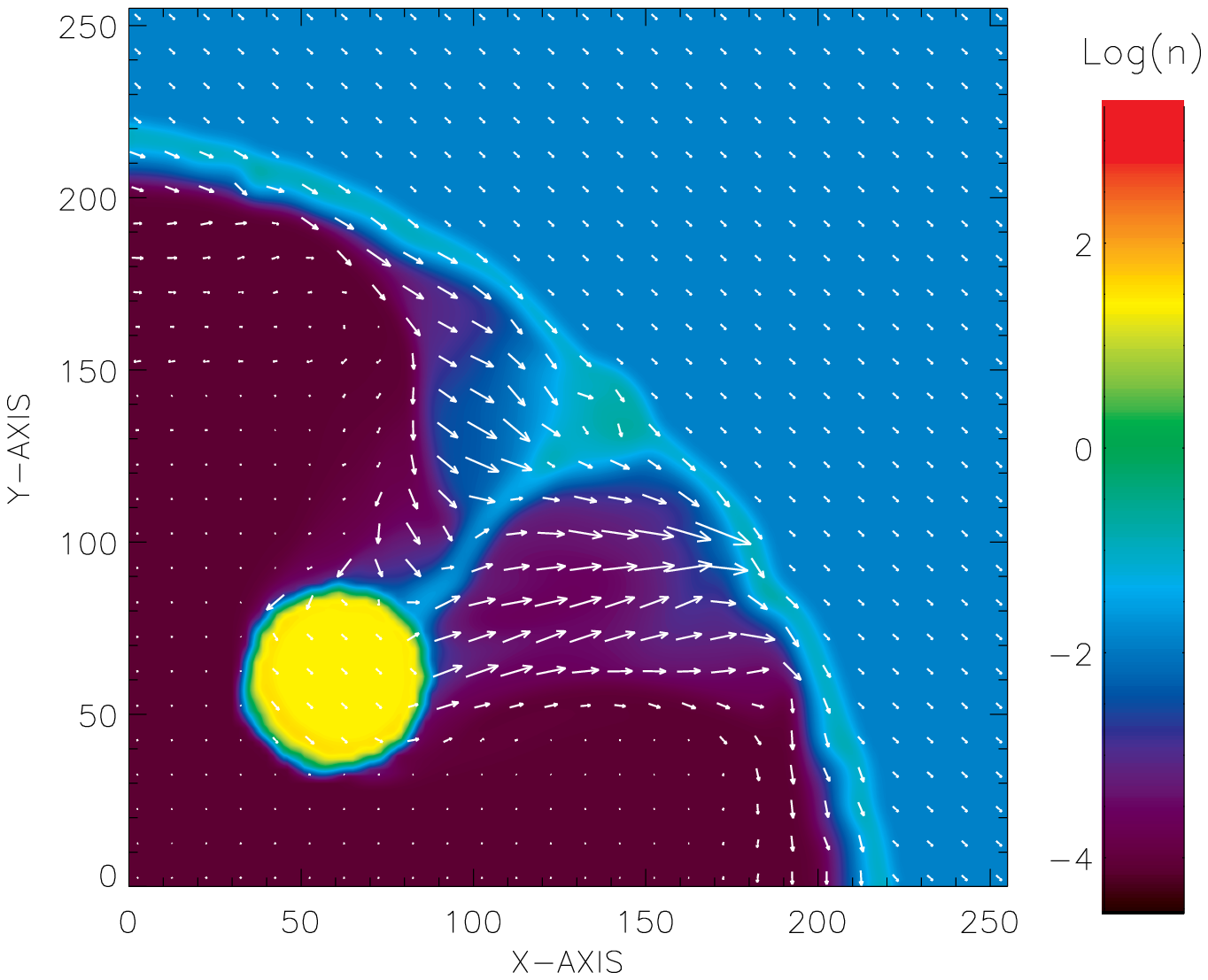}\\
\includegraphics[width=6.7cm]{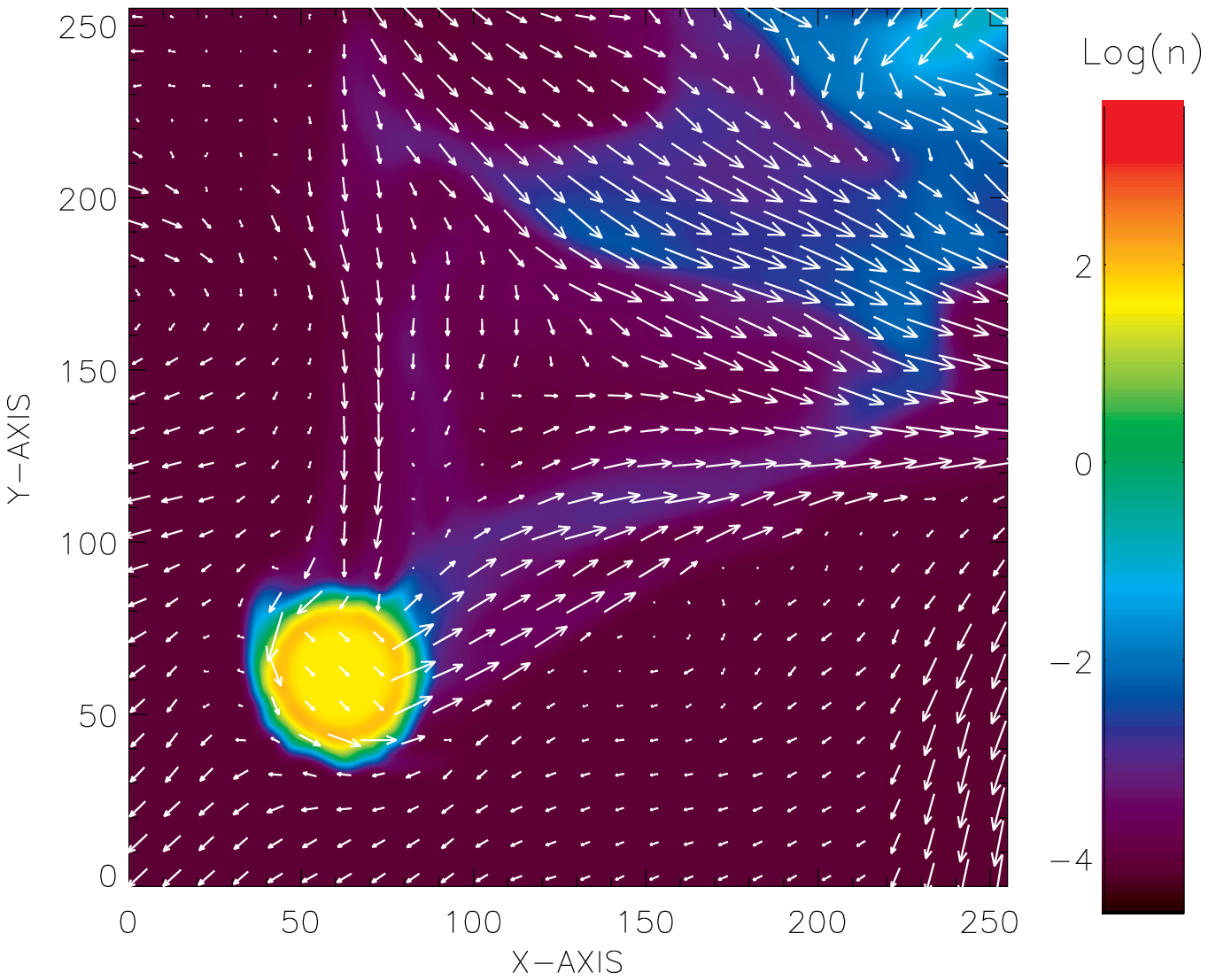}
\includegraphics[width=6.7cm]{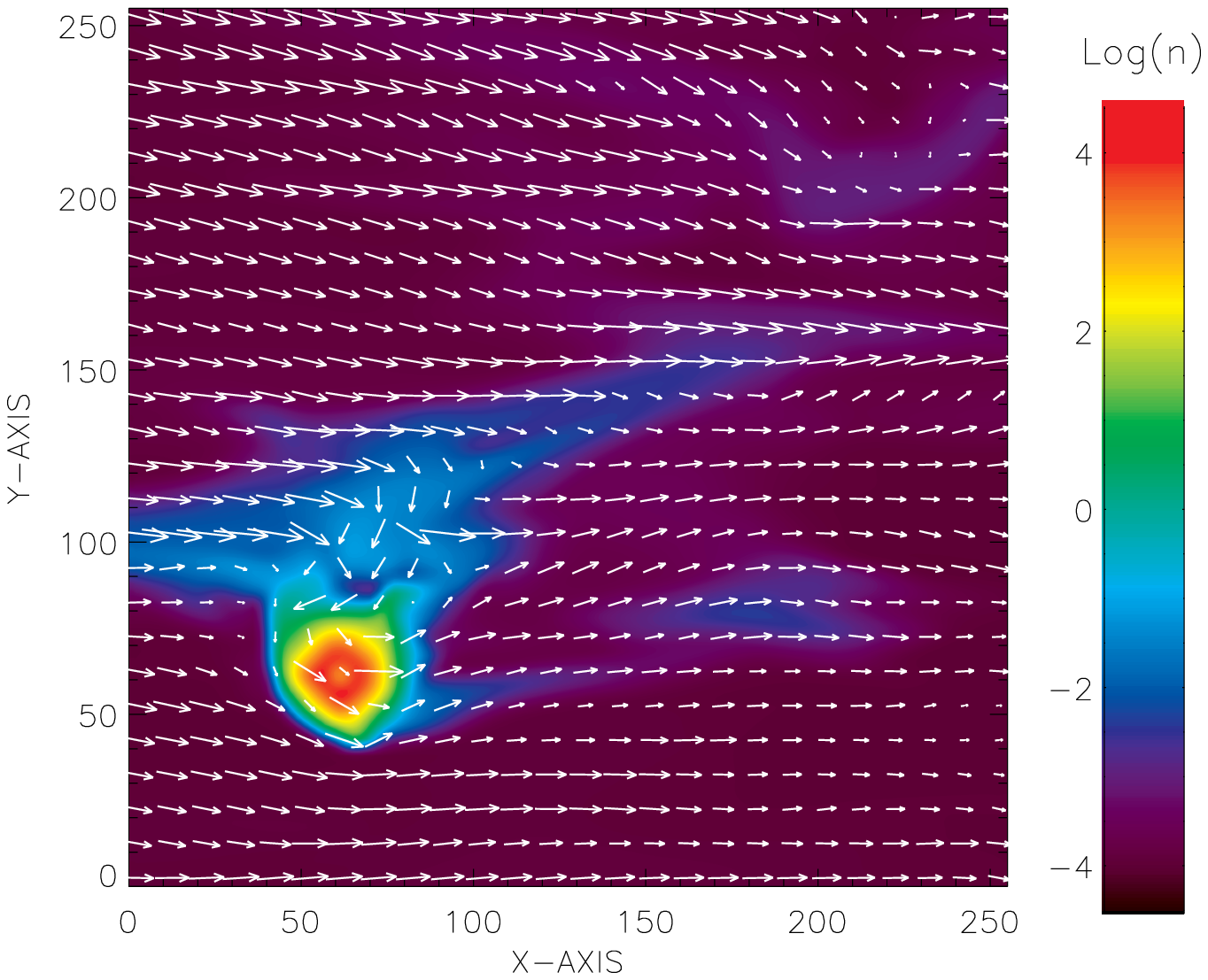}
\caption{Colour-scale maps of the mid-plane density distribution and magnetic field vectors for the model with  $n_c=100$ cm$^{-3}$, $r_c=10$ pc, $R_{SNR}=25$ pc, and $B=1 \mu$G. The evolutionary times are $t=2 \times 10^5$ yr (top left-hand frame); $5 \times 10^5$ yr (top right-hand frame);  $1.2 \times 10^6$yr (bottom left-hand frame); and $2.5 \times  10^6$ yr (bottom right-hand frame). The vectors are normalized by the maximum at each frame.}
\label{fig:dmag100}
\end{figure*}

Figure \ref{fig:dmag100} shows the mid-plane maps of the density distribution  and the magnetic field vectors for the model of Table \ref{tab:tabela} with a cloud with $n_c = 100$ cm$^{-3}$ and $r_c = 10$ pc, and a SNR with $R_{SNR} = 25$pc. The evolution in this case is similar to the previous model. However, after $t = 1.2 \times 10^6$yr, as the cloud collapses the magnetic pressure is not able to counter-balance  gravity and the cloud keeps contracting. The higher cloud density in this case causes an efficient radiative cooling of the shock compressed material in the cloud that keeps the thermal energy low. Thus the collapse simply drags the magnetic field lines that increase the magnetic energy density, but this will never be able to stop the collapse. This result is consistent with the predictions of the SF formation diagram. The initial conditions of this system of Fig. \ref{fig:dmag100} correspond to the star symbol at the third panel of Fig. \ref{fig:diag1muG} (from top to bottom) which lies inside the SF shaded zone.

As remarked before, the stability of a cloud supported by the magnetic  pressure may be quantified by the  mass-to-magnetic flux ratio, $M/\Phi \simeq N/B$, where N is the cloud column density (e.g., Crutcher 1999). We have computed this mass-to-flux ratio for the simulated clouds above at several time-steps. The  results are shown in Figure 7. The open dots represent the maximum column density  for each snapshot as a function of the magnetic field averaged along the given line of sight (with  maximum $N$). We notice  that for the model with $n_c=10$ cm$^{-3}$, $r_c=10$ pc and $R_{SNR}=50$ pc (of Fig. 4), the cloud starts collapsing right after the interaction with the SNR. The  column density increases and reaches the unstable regime. The cloud contracts and the  total energy increases. Both the internal and the magnetic energy densities suppress further collapse and then the cloud  re-expands. {\bf Despite of the probably unrealistic re-expansion, as discussed before, the final stage is a stable cloud. For the case of $n_c = 100$ cm$^{-3}$, $r_c = 10$ pc, and $R_{SNR} = 25$ pc, the same process occurs initially. However, the increase of the magnetic and internal energy densities are not enough to avoid the continuous collapse.}

\begin{figure*}
  \parbox[t]{\textwidth}{%
     \vspace{0pt}
     \includegraphics[width=\columnwidth]{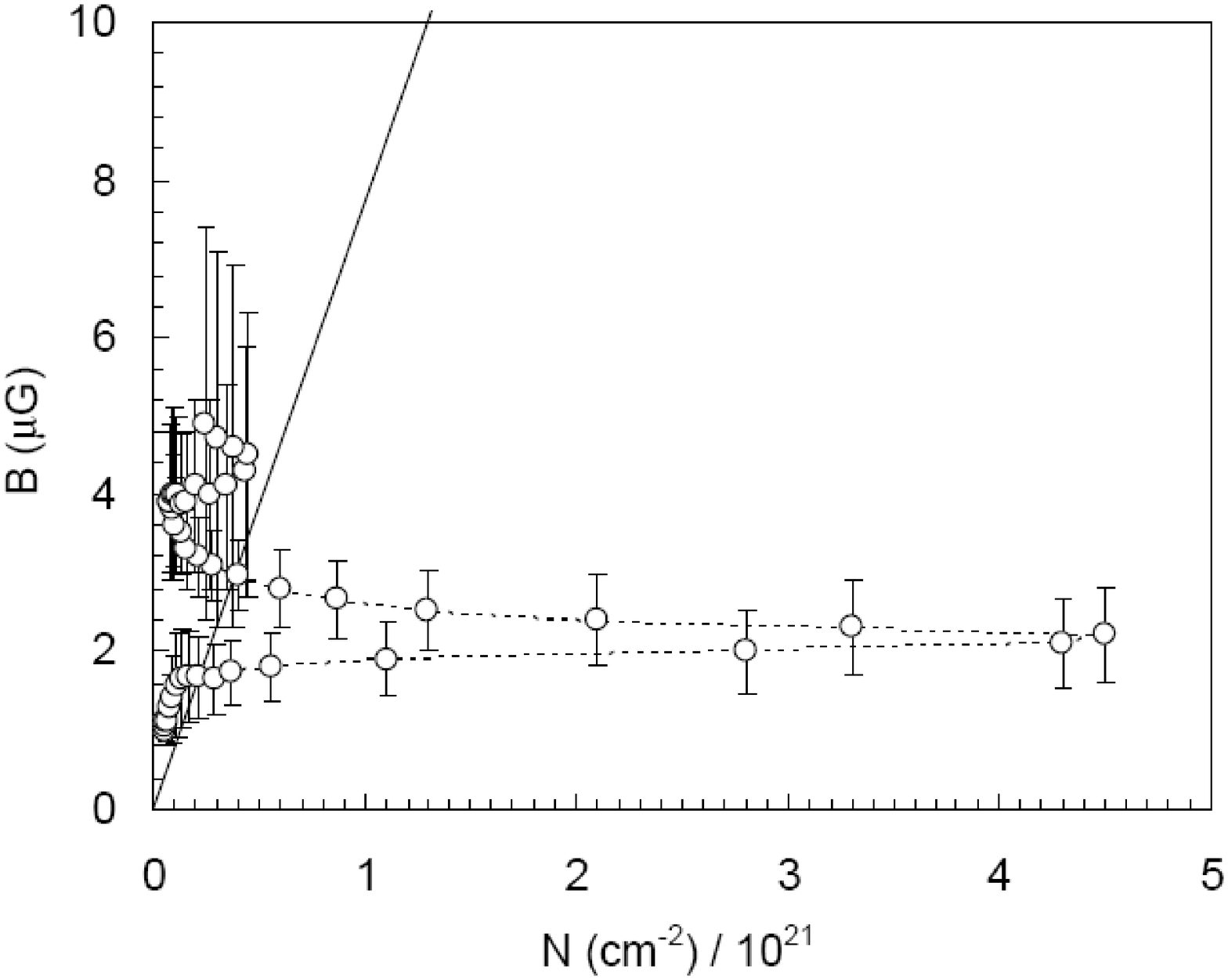}%
     \hfill%
     \includegraphics[width=\columnwidth]{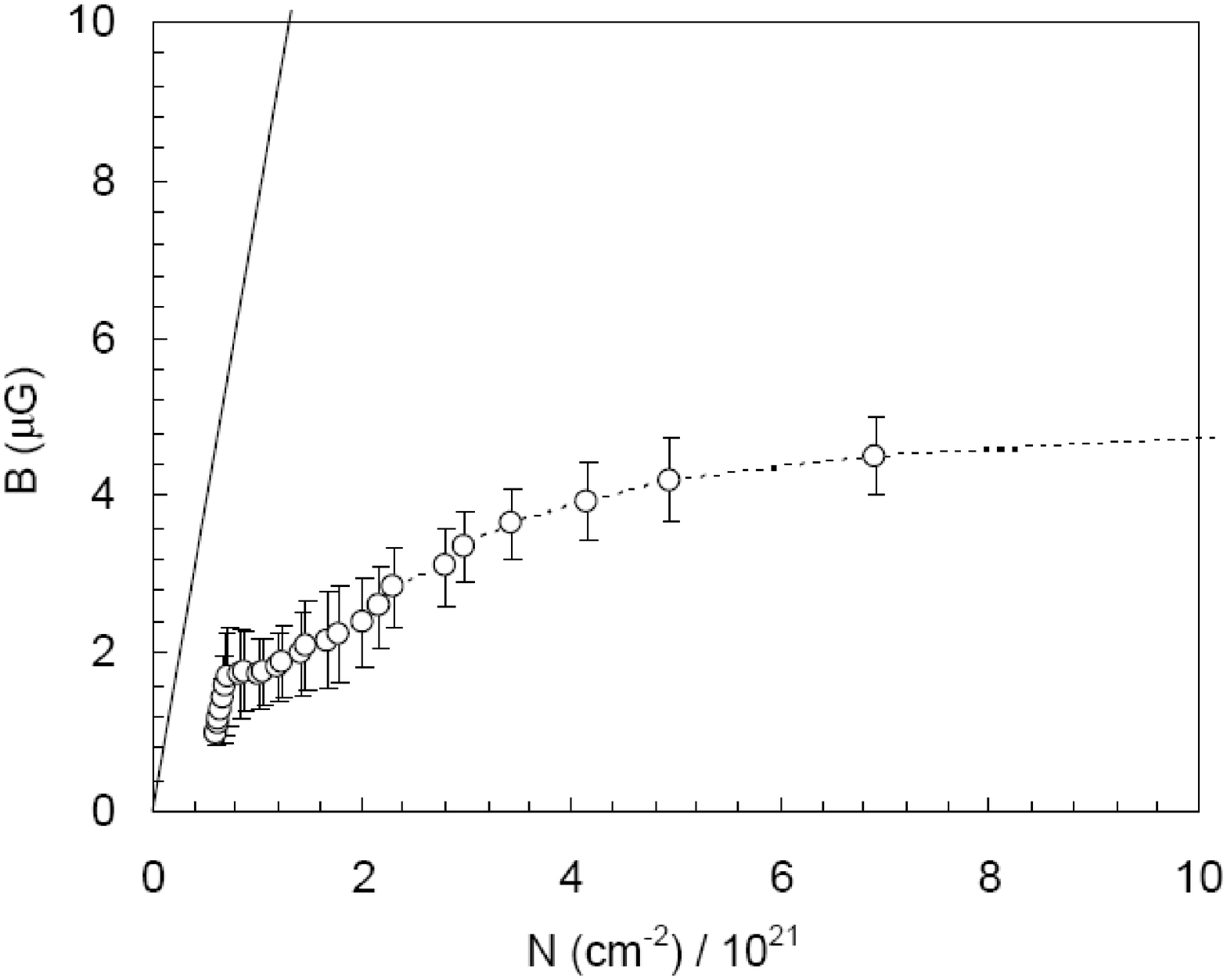}
     }
\caption{Mass-to-flux relation obtained for several time-steps of the  simulations described in the text, for cloud conditions $n=10$ cm$^{-3}$, $r_c=10$ pc,  $R_{SN}=50$ pc (left), and $n=100$ cm$^{-3}$, $r_c=10$ pc, $R_{SN}=25$ pc (right). The solid  line in each diagram gives  the critical mass to flux ratio  which separates the gravitational unstable (on the right-hand side  of the line)  from the stable domain (left-hand side of the line).}
\label{fig:masstoflux}
\end{figure*}

The other simulated models of Table \ref{tab:tabela} have presented results which are also consistent with the SF diagrams. The first model of the table is a pure hydrodynamical system with no magnetic field whose initial conditions correspond to the  cross labeled in the bottom panel of Fig. \ref{fig:diagrama}. According to Table \ref{tab:tabela}, the numerical simulation of the evolution of this SNR-cloud interaction including the effects of self-gravity leads to the gravitational  collapse of the cloud, which is in agreement with the prediction of the SF diagram. When a magnetic field of 1 $\mu$G is included in a system with the same initial conditions, the MHD numerical simulation  shows that the magnetic pressure prevents the cloud collapse  (second model of Table \ref{tab:tabela}). This model is  represented by the cross in the bottom panel of Fig. \ref{fig:diag1muG} which consistently lies outside the gravitational unstable zone of the diagram.

The third model of Table \ref{tab:tabela} also lies outside the unstable shaded zone of the SF diagram  (see the $X$ symbol in the second panel of Fig. \ref{fig:diag1muG}), and the MHD numerical simulation of the evolution of this system  shows that the cloud ends up evaporating due to the strong SNR shock interaction.

\section{Application to isolated regions of the ISM}
\label{Applications}

We can apply the simple analytical study above to isolated star formation regions of our own galaxy. In paper I, we focused on the formation of the young stellar association of $\beta-$Pictoris induced by a SNR-cloud interaction. Here, we will address  few other examples in our ISM that present some evidence of recent past interactions with SNRs, like the Large CO Shell in the direction of Cassiopea (Reynoso \& Mangum 2001) and the so called Edge Cloud 2 in the direction of Scorpious (Ruffle et al. 2007, Yasui et al. 2006). A counter example is the region apparently without star formation  around the Vela SNR. In fact, we will see below that the conditions of this region correspond to a point in our diagrams that lies outside the shaded SF zone.

\begin{figure}
    \begin{center}
        \includegraphics[width=0.95\columnwidth]{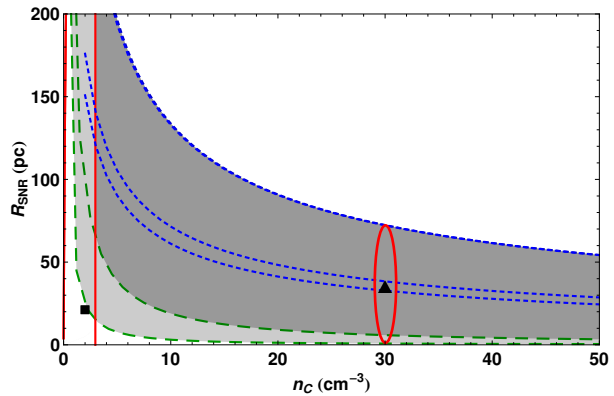}
    \end{center}
    \caption{Diagram presenting the parameter space domain for SF for a cloud interaction with a SNR  in the adiabatic regime. The light-gray shaded zone corresponds to the allowed SF zone for the interaction without magnetic field while the dark-gray corresponds to the interaction with the magnetized cloud with $B=1\;\mu$G. As in the previous figures, the dotted (blue) lines from top to bottom represent the shock penetration constraint for three different values of $\Lambda = 3\times 10^{-27} \;\rm erg\; cm^{3}\; s^{-1}$; $5\times 10^{-27} \;\rm erg\; cm^{3}\; s^{-1}$; and  $1\times 10^{-25} \; \rm erg\; cm^{3}\; s^{-1}$, respectively. The other initial conditions are  $r_c=$ 50 pc and $T_c=100\;\rm K$. {\bf The ambient medium density is $n=$ 1 cm$^{-3}$}. The triangle represents the average conditions for the interaction involving the Large CO Shell,
    {\bf centered at  $n_c=$ 30 cm$^{-3}$ and $R_{SNR}=$ 36 pc}, while the (red) ellipse indicates a possible range of values for the interacting SNR radius. The square represents the initial conditions for Vela system (see text for details).}
\label{fig:compdiag50pcpts}
\end{figure}

The Large CO shell is an expanding structure with a velocity  $\sim
3\; km/s$, a mass of $9.3 \times 10^{5}\; \rm M_{\odot}$ and a
density of $\sim 35 \; \rm cm^{-3}$. Reynoso \& Mangum (2001)
suggest that this expanding structure has probably originated from
the explosion of a SN about $\sim 4 \times 10^6$ yr ago. Assuming
that the cloud mass was originally uniformly distributed within a
sphere of radius of 435'' ($\sim 50$ pc), the initial density could
be $n_c \simeq 30\; \rm cm^{-3}$. The SN shock front possibly
induced the formation of the O 9.5 type star that has been detected
as an IR source (IRAS 17146-3723). The Large CO Shell has an
external radius of $50$ pc and an inner radius of $\sim 28$ pc
(Reynoso \& Mangum, 2001). The age and small expansion velocity
suggest that it is now a fainting evolved SNR. If we consider a
cloud with the above density and radius ($\sim $ 50 pc) at the time
of the potential interaction with SNR in the adiabatic regime, we
can identify this system in the SF diagram within the shaded zone,
as indicated in Figure \ref{fig:compdiag50pcpts}, if the SNR had a
radius  between $2.5 - 72$ pc {\bf and an ambient medium density
$n\simeq$ 1 cm$^{-3}$}. However, when we include a magnetic field in
the cloud of $1\; \mu$G, the range of possible radii for the SNR is
reduced to $R_{SNR} \sim 7.8 - 72$ pc if the maximum radius is
calculated using a radiative cooling function  $\Lambda = 3\times
10^{-27}$ erg cm$^{3}$ s$^{-1}$. For an average  $\Lambda$ $=5\times
10^{-26}$ erg cm$^{3}$ s$^{-1}$, the maximum possible radius of the
SNR is reduced to 38.4 pc. Considering that the present radius of
the evolved SNR is probably around 50 pc, the range above of initial
conditions for the SNR-cloud interaction is quite plausible.

The SNR Vela has an almost spherical, thin HI shell expanding at a velocity of  $\sim 30\; km/s$. Instead of impinging on an interstellar cloud, it is expanding in a fairly dense environment with evidence of some structure formation. Assuming that Vela is at a distance of $\sim 350$ pc from the Sun, its shell radius is of the order of 22 pc. The ambient density is $\sim$ 1 to 2 $\rm cm^{-3}$ and the initial energy of the SN was around  $1 - 2.5\times 10^{51}\;\rm erg$ (Dubner et al. 1998). These initial conditions correspond to the square symbol in the diagram of Figure \ref{fig:compdiag50pcpts} and it lies outside the SF shaded zone. This is consistent with the absence of dense clouds, clumps, filaments, or new born stars in the neighborhood of this SNR.

The SF region in the neighborhood of the Edge Cloud 2 is a possible example involving an interaction of a SNR in the radiative phase with a cloud. This is actually a giant molecular cloud with a diameter $\sim 30-40$ pc. It is one of the most distant cloud complexes from the galactic center of the Milky Way ($d\simeq$  $22-28$ kpc; Ruffle et al. 2007). For this reason, it is embedded in a region where the gas pressure is extremely small and the presence of external SF agents like spiral arm perturbations is improbable. This could be an indication that this cloud complex is very stable, except for the recent detection of two associations of T-Tauri stars with ages of $\sim 10^6$ yr (Kobayashi \& Tokunaga 2000, Yasui et al. 2006). The Edge 2 cloud has a temperature of  $20$ K, a density  $n_{H_2}\sim 10^4$ cm$^{-3}$, and an estimated mass of  $\sim 10^4\; M_{\odot}$.  There is an old and large SNR associated to this cloud,  GSH 138-01-94, which consists of an HI shell with a radius of  $180$ pc  expanding into the ISM  with a speed of  $11.8\pm 0.9$ km/s and  an age of about $4.3\times 10^6$ yr (Ruffle et al. 2007).  According to Ruffle et al. (2007), the formation of the present structure and chemical composition of the Edge 2 Cloud  is possibly a result of interactions of this SNR with the IS gas.

\begin{figure}
    \begin{center}
        \includegraphics[width=0.95\columnwidth]{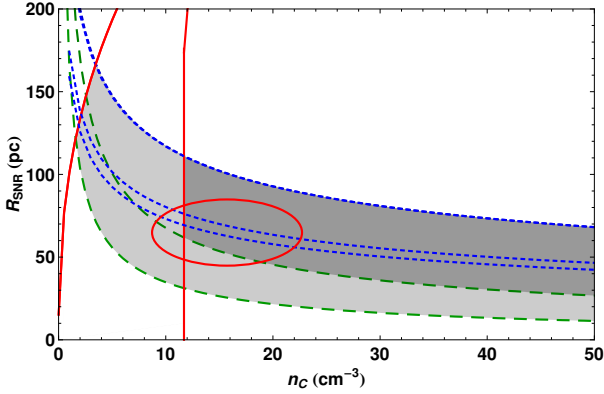}
    \end{center}
    \caption{Diagram presenting the parameter space domain for SF for a cloud interaction with a SNR  in the radiative regime expanding in an ambient medium with $n\sim 0.14$ cm$^{-3}$. The light-gray shaded zone corresponds to the allowed SF zone for the interaction without magnetic field while the dark-gray corresponds to the allowed SF zone for the interaction with the magnetized cloud with $B=1 \; \mu$G. As in the previous Figures, the dotted (blue) lines from top to bottom represent the shock penetration constraint for three different values of $\Lambda = 3\times 10^{-27} \;\rm erg\; cm^{3}\; s^{-1}$; $5\times 10^{-27} \;\rm erg\; cm^{3}\; s^{-1}$; and  $1\times 10^{-25} \; \rm erg\; cm^{3}\; s^{-1}$, respectively. The other initial conditions are $r_c = 20$ pc and $T_c=100$ K which are appropriate for the Cloud 2$-$SNR GSH 138-01-94 system. The  ellipse indicates a possible range in the parameter space for the interaction between the Edge Cloud 2 and the SNR to result in gravitational collapse. It spans  the  ranges $n_c\simeq 9.4 - 22$ cm$^{-3}$ and $R_{SNR}\simeq 46-84$ pc.}
\label{fig:compdiag20pcRadN2c}
\end{figure}

Considering the SNR characteristics above (i.e., its velocity, radius and an energy $\sim 10^{51}$ erg) and using Eq. 8 of the Paper I, we can estimate an ambient density $n\sim 0.14$ cm$^{-3}$. Also, assuming that the mass of the cloud was originally distributed uniformly in a sphere of average radius $\sim 17.5$ pc, we find that the density of the cloud before the  compression was $n_c\sim14\; \rm cm^{-3}$. With these initial conditions, the SNR-cloud interaction would lie within the allowed SF zone of the diagram of Figure \ref{fig:compdiag20pcRadN2c} for a SNR with a radius between $\sim$ 31 and 102 pc at the time of the interaction and for a value of  $\Lambda=3\times 10^{-27}$erg cm$^{3}$ s$^{-1}$. We can try to constrain further the possible value of the radius of the SNR assuming that the interaction started sometime after the remnant had become radiative and before $10^6$ yr, which is the approximate age of the observed stars. These limits imply $R_{SNR} \simeq $  $46$ pc - $84$ pc, as indicated by the region within the ellipse in the diagram of Figure \ref{fig:compdiag20pcRadN2c}.

\section{Estimating the efficiency of star formation from SNR-cloud interactions}
\label{SFE}

In the study we have carried out here, we concentrated on isolated interactions between diffuse clouds and  SNRs without focusing on the  effects that such interactions can have upon the global SF in the Galaxy. The present star formation efficiency is typically observed to be very small, of the order of a few 0.01 in dispersed regions, but it can attain a  maximum of $\sim 0.3$ in cluster-forming regions (Lada \& Lada 2003; see also Nakamura \& Li 2006 for a review).

From the present analysis, we can try to estimate the star formation efficiency that interactions between SNRs and diffuse clouds produce and compare with the observed values in order to see the contribution of this mechanism on the overall SFE in the Galaxy. The diagrams built in this work provide a SF domain for these interactions, in other words, they establish a set of conditions that isolated interactions must satisfy in order to result in successful gravitational collapse of the compressed cloud material. In order to evaluate the corresponding global SFE of these interactions we have to calculate first their probability of occurrence in the Galaxy. Considering that once a SNR is formed it will propagate and compress the diffuse medium around it providing the sort of  interactions  we are examining,  the probability of these interactions to occur must be proportional to:

\begin{eqnarray}
f_{SNR} \simeq N_{SNII}\Delta t_{SNR}(R_{SNR}) \frac{A_{SNR}(R_{SNR})}{A_G} \nonumber\\
\end{eqnarray}

\begin{equation}
\simeq 10^{-1}\ {{ N_{SNII}} \over {1.7 \times 10^{-2} \rm s}} \; {{\Delta t_{SNR}} \over {5 \times 10^5 \rm s }} \; {{A_{SNR}} \over {2.5\times 10^3 pc^2}} \; {{A_{G}} \over {4\times 10^8 pc^2}}
\label{eq:filling factor}
\end{equation}

\noindent Where we have assumed a homogeneous galactic thin disk with a radius of 20 kpc to compute the galactic area ($A_G$), and where $N_{SNII}$ is the rate of SNII explosions (e.g. Cappellaro, Evans \& Turatto 1999), $\Delta t_{SNR}(R_{SNR})$  is the lifetime and $A_{SNR}$  is the area of a SNR  and both depend on $R_{SNR}$. Since not all  the galactic volume is filled up with clouds, this quantity above must be multiplied by the diffuse neutral clouds filling factor in order to give the approximate probability of occurrence of SNR-clouds interactions. If we consider the amount of gas that is concentrated within  cloud complexes  in the cold phase of the ISM, the corresponding  volume filling factor of the clouds is $f_c \simeq 5 \%$ (e.g., de Avillez \& Breitschwerdt 2005), and then the probability of occurrence of the interactions will be given by $f_{SNR-c} \simeq f_{SNR}\times f_c$. Substituting Eq. (2) and (6) for $\Delta t_{SNR}(R_{SNR})$ of Paper I in Eq. (\ref{eq:filling factor}) we find:

\[f_{SNR-c}(R_{SNR},a)=f_{SNR}\times f_c \simeq 9.5\times 10^{-3}\;R_{SNR,50}^{9/2}\] for an interaction with a SNR in the adiabatic regime and \[f_{SNR-c}(R_{SNR},r)=2.9\times 10^{-3}\;R_{SNR,50}^{11/2}\] for an interaction with a SNR in the radiative regime.

This probability can be multiplied by the mass fraction of the shocked gas that is gravitationally unstable within the SF domain of our SNR-cloud interaction diagrams in order to give an effective global star formation efficiency for these interactions. Using the calculated Jeans mass for the shocked gas (Eqs. \ref{eq:mjeansad}, \ref{eq:mjeansrad} and \ref{eq:mJmag}) as an approximate lower limit for the mass fraction of the cloud that should collapse to form stars, we obtain:

\[sfe_{SNR-c}  \simeq f_{SNR-c} \frac{m_{J}}{m_{c}}\]

\noindent which in the case of a non-magnetized cloud interacting with a SNR in the adiabatic regime gives (according to Eq. \ref{eq:mjeansad}):

\begin{equation}
    sfe_{SNR-c,a}\simeq 5.4\times 10^{-3} \frac{T_{c,100}^2\; R_{SNR,50}^{6}}{\; n_{c,10}\;r_{c,10}^{3}\; E_{51}^{1/2}\;I_5}
    \label{eq:sfead}
\end{equation}

And for the case of the interaction of a non-magnetized cloud with a SNR in the radiative phase (using Eq. \ref{eq:mjeansrad}):

\begin{equation}
   sfe_{SNR-c,r} \simeq 1.5\times 10^{-3} \frac{T_{c,100}^{2}\;R_{SNR,50}^{6}\;n^{0.41}}{\; n_{c,10}\;r_{c,10}^{3}\;I_5\; E_{51}^{0.8}f_{10}^{1/2}}
    \label{eq:sferad}
\end{equation}

For the interaction involving a magnetized cloud we have:

\begin{equation}
    sfe_{SNR-c,B} \simeq   \frac{5\; f_{SNR-c}(R_{SNR})}{\;y^{1/2}\; n_{c,10}^{3/2}\;r_{c,10}^{3}} \left[4.14 \; T_{c,100} + \frac{4\;y\; B_{6}^2}{n_{c,10}}\right]^{3/2}
    \label{eq:sfemag}
\end{equation}

\noindent where $y$ must be substituted by Eq. \ref{eq:y}. Using Eqs. \ref{eq:machAlfad} and \ref{eq:machad} it gives the sfe for interactions with adiabatic SNRs and using Eqs. \ref{eq:machAlfrad} and \ref{eq:machrad} it gives the sfe for interactions with radiative SNRs.

\begin{figure}
    \centering
        \includegraphics[width=\columnwidth]{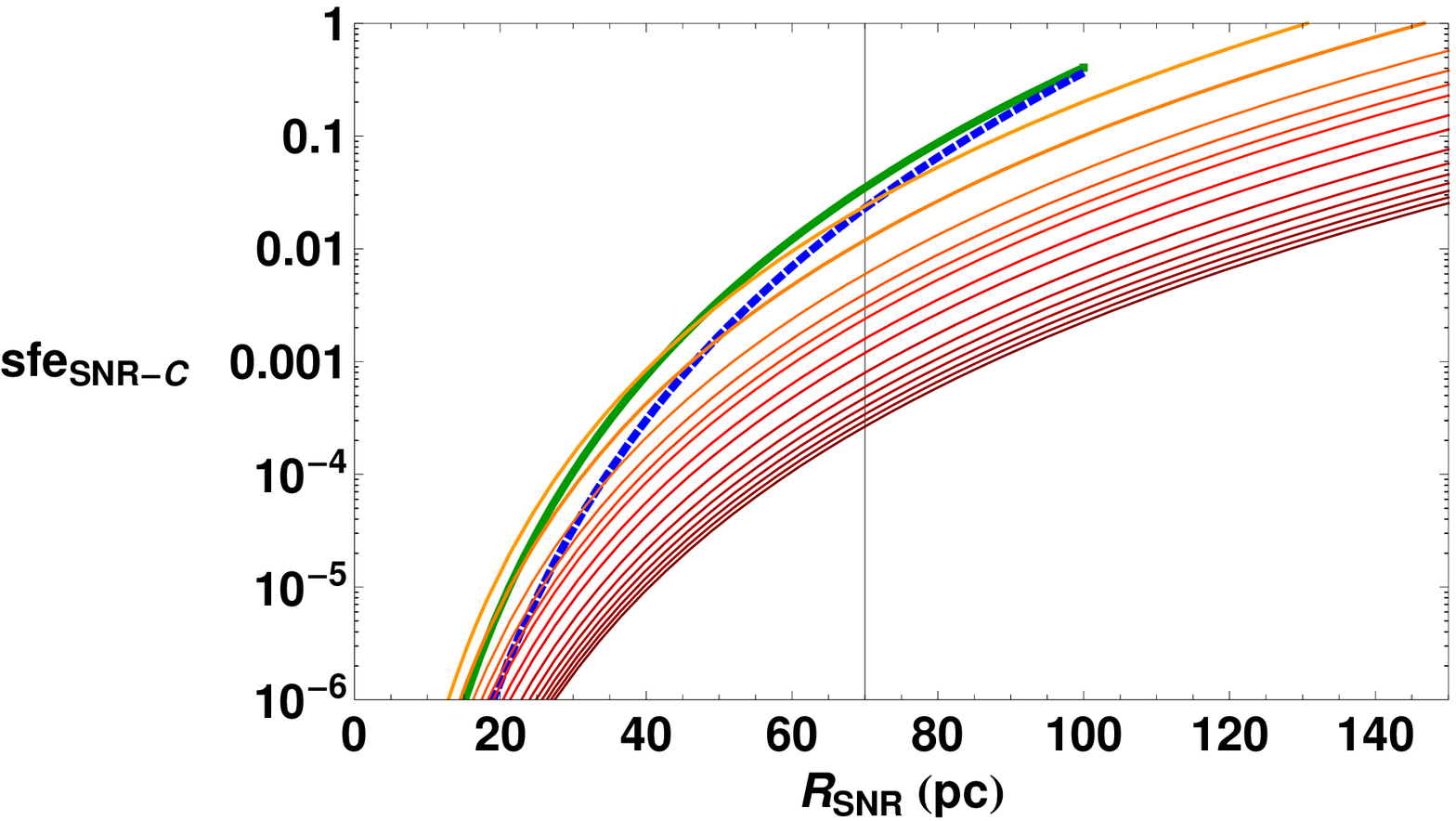}
        \includegraphics[width=\columnwidth]{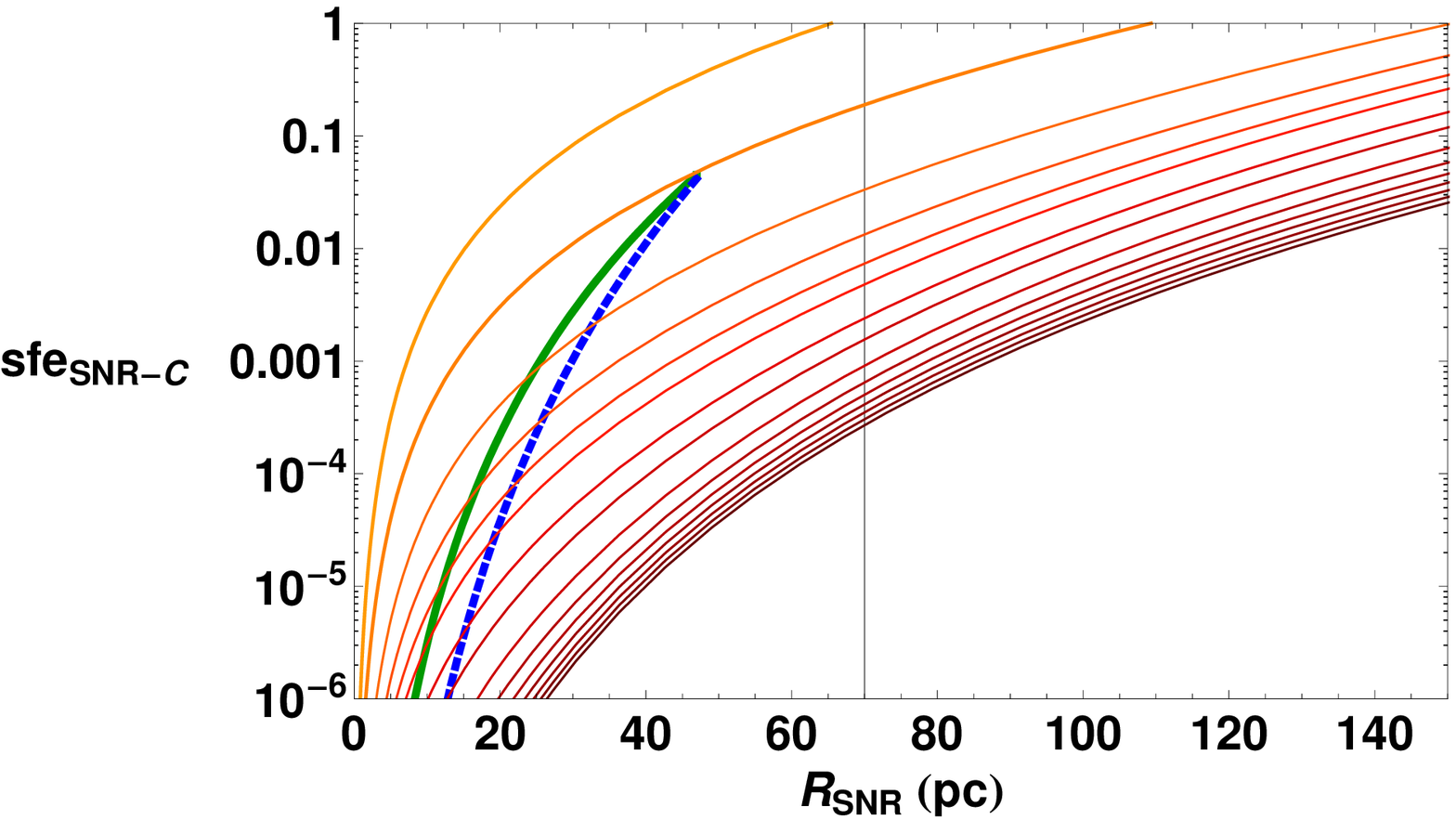}
        \includegraphics[width=\columnwidth]{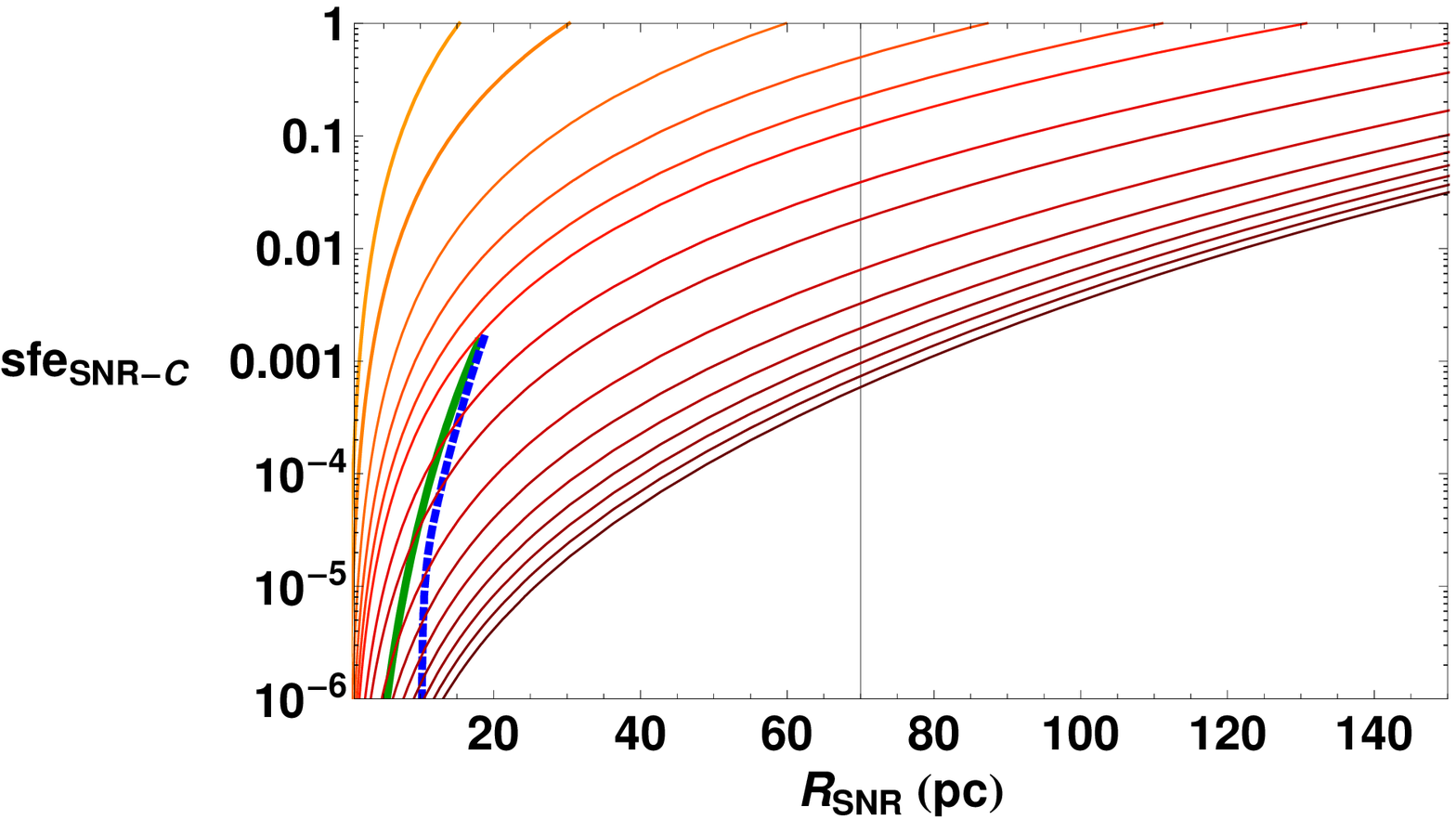}
 \caption{Plots of the calculated star formation efficiency as a function of the SNR radius in the adiabatic regime for several values of the cloud density represented by the continuous (reddish) lines (from top to bottom: 10, 20, 40, 60, 80, 100, 150, 200, 300, 400, 500, 600, 700, 800, 900 $\rm cm^{-3}$) for an interaction with a cloud with: (a) $B_c=0$ (top panel); (b) $B_c=1\;\mu$G ( middle panel); and (c) $B_c=10 \;\mu$G (bottom panel). The other initial conditions are $r_c=10\;\rm pc$, $T_c=100\;\rm K$,  and $\Lambda=5\times 10^{-26}$ erg cm$^{3}$ s$^{-1}$. {\bf The ambient medium density used is $n= 0.05$ cm$^{-3}$}. The two constraints, i.e., the non-destruction condition of the cloud (solid, green line) and the shock  penetration condition (dotted, blue line) delineate the zone for allowed SF in these sfe versus $R_{SNR}$ diagrams. The continuous vertical (black) line represents the transition radius of the SNR from the adiabatic to the radiative regime so that only the values of $R_{SNR}$ on the left-hand side of this line are relevant in these diagrams (see text for details).}
\label{fig:sfe10pc}
\end{figure}

\begin{figure}
    \centering
        \includegraphics[width=\columnwidth]{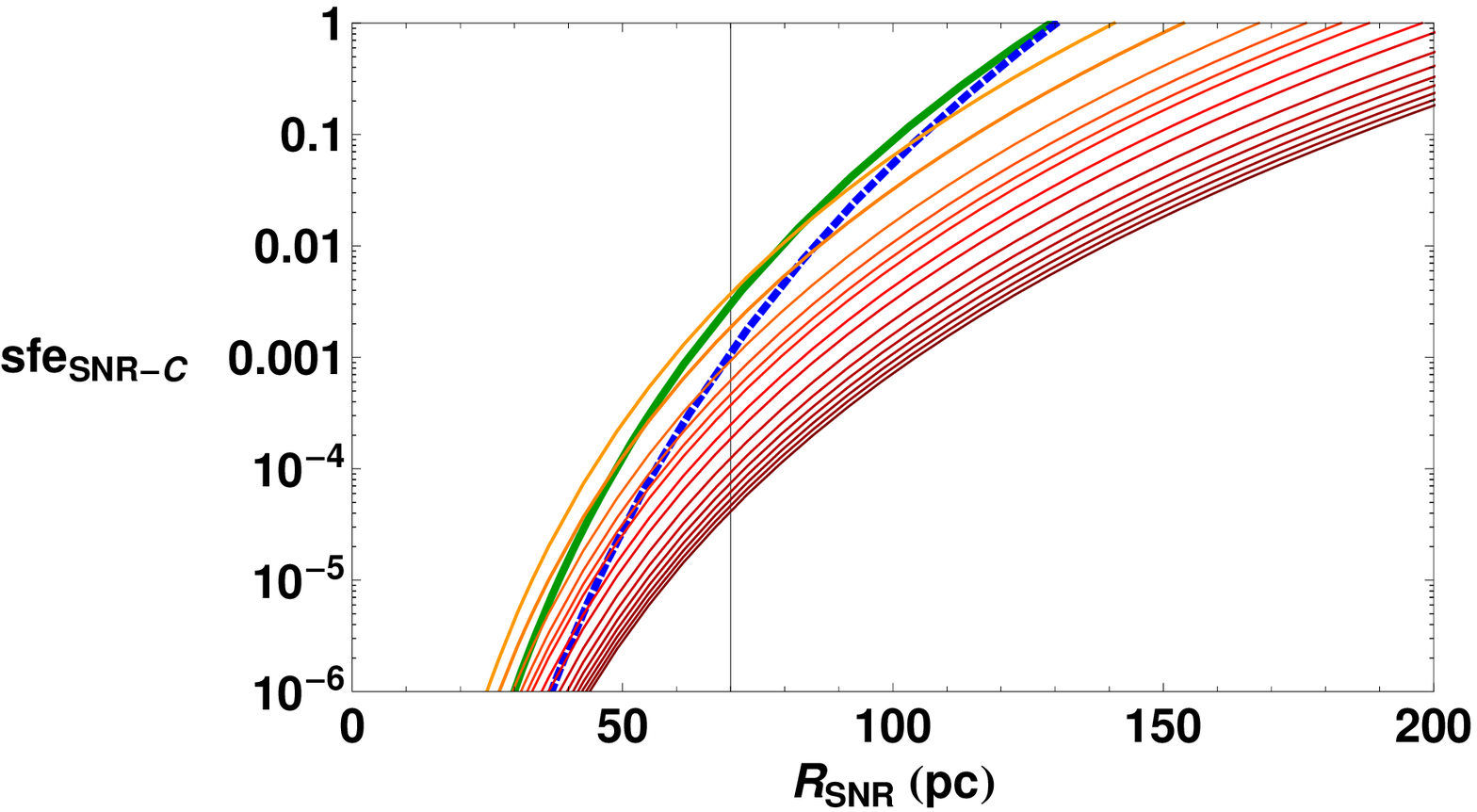}
        \includegraphics[width=\columnwidth]{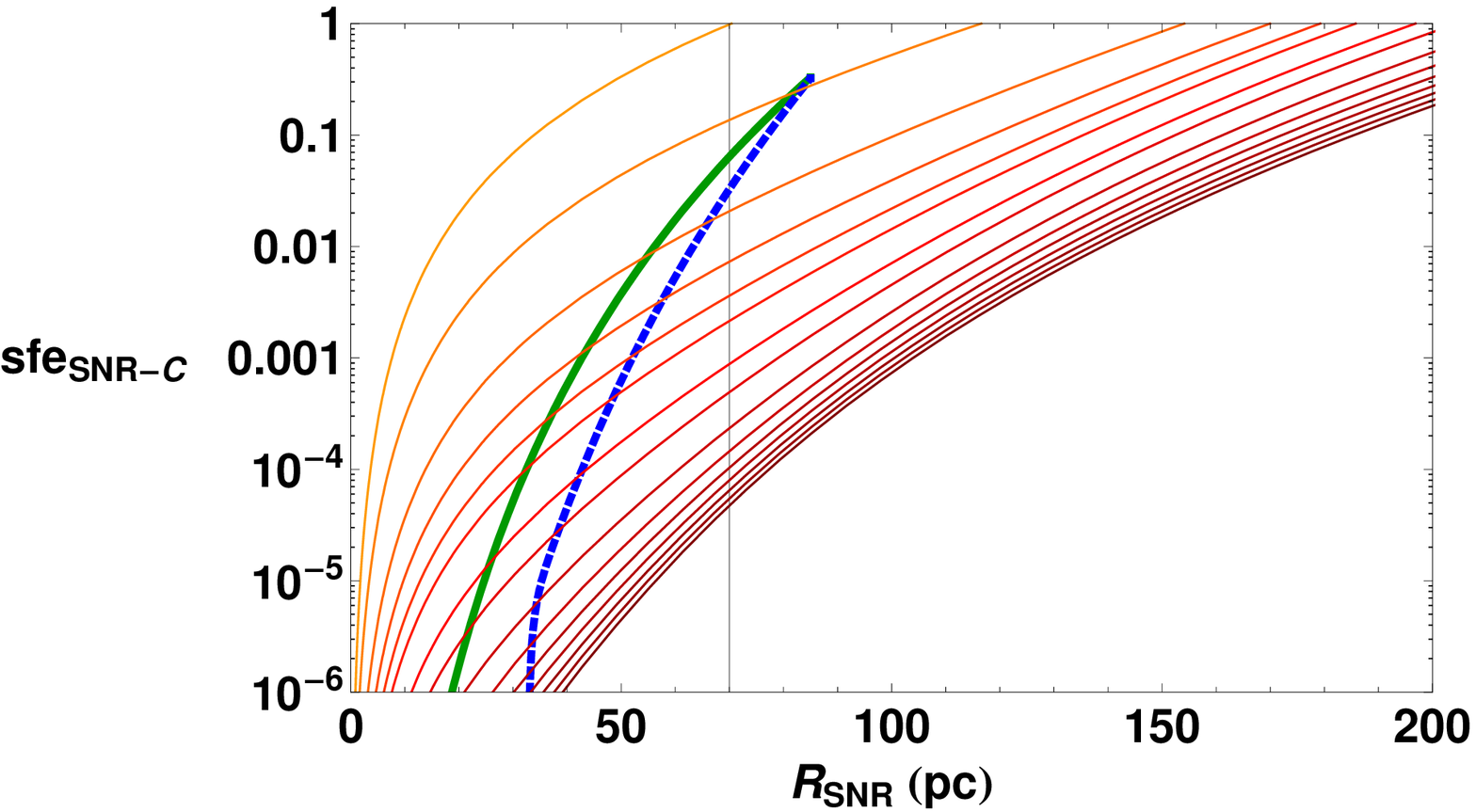}
        \includegraphics[width=\columnwidth]{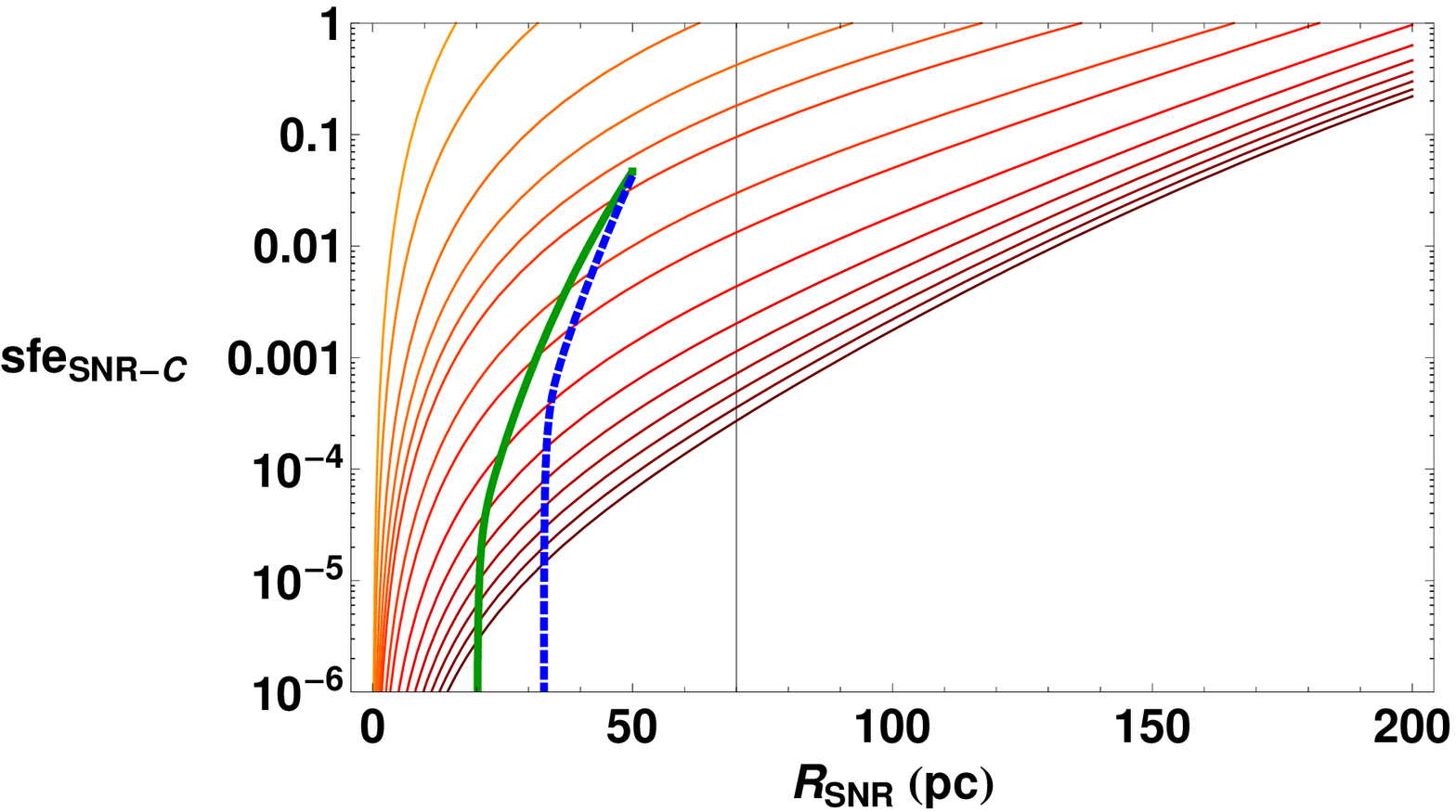}
    \caption{The same as in Figure \ref{fig:sfe10pc}, but considering the interaction with a SNR in the radiative phase. In these diagrams, only the values of $R_{SNR}$ on the right-hand side of the vertical (black) line are relevant in this phase.}
    \label{fig:sfe10pcrad}
\end{figure}

As examples, Figures \ref{fig:sfe10pc} and \ref{fig:sfe10pcrad} show plots of the approximate sfe computed for  SNR-clouds interactions as a function of the SNR radius for  different values of the cloud magnetic field and for different values of the cloud density. Since the computed sfe is already dependent of the Jeans mass constraint (see Eqs. \ref{eq:sfead} and \ref{eq:sferad}), the zone that defines the allowed SF domain in these new plots is constrained by the other two conditions over $R_{SNR}$, i.e., the shock penetration into the cloud and the cloud non-destruction conditions calculated in Section 3. The increase of the magnetic field tends to shift the SF zone to smaller values of the radius of the SNR both for SNRs in the adiabatic and in the radiative regimes.\footnote{\bf As remarked in Sec. 3.4, this shift that is imprinted in the Jeans mass constraint when considering a non-null magnetic field normal to the shock front is expected to affect only the initial compression and collapse of the shocked cloud since the later collapse will occur mostly in the direction parallel to $\vec B$. Thus, in Figs. 10 and 11, the top diagrams should be considered as the most realistic ones when comparing with the observations.} The vertical line in the diagrams establishes the transition radius of the SNR from the adiabatic (left) to the radiative (right) regime for the specific initial conditions of the diagrams. In other words, the relevant domain in the diagrams of Figure \ref{fig:sfe10pc} is on the left-hand side of this line, while in Fig. \ref{fig:sfe10pcrad}, it is on the right-hand side.

We note that the effective sfe found for these interactions is generally smaller than the typical values observed for the Galaxy and the range of possible values for sfe decreases with the increase of the magnetic field in the cloud. We also see that the parameter space that allows sfe values close to the observed values (0.01$-$0.3) is very reduced. The diagrams of Fig. \ref{fig:sfe10pcrad}, which appy to interactions involving more evolved SNRs already in the radiative phase, shows interesting results. In this phase the SNR is less powerful and therefore, less destructive than in the adiabatic phase. Besides, it is much more expanded increasing the probability of encounters with clouds. This explains the larger attained values of the calculated $sfe_{SNR-c}$. However, the parameter space that allows successful interactions for SF on the right-hand side of the diagrams is even thinner than in the interactions with adiabatic SNRs (Fig. \ref{fig:sfe10pc}) and disappears when $B_c=10\;\mu$G.

The results above suggest that these interactions are not sufficient to explain the observed sfe of the Galaxy either in the presence or in the absence of the magnetic field in the cloud. They are consistent with previous analysis performed by Joung \& Mac Low (2006) where these authors have concluded that Supernova-driven turbulence tends to inhibit global star formation rather than triggering it. We should note however, that they have based their conclusion on the computation of the star formation rate (SFR), rather than the sfe, from box simulations of the ISM with SN turbulence injection and their computed SFR has been weighed by a fixed value of the sfe taken from the observations (sfe $\sim$ 0.3).

\section{Conclusions}
\label{Conclusions}

We have presented here a study of isolated interactions between SNRs
and diffuse neutral clouds focusing on the determination of the
conditions that these interactions must satisfy in order to lead to
gravitational collapse of the shocked cloud material and to star
formation, rather than to cloud destruction. A preliminary study of
these interactions neglecting the effects of the magnetic field in
the cloud had been  performed previously by Melioli et al. (2006).
We have presently incorporated these effects and derived a new set
of conditions for the interactions. A first condition determines the
Jeans mass limit for the compressed cloud material in the impact. A
second one establishes the penetration extent of the SNR shock front
inside the cloud before being stalled due to radiative losses. It
must have energy enough  to compress as much cloud material as
possible before fainting.  A third constraint establishes the
condition upon the same shock front under which it will $not$ become
too strong to destroy the cloud completely. We have then built
diagrams of the radius of the SNR as a function of the cloud density
where this set of constraints delineate a domain within which star
formation may result from these SNR-cloud interactions (Section 3).
As expected, we find that an embedded magnetic field in the cloud
normal to the shock front with an intensity of 1 $\mu$G inhibits
slightly the domain of SF in the diagram when compared to the
non-magnetized case. The magnetic field plays a dominant role over
the Jeans constraint causing a drift of the allowed  SF zone to
higher cloud densities in the diagram. When larger intensities of
magnetic fields are considered (5-10 $\mu$G), the shrinking of the
allowed SF zone in the diagrams is much more significant. We must
emphasize however that, though observations indicate typical values
of $B_c \simeq 5-10 \mu$G for these neutral clouds, the fact that we
have assumed uniform, normal fields in the interactions have
maximized their effects against gravitational collapse. We should
thus consider as  more realistic the result obtained when an
effective $B_c \simeq 1 \mu$G was employed. These diagrams derived
from simple analytical considerations provide a useful tool for
identifying sites where star formation could be triggered by the
impact of a SN blast wave.

{\bf We have also performed fully 3D MHD
numerical simulations of the impact between a SNR and a
self-gravitating cloud for different initial conditions (Section
4) tracking the evolution of these interactions and identifying
the conditions that have led either to cloud collapse
and star formation or to complete cloud destruction and
mixing with the ambient medium.
We have found the numerical results to be consistent with those established by the SNR-cloud
density diagrams, in spite of the fact that
the later have been derived from simplified analytic theory.
We remark that the radiative cooling in the MHD numerical simulations has been
considered through the adoption of an approximate polytropic pressure equation with
$\gamma_{eff} \sim 1.2$ (see Spaans \& Silk 2000).
We are presently implementing a more realistic cooling function in our  Godunov-MHD
code in order to test, e.g.,  the re-expanding cloud case and its validity under more realistic situations.}

We have applied the results above to a few examples of regions in
the ISM with some evidence of  interactions of the sort examined in
this work. In paper I, the application of the results of the $SF$
diagram for a non-magnetized cloud (as in Figure \ref{fig:diag1muG})
to the conditions around the young stellar association of
$\beta-$Pictoris in our local ISM had led us to conclude that this
stellar association could have originated from past cloud-SNR
interaction only under very restrict conditions, i.e., with  a cloud
with radius $\sim$ 10 pc and density $\sim$ 20 cm$^{-3}$ and a SNR
with a radius $\sim$ 42 pc. However, in the present work we find
that  with the inclusion of an effective magnetic field in the cloud
with an intensity of only 1 $\mu$G this interaction is unlikely to
produce that stellar association (Figure 2, cross in the third panel
from top to bottom), at least not for the set of initial conditions
proposed in the literature for that system (see also Melioli et al.
2006). In the case of the expanding Great CO Shell$-$O9.5 star
system, we find that local star formation could have been induced in
this region if, at the time of the interaction, the SNR that
probably originated this expanding shell was still in the adiabatic
phase and a radius between $\sim$ 8 pc $-$ 29  pc impinged a
magnetized cloud with  density around 30 $ \rm cm^{-3}$ (Figure
\ref{fig:compdiag50pcpts}). Another example is the SF region near
the Edge Cloud 2. This is one of the most distant cloud complexes
from the galactic center where external perturbations should thus be
rare. But the recent detection of two young associations of T-Tauri
stars in this region could have been formed from the interaction of
a SNR in the radiative phase with a cloud,  if the interaction
started $\lesssim 10^6$ yr,  the SNR had a radius $R_{SNR} \simeq$
$46$ pc $-$ $84$ pc and the magnetized cloud a density around
$n_c\sim14\; \rm cm^{-3}$ (Figure \ref{fig:compdiag20pcRadN2c}).

Finally, though in this study we have focused on isolated
interactions involving SNRs and clouds, we  used the results of the
diagrams to  estimate the contribution of these interactions to
global star formation. Our evaluated effective star formation
efficiency for this sort of interaction is generally smaller
than the observed values in our own Galaxy (sfe $\sim$ 0.01$-$0.3)
 (Figures \ref{fig:sfe10pc} and
\ref{fig:sfe10pcrad}). This result seems to be consistent  with
previous analysis (e.g., Joung \& Mac Low 2006) and suggests that
these interactions are powerful enough to drive structure formation,
supersonic turbulence (see, e.g., simulation of Figure
\ref{fig:dmag10}) and eventually "local" star formation, but they do
not seem to be sufficient to drive $global$ star formation in our
galaxy or in other normal star forming galaxies, not even when the
magnetic field in the neutral cloud is neglected. In conclusion, the
small size of the allowed SF domain in the diagrams and the results
for the estimated sfe indicate that these interactions must lead
more frequently to the destruction of the clouds, rather than to
their gravitational collapse.

\section*{Acknowledgments}

We are indebted to the referee M.-M. Mac Low for his very useful
comments and suggestions which we believe have helped to improve
this manuscript. E.M.G.D.P., M.R.M.L., D.F.G. and F.G.G. acknowledge
financial support from grants from the Brazilian Agencies FAPESP,
CNPq and CAPES. M.R.M.L. also acknowleges R. F. Le\~ao and R. A.
Mosna for insightful discussions on this work.

\appendix

\section{}
\label{Apendice}

\begin{figure}
    \begin{center}
        \includegraphics[width=6cm]{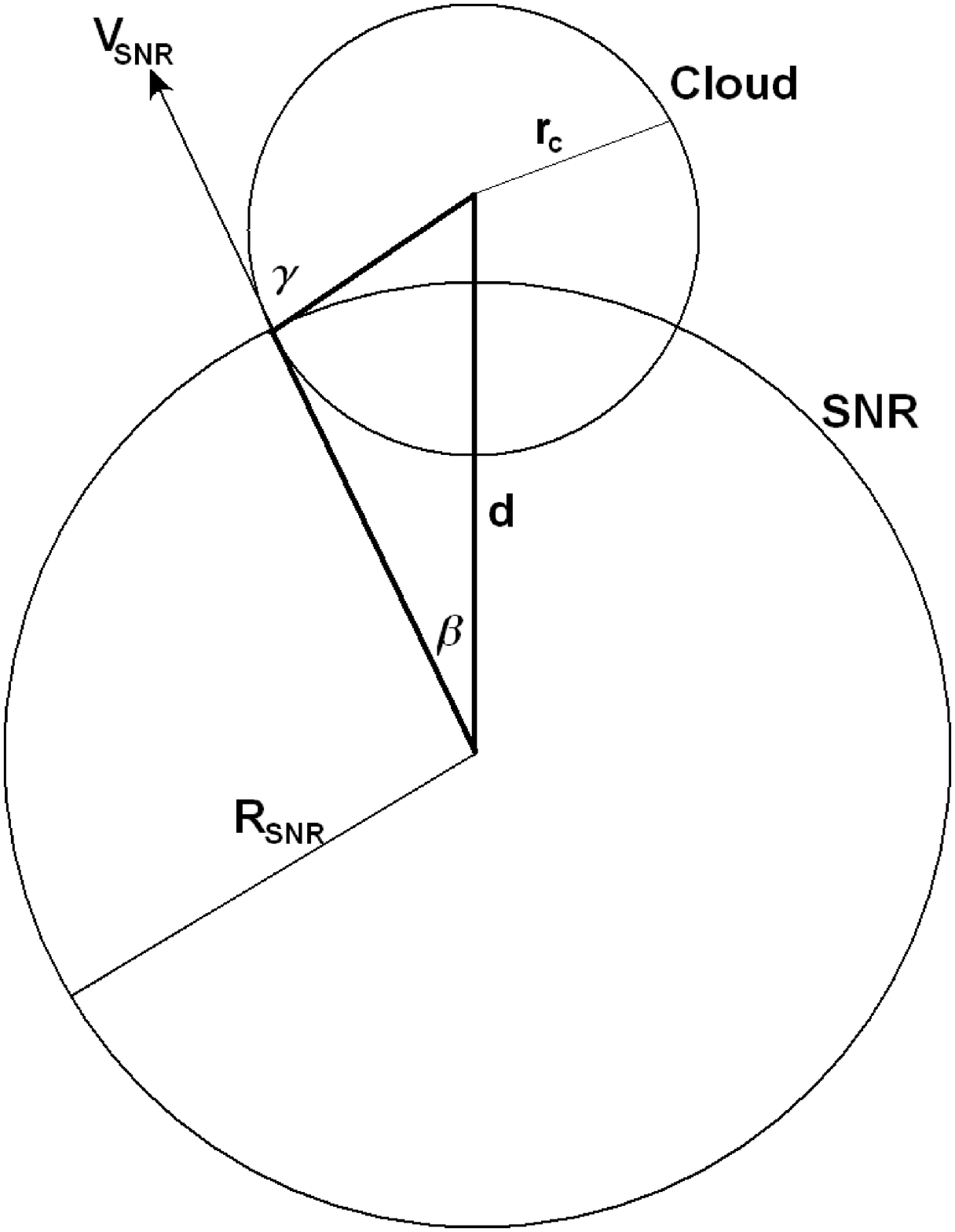}
    \end{center}
      \caption{Schematic diagram of the interaction
      between an SNR and a cloud. The SNR expands and impacts the cloud.
      The angles $\beta$ and $\gamma$ are functions of the time, the SNR
      velocity, and the cloud and SNR radii, as indicated by the equations of the text.}
    \label{fig:diagram}
\end{figure}

For spherical SNR$-$cloud interactions we need to consider the effects of curvature in the cloud-SNR interactions. The instantaneous velocity of the shocked gas moving towards the center of the cloud, $v_{cs}$, is only a fraction of the SNR velocity and depends on the density contrast, $\chi$, between the shell and the cloud (as in the planar shock case) and the angle $\gamma$ between the SNR velocity vector and the line that links the center of the cloud and the instantaneous contact point between the cloud and the SNR (Fig. \ref{fig:diagram}). We see that at $t=0$, i.e, when the SNR touches the cloud, these two lines are coincident and $\gamma = 0$, then $v_{cs} = \chi^{0.5}v_{SNR}$. {\bf Later, at a time $t_{c,SNR}$ when the SNR approaches the center of the cloud and the SNR shock energy input ends, $\gamma = \pi/2$ and $v_{cs} = 0$ \footnote{\bf We note that only for a planar shock, $\gamma$ will be equal to $\pi/2$ when the shock reaches exactly the center of the cloud. For $r_c \leq R_{SNR}$, $\gamma$ will be $= \pi/2$ a little before it reaches the center.}.} The average value of the velocity integrated over this SNR crossing time, is

\begin{eqnarray}
    {\hat{v}}_{cs} \simeq v_{SNR} \ \left({{n_{sh}} \over {n_c}}\right)^{0.5}\ {2 \over t_{c,SNR}} \ \int^{t_{c,SNR}}_0 {\cos \gamma(t) \ dt}
\label{eq:integral}
\end{eqnarray}

\noindent where
\begin{equation}
 \cos\gamma(t)\;= \frac{d^2 - R_{SNR}^2(t) - r_c^2}{2 R_{SNR}(t) \; r_c} \; ,
\end{equation}

\noindent $d$ is the distance between the SNR center and the cloud center, $R_{SNR}$ is the SNR radius, $r_c$ is the cloud radius, $n_c$ is the initial cloud density, $n_{sh}$ is the SNR shell density and $v_{SNR}$ is the expansion velocity of the SNR.

{\bf We assume that the distance between the center of the cloud and that of SNR remains constant:

\begin{equation}
 d=R_{SNR} + r_c
\end{equation}

\noindent After a time $t_{c,SNR}$, this distance can be written as

\begin{equation}
d= \sqrt{R_{SNR}(t)^2 + r_c^2}
\end{equation}

\noindent At this time, the SNR radius is approximately:
\[
R_{SNR}(t)\simeq R_{SNR}+ v_{SNR} \;\;t_{c,SNR}\;\;.\] Thus\footnote{\bf We note that in Paper I, it was assumed that the shock always reaches the center so that $t_{cs,SNR} = r_c/v_{SNR}$,  The factor 2 that appears below eq. (12) in Paper I is a typo.}: }

\begin{equation}
t_{c,SNR} \simeq \frac{\sqrt{R_{SNR}^2+2r_c\;R_{SNR}}-R_{SNR}}{v_{SNR}}\;,
\end{equation}

\noindent where \[R_{SNR}=R_{SNR}(0)\; .\] 

With the integration limits above, eq. (\ref{eq:integral}) has {\bf a solution} that should replace the {\bf more} approximate one given in Paper I.\footnote{\bf We note that the multiplying factor 2 to the integrand of eq. (\ref{eq:integral}), which is due to the fact that the total aperture angle of the SNR shock is actually $2 \beta$, rather than $\beta$, was not considered in Paper I.}


\begin{figure}
    \begin{center}
        \includegraphics[width=6.9cm]{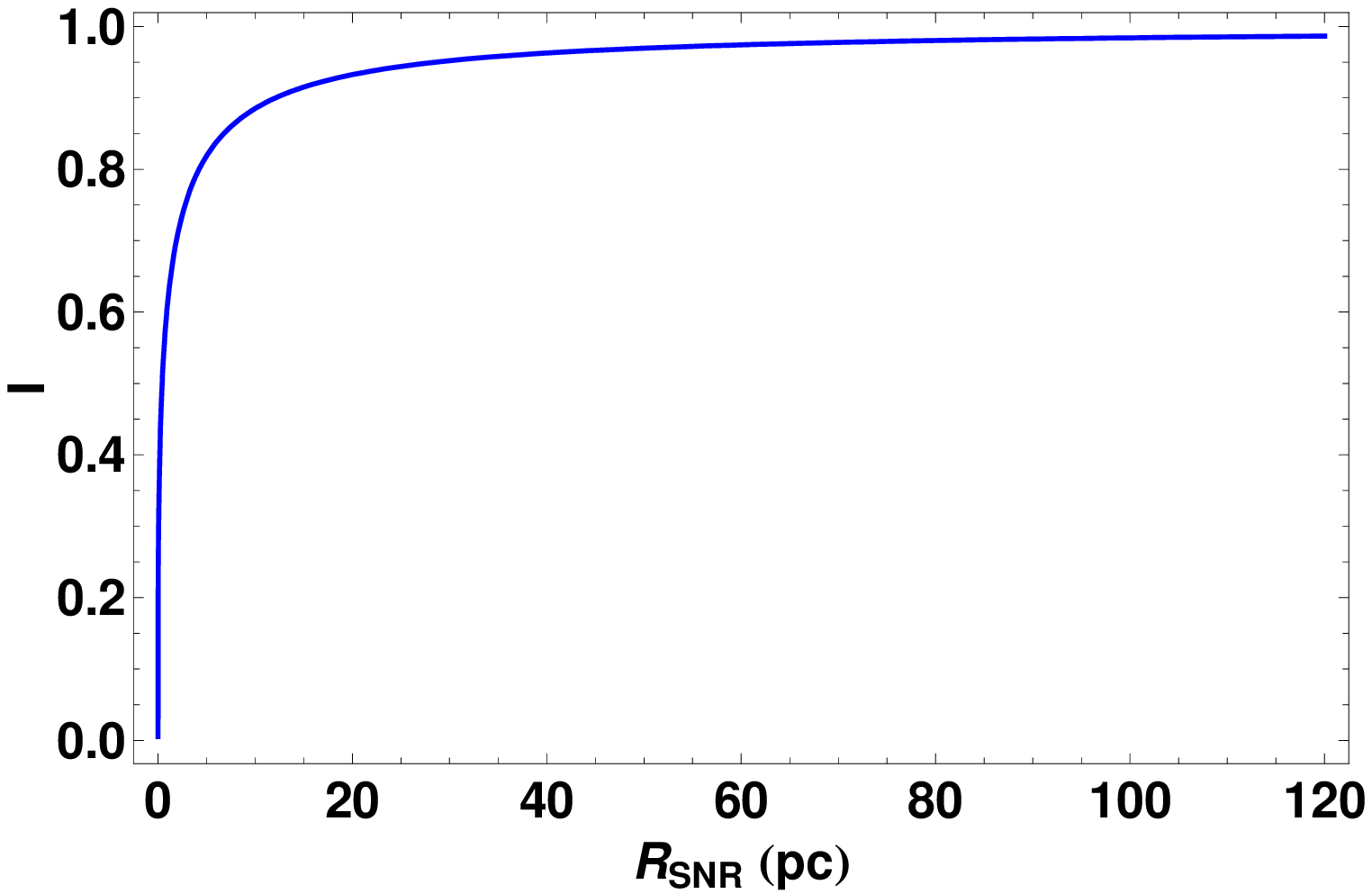}
    \end{center}
    \caption{Values of I for a cloud with $r_c=$ 10 pc and for different values of $R_{SNR}$}
    \label{fig:integral}
\end{figure}

{\bf Figure \ref{fig:integral} depicts the plot of the integral term of eq. (\ref{eq:integral}), $I= {2 \over t_{c,SNR}} \ \int^{t_{c,SNR}}_0 {\cos \gamma(t) \ dt}$, for a cloud with $r_c=10$ pc as a function of $R_{SNR}$. It clearly shows that the effect of curvature will be relevant only for values of $R_{SNR}/ r_c$ near unity}. The corrections in the solution above to $\hat{v}_{cs}$ will produce a few changes in the multiplying factors that appear in Equations (13) to (26) of Paper I (in the absence of magnetic field), as indicated below:

\begin{equation}
t_{cc,a}\sim 4.7 \times 10^5 \; \frac{n_{c,10}^{0.5}\; r_{c,10} \;
R_{SNR,50}^{1.5}}{I_5 \; E_{51}^{0.5}} \ \ \ {\rm yr}
\end{equation}

\begin{equation}
t_{cc,r} \sim 4.3 \times 10^5 \; \frac{ r_{c,10}
\;R_{SNR,50}^{2.5}\; n_{c,10}^{0.5}\; n^{0.41}}{I_5\; f_{10}^{0.5}\;
E_{51}^{0.8}} \ \ \ {\rm yr}
\end{equation}

\begin{equation}
M_a \approx 40.3\;\frac{E_{51}^{0.5}\;I_5}{T_{c,100}^{0.5}\;
R_{SNR,50}^{1.5}\; n_{c,10}^{0.5}}
\label{eq:machad}
\end{equation}

\begin{equation}
M_r \approx 44.1\;\frac{f_{10}^{0.5} \;E_{51}^{0.8}\;I_5}{n_{c,10}^{0.5}\; T_{c,100}^{0.5}\;R_{SNR,50}^{2.5}\; n^{0.41}}
\label{eq:machrad}
\end{equation}

\begin{equation}
    n_{c,sh,a} \sim \frac{1.6\times 10^4}{R_{SNR,50}^3}\; \frac{E_{51}\;I_5^2}{T_{c,100}} \ \ \ {\rm cm^{-3}}
\label{eq:densadiabatica}
\end{equation}

\begin{equation}
    n_{c,sh,r} \sim \frac{1.9\times 10^4}{R_{SNR,50}^5}\; \frac{E_{51}^{1.6}\;I_5^2\;f_{10}}{T_{c,100}\; n^{0.82}} \ \ \ {\rm  cm^{-3}}
\label{eq:densradiativa}
\end{equation}

\noindent {\bf where $t_{cc}$ is the cloud crushing time
\begin{equation}
t_{cc}=\frac{2r_c}{v_{SNR} \;I}\left(\frac{\rho_c}{\rho_{SNR}}\right)^{1/2}=\frac{2 r_c}{\hat{v}_{cs}}\;\;,
\end{equation}
\noindent i.e., the time the internal shock takes to cross the
cloud}, $M=\hat{v}_{cs}/\gamma^{1/2} c_s$ is the Mach number of the
shock into the cloud and $c_s$ is the cloud sound speed, $n_{c,sh}$
is the shocked cloud gas density, $r_{c,10}$ is the cloud radius in
units of 10 pc, $n_{c,10}$ is the unshocked cloud density in units
of 10 $\rm cm^{-3}$, $E_{51}$ is the SN energy in units of
$10^{51}\;\rm erg$, $R_{SNR,50}$ is the SNR shell radius in units of
50 pc, $f_{10}$ is the density contrast between the SNR shell and
the ISM density in units of 10, $n$ is the ambient medium density,
$T_{c,100}$ is the cloud temperature in units of 100 K, and $I_5$ is
the I factor calculated for $R_{SNR}/r_c=5$. In the equations above
the indices "a" and "r" refer to interactions involving SNRs in the
adiabatic and in the radiative phase, respectively.

\bsp

\label{lastpage}

\end{document}